\documentclass[twoside,journey]{IEEEtran}

\usepackage{makecell}
\usepackage{hyperref}
\usepackage{array}
\usepackage{float}
\usepackage{graphicx,amssymb,amsmath}
\usepackage{multicol}
\usepackage[noadjust]{cite}
\usepackage{setspace}
\usepackage{subfig}
\usepackage{graphicx}
\usepackage{url}
\usepackage{stfloats}
\usepackage{amsthm,pifont}
\usepackage{flushend}
\usepackage{cases,subeqnarray}
\usepackage{bm,multirow,bigstrut}
\usepackage{amsmath, amsthm, amssymb}
\usepackage{textcomp}
\usepackage{latexsym,bm}
\usepackage{booktabs}
\usepackage{xcolor}
\usepackage{mathtools}
\usepackage{dsfont}
\usepackage{extarrows}
\usepackage{epsfig}
\usepackage{epsfig}
\usepackage{epstopdf}
\usepackage[noend]{algpseudocode}
\usepackage{algorithmicx,algorithm}
\theoremstyle{plain}

\theoremstyle{plain}
\newtheorem{rem}{Remark}

\usepackage{amsmath}

\IEEEoverridecommandlockouts
\begin{document}
\title{Enhancing Deep Reinforcement Learning: A Tutorial on Generative Diffusion Models in Network Optimization}
\author{Hongyang~Du$^*$, Ruichen~Zhang$^*$, Yinqiu~Liu$^*$, Jiacheng~Wang$^*$, Yijing~Lin$^*$, Zonghang~Li$^*$, Dusit~Niyato,~\IEEEmembership{Fellow,~IEEE}, Jiawen~Kang, Zehui~Xiong, Shuguang~Cui,~\IEEEmembership{Fellow,~IEEE}, Bo~Ai,~\IEEEmembership{Fellow,~IEEE}, Haibo~Zhou, Dong~In~Kim,~\IEEEmembership{Fellow,~IEEE}
\thanks{H. Du is with the School of Computer Science and Engineering, the Energy Research Institute @ NTU, Interdisciplinary Graduate Program, Nanyang Technological University, Singapore (e-mail: hongyang001@e.ntu.edu.sg). R.~Zhang, Y.~Liu, J.~Wang and D.~Niyato are with the School of Computer Science and Engineering, Nanyang Technological University, Singapore (e-mail: ruichen.zhang@ntu.edu.sg, yinqiu001@e.ntu.edu.sg, jiacheng.wang@ntu.edu.sg, dniyato@ntu.edu.sg). Y. Lin is with the State Key Laboratory of Networking and Switching Technology, Beijing University of Posts and Telecommunications, China (e-mail: yjlin@bupt.edu.cn). Z. Li is with the School of Information and Communication Engineering, University of Electronic Sciences and Technology of China, Chengdu, China. (Email: lizhuestc@gmail.com). J. Kang is with the School of Automation, Guangdong University of Technology, China. (e-mail: kavinkang@gdut.edu.cn). Z. Xiong is with the Pillar of Information Systems Technology and Design, Singapore University of Technology and Design, Singapore (e-mail: zehui\_xiong@sutd.edu.sg). S. Cui is with the School of Science and Engineering (SSE) and the Future Network of Intelligence Institute (FNii), The Chinese University of Hong Kong (Shenzhen), Shenzhen, China (e-mail: shuguangcui@cuhk.edu.cn). B. Ai is with the State Key Laboratory of Rail Traffic Control and Safety, Beijing Jiaotong University, Beijing 100044, China (email: boai@bjtu.edu.cn). Haibo Zhou is with School of Electronic Science and Engineering, Nanjing University, Nanjing, Jiangsu 210093, China (email: haibozhou@nju.edu.cn). D.~I.~Kim is with the Department of Electrical and Computer Engineering, Sungkyunkwan University, Suwon 16419, South Korea (email:dikim@skku.ac.kr). $*$ means equal contribution.}
}
\maketitle
\vspace{-1cm}

\begin{abstract}
Generative Diffusion Models (GDMs) have emerged as a transformative force in the realm of Generative Artificial Intelligence (GenAI), demonstrating their versatility and efficacy across various applications. The ability to model complex data distributions and generate high-quality samples has made GDMs particularly effective in tasks such as image generation and reinforcement learning. Furthermore, their iterative nature, which involves a series of noise addition and denoising steps, is a powerful and unique approach to learning and generating data. This paper serves as a comprehensive tutorial on applying GDMs in network optimization tasks. We delve into the strengths of GDMs, emphasizing their wide applicability across various domains, such as vision, text, and audio generation. We detail how GDMs can be effectively harnessed to solve complex optimization problems inherent in networks. The paper first provides a basic background of GDMs and their applications in network optimization. This is followed by a series of case studies, showcasing the integration of GDMs with Deep Reinforcement Learning (DRL), incentive mechanism design, Semantic Communications (SemCom), Internet of Vehicles (IoV) networks, etc. These case studies underscore the practicality and efficacy of GDMs in real-world scenarios, offering insights into network design. We conclude with a discussion on potential future directions for GDM research and applications, providing major insights into how they can continue to shape the future of network optimization.
\end{abstract}


\section{Introduction}
\subsection{Background}
The emergence of Generative Artificial Intelligence (GenAI) has marked a significant milestone, offering a transformative potential that extends beyond the traditional boundaries of Artificial Intelligence (AI)~\cite{jovanovic2022generative}. Unlike conventional AI (also so-called discriminative AI) models that focus primarily on analyzing or classifying existing data, GenAI can create new data, including text, image, audio, synthetic time-series data, and more~\cite{jovanovic2022generative}. This potential of GenAI has far-reaching implications across diverse sectors, from business and science to society at large~\cite{korzynski2023generative,peres2023chatgpt}. For instance, in the business sector, GenAI can power customer service bots or generate product designs, thereby maximizing efficiency and boosting competitive advantages~\cite{van2023processgan}. According to Accenture's 2023 Technology Vision report~\cite{Accenture}, 97\% of global executives agree that GenAI will revolutionize how AI is used, enabling connections across data types and industries. In the natural science research community, GenAI can aid in generating synthetic data for research, e.g., protein sequences for disease prediction models~\cite{ni2023generative}, and accelerating the pace of discoveries~\cite{peres2023chatgpt}. Furthermore, GenAI can augment human creativity in our society, enabling the creation of new art, music, and literary work, thereby enriching our cultural heritage~\cite{srinivasan2021biases}.

GenAI is not a singular technique but a collection of various models and methods, each of which is with its unique strengths and applications. Each of these models has contributed to the advancement of AI in different ways, forming the backbone of the current GenAI landscape, in which major examples include:
\begin{itemize}
\item \textbf{Transformers:} Transformers~\cite{vaswani2017attention} have revolutionized Natural Language Generation (NLG) tasks, as exemplified by OpenAI's ChatGPT~\cite{GPTReport}. They excel in applying context, a critical aspect of language understanding, and allow for greater parallelization of computing during training and inference.
\item \textbf{Generative Adversarial Networks (GANs):} GANs~\cite{goodfellow2020generative} have been instrumental in the field of image synthesis. They consist of a generative model and a discriminative model that interact and compete against each other, leading to continuous improvement in performance.
\item \textbf{Variational Autoencoders (VAEs):} VAEs~\cite{kingma2019introduction} transform input data into a set of parameters in a latent space, which are then used to generate new data that closely aligns with the original distribution.
\item \textbf{Flow-based Generative Models:} Flow-based models~\cite{rezende2015variational} use probabilistic flows for data generation. They employ back-propagation for gradient computation, enhancing learning efficiency. Their ability to directly compute the probability density function during generation makes them computationally efficient, especially in mobile edge networks.
\item \textbf{Energy-based Generative Models:} Energy-based models~\cite{zhao2017energy} represent data using energy values. They define an energy function and optimize it to minimize the input data's energy value. These models are intuitive, flexible, and capable of capturing dependencies by associating an non-normalized probability scalar with each configuration of observed and latent variables.
\item \textbf{Generative Diffusion Models (GDMs):} Initially proposed in~\cite{sohl2015deep}, the concept of GDMs drew inspiration from the thermodynamic diffusion process. This thermodynamic correlation not only sets GDMs apart from other generative models but also establishes intriguing associations with score-based models~\cite{songscore} and stochastic differential equations~\cite{peng2019stochastic}, thereby enabling unique avenues for further research and applications.
\end{itemize}
\begin{figure}[t]
\centering
\includegraphics[width=0.45\textwidth]{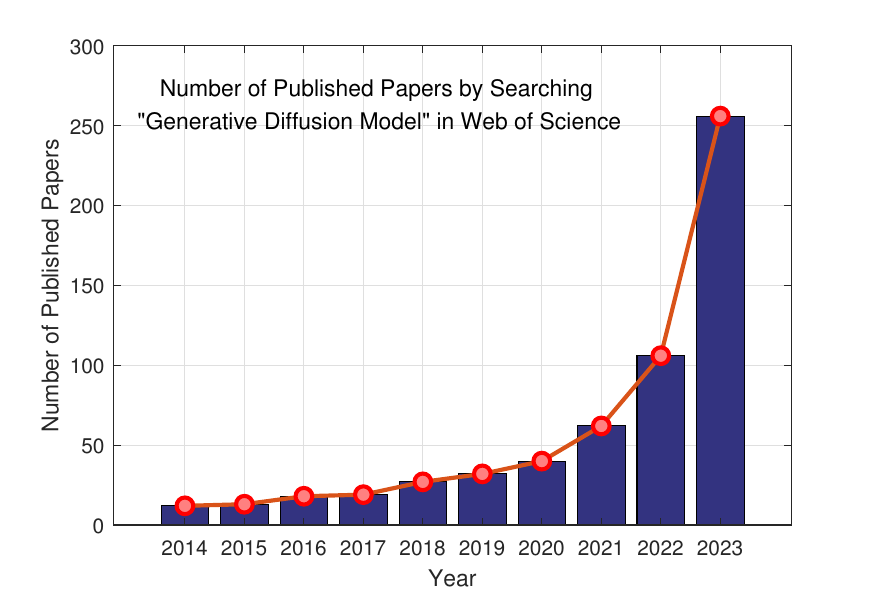}
\caption{The number of published papers by searching {\textit{"Generative Diffusion Model"}} in Web of Science (Access date: Jan-01-2024).}
\label{paperyear}
\end{figure}
Amidst these techniques, GDMs stand out due to their unique approach to data generation and their ability to model complex data distributions~\cite{cao2022survey}. As shown in Fig.~\ref{paperyear}, recently, the versatility and potency of GDMs have been demonstrated in numerous applications, particularly in AI-Generated Content (AIGC) domains. For instance, Stable Diffusion~\cite{stabdiff}, a diffusion model-based image generation application, has amassed over 10 million daily users, showcasing the practical utility and popularity of diffusion models. Furthermore, GDMs have been leveraged in various fields. In Computer Vision (CV), they have been used to generate high-quality images from noise, with models such as Denoising Diffusion Probabilistic Models (DDPM)~\cite{ho2020denoising} and Denoising Diffusion Implicit Models (DDIM)~\cite{song2020denoising}. They have also been employed in text generation tasks, enhancing the controllability and coherence of the generated text~\cite{li2022diffusion}. In the audio domain, GDMs have been used for tasks like symbolic music generation and text-to-speech conversion~\cite{mittal2021symbolic, huang2022prodiff}. Beyond traditional domains, GDMs have been utilized in graph generation~\cite{niu2020permutation, vignac2022digress,chen2023efficient}, molecular and material generation~\cite{peng2023moldiff, ketata2023diffdock, huang2022mdm}, and in synthesizing tabular data to electrocardiogram signal synthesis~\cite{lee2023codi, kotelnikov2022tabddpm, neifar2023diffecg}.

The widespread adoption of GDMs can be attributed to several key advantages over other GenAI methods.
\begin{itemize}
\item {\bf{High-quality data generation ability.}} GDMs employ a forward and reverse diffusion process~\cite{yang2022diffusion}, enabling them to accurately capture complex data distributions and embrace high-quality. This stands in contrast to GANs, which can suffer from mode collapse, and VAEs, which can yield blurry results due to their Gaussian assumption~\cite{croitoru2023diffusion}.
\item {\bf{Flexibility.}} GDMs are adaptable to various types of data and applications due to their reliance on stochastic differential equations~\cite{cao2022survey}. This flexibility is a significant advantage over Transformer-based models, which, while powerful, are primarily designed for sequence data.
\item {\bf{Simplicity of Implementation.}} GDMs' structure, featuring a fixed bottom-up path defined by a diffusion process and a top-down path parameterized by Deep Neural Networks (DNNs), simplifies their implementation~\cite{reuss2023goal,10108002}. This is a notable advantage over GANs and VAEs, which often require complex architectures and training procedures~\cite{lu2023contrastive}.
\end{itemize}
\subsection{Motivations}
The significant success of diffusion models has been demonstrated across various domains, which suggests their potential utility in optimization scenarios. Recently, the authors in~\cite{krishnamoorthy2023diffusion} introduce the Denoising Diffusion Optimization Models (DDOM), which employ an inverse mapping from function values back to input domains, utilizing the GDM's ability to refine solutions towards optimal outcomes iteratively. Meanwhile, the authors in~\cite{liu2024graph} develop the Graph Diffusion Policy Optimization (GDPO) method, integrating reinforcement learning with diffusion processes to address optimization in graph structures for non-differentiable reward signals. As shown in Table~\ref{fjlae}, these studies exemplify the expanding role of diffusion models in tackling complex problems beyond their traditional generative contexts, inspiring us to support intelligent network optimization~\cite{du2023age, du2023ai,du2023yolo,du2023user}. 
Moreover, future intelligent networks such as Integrated Sensing and Communications (ISAC)~\cite{cheng2022integrated,wang2023generative}, Semantic Communications (SemCom)~\cite{yang2022semantic,du2023generativeica}, and Internet of Vehicles (IoV)~\cite{ang2018deployment} are characterized by high-dimensional configurations, non-linear relationships, and intricate decision-making processes that are tightly linked with semantics and interpretations~\cite{zhou2023heterogeneous}. For example, SemCom networks require a deep understanding of semantic information to facilitate efficient and accurate communication~\cite{lin2023unified}, and IoV networks involve the interaction of numerous highly mobile entities with heterogeneous communication capabilities~\cite{ang2018deployment,zhou2014chaincluster}. In all these cases, they exhibit complex dynamics with significant dependencies on prior and current states and the environment, leading to high dimensional and multimodal state distributions \cite{9798257}. GDMs in this context are capable of capturing such high-dimensional and complex structures and effectively dealing with numerous decision-making processes and optimization problems, understanding and capturing the nuances of the complex trade-offs involved in the operation and optimization of intelligent networks~\cite{du2023spear}.

The roles of GDMs in optimization can be categorized into enhancing {\textit{decision making}} and {\textit{Deep Reinforcement Learning (DRL)}}. 
In decision-making scenarios, GDMs have been adopted to represent complex dynamics, incorporating additional conditioning variables such as constraints and demonstrating scalability over long time horizons~\cite{ajay2022conditional, janner2022planning}. 
Specifically, the authors in \cite{janner2022planning} introduce a diffusion probabilistic model that subsumes much of the trajectory optimization process, effectively aligning sampling with planning strategies for long-horizon and complex control settings. Meanwhile, the authors in \cite{ajay2022conditional} show return-conditional diffusion models' ability to exceed the performance of traditional offline DRL methods by modeling policies with additional variables like constraints to simplify the complexities.
In the framework of DRL, GDMs have been employed as policy representations, capturing multi-modal action distributions and improving performance in offline RL tasks~\cite{wang2022diffusion}. Furthermore, the authors in \cite{chen2022offline} pioneer a generative approach by decoupling the learned policy into a generative behavior model and an action evaluation model, utilizing GDM-based methods to model diverse behaviors and significantly enhancing the expressiveness and effectiveness of policies in offline RL scenarios.
These developments underscore GDMs' potential to innovate and enrich optimization in complex, high-dimensional spaces, setting the stage for more detailed discussions in Section~\ref{section2} and Section~\ref{section3}.

Despite the promising advantages of GDMs in network optimization, we acknowledge that GDMs also come with their own set of challenges, e.g., the computational complexity introduced by the iterative nature of GDMs. This complexity could potentially pose difficulties in large-scale DRL tasks, such as those involving the optimization of extensive communication networks~\cite{wang2023diffusion}. Additionally, GDMs might face challenges when dealing with data distributions that are characterized by high levels of noise or irregularities. This is particularly relevant in the context of real-world network traffic data~\cite{yang2022diffusion}. Nevertheless, these challenges should not overshadow the potential of GDMs in network optimization. Instead, the challenges should be viewed as areas of opportunity for further research and development. The refinement and adaptation of traditional GDMs to address these issues effectively could pave the way for significant advancements in the field of network optimization.

\renewcommand{\arraystretch}{1.1}
\begin{table*}[t]
\centering
\begin{tabular}{m{1.5cm}|m{9.5cm}|m{5cm}}
\toprule[1pt]
\hline
\textbf{Survey} & \textbf{Contributions} & \textbf{Emphasis} \\ \hline
\cite{cao2022survey} & Discuss generative diffusion models and their applications in CV, speech, bioinformatics, and NLP & \multirow{2}{*}{General review of GDMs} \\ \cline{1-2}
\cite{yang2022diffusion} & Provide an overview of diffusion models research, categorized into efficient sampling, improved likelihood estimation, and handling data with special structures & \\ \hline
\cite{kazerouni2022diffusion} & Discuss use of diffusion models for medical image analysis and various applications &  \\ \cline{1-2}
\cite{zhang2023text} & Discuss diffusion models in image generation from text and recent advancements in GenAI models & Focus on the applications of GDMs on CV \\ \cline{1-2}
\cite{ulhaq2022efficient} & Survey efficient diffusion models for vision and their applications in CV tasks &  \\ \cline{1-2}
\cite{croitoru2023diffusion} & Survey diffusion models in vision and their applications in various vision tasks &   \\ \hline
\cite{zou2023diffusion} & Provide an overview of diffusion models in NLP, discussing text generation, translation, and summarization & Focus on NLP \\ \hline
\cite{li2023diffusion} & Discuss diffusion models in non-autoregressive text generation for improving text generation efficiency & Focus on non-autoregressive text generation \\ \hline
\cite{lin2023diffusion} & Analyze the applications of diffusion models for time series data crucial in finance, weather, and healthcare & Focus on time series data \\ \hline
\cite{luo2023comprehensive} & Discuss knowledge distillation in diffusion models, transferring complex knowledge to simplify models & Focuses on knowledge distillation\\ \hline
\cite{zhang2023survey} & Focuse on using diffusion models for generating molecules, proteins, and materials in drug discovery and materials science & Focus on several specific scientific applications \\ \hline
\cite{zhang2023audio} & Discuss audio diffusion models in speech synthesis and recent advancements in GenAI models & Focus on audio and speech \\ \hline
\cite{guo2023diffusion} & Provide an overview of diffusion models in bioinformatics, including key concepts and various applications & Focus on the applications in bioinformatics \\ \hline
\cite{fan2023generative} & Present a survey on generative diffusion models on graphs, providing a state-of-the-art overview & Focus on the applications of GDMs on graphs \\ \hline
\bottomrule[1pt]
\end{tabular}
\caption{Overview of survey papers on GDMs with different applications.}
\label{tab:eafe}
\end{table*}
\renewcommand{\arraystretch}{1}
\subsection{Contributions}
The continuous advancements of GDMs in addressing optimization problems have inspired researchers to use them in specific design challenges within intelligent networks, such as optimizing incentive mechanisms~\cite{du2023ai} and selecting service providers~\cite{du2023generative}. Despite these developments, we believe that the full potential of GDMs has yet to be explored, in which GDMs are expected to revolutionize the paradigm of AI-driven intelligent network management. 
In this tutorial paper, we aim to expand the discourse within the network optimization community by presenting the application of GDMs. The value of this tutorial lies in its potential to broaden the existing toolkit for researchers and practitioners in the networking area, introducing new possibilities for integrating GDMs with traditional optimization methods.

\begin{figure*}[!t]
\centering
\includegraphics[width=0.85\textwidth]{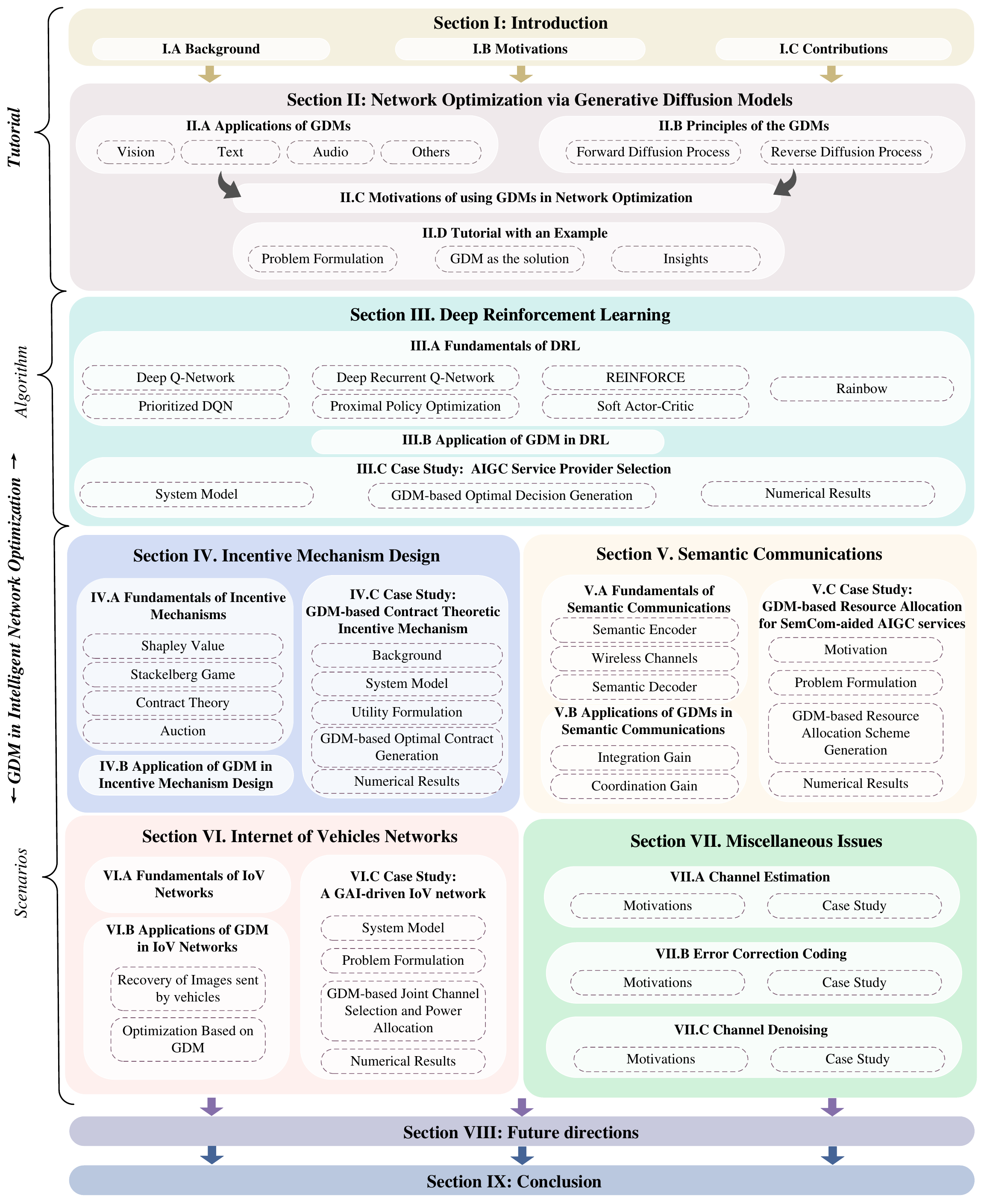}
\caption{Structure of Our Tutorial: We initiate our discussion with the foundational knowledge of GDM and the motivation behind their applications in network optimization. This is followed by exploring GDM's wide applications and fundamental principles and a comprehensive tutorial outlining the steps for using GDM in network optimization. In the context of intelligent networks, we study the impact of GDM on algorithms, e.g., DRL, and its implications for key scenarios, e.g., incentive mechanism design, SemCom, IoV networks, channel estimation, error correction coding, and channel denoising. We conclude our tutorial by discussing potential future research directions and summarizing the key contributions.}
\label{fig:frame}
\end{figure*}

While there are several surveys on GDMs, as shown in Table~\ref{tab:eafe}, these works either provide a broad overview or focus on a specific area, such as CV or Natural Language Processing (NLP), leaving a gap in the comprehensive understanding of GDMs in the context of network optimization. This tutorial bridges this gap by providing an extensive introduction to GDMs, emphasizing their applications in network optimization challenges. Crucially, we present specific case studies drawn from several significant intelligent network scenarios. The contributions of our tutorial are listed below:
\begin{itemize}
\item We provide a comprehensive tutorial on the applications of GDMs, particularly in intelligent network optimization. This tutorial aims to offer a broad understanding of the origin, development, and major strength of GDMs, and to detail how the GDMs can be effectively implemented to solve complex optimization problems in the dynamic wireless environment.
\item We provide several case studies regarding the integration of GDMs with future intelligent network scenarios, e.g., \textit{DRL}, \textit{Incentive Mechanism Design}, \textit{ISAC}, \textit{SemCom}, and \textit{IoV Networks}. These case studies demonstrate the practicality and efficacy of GDMs in emerging network technologies.
\item We discuss potential directions for GDM research and applications, providing insights into how GDMs can evolve and continue to influence future intelligent network design.
\end{itemize}
As shown in Fig.~\ref{fig:frame}, the rest of the tutorial is structured as follows: We first study the applications of GDM in network optimization in Section~\ref{section2}. The role of GDM in DRL is then explored in Section~\ref{section3}. In Section~\ref{section4}, we present GDM's role in incentive mechanism design. SemCom enhanced by GDMs are discussed in Section~\ref{section6}, and Section~\ref{section7} focuses on applying GDMs in IoV Networks. 
In Section~\ref{section77}, we discuss the applications of GDM to several other network issues, i.e., channel estimation, error correction coding, and channel denoising.
Furthermore, we outline potential research directions in Section~\ref{section8}. Section~\ref{section9} concludes this tutorial.

\section{Network Optimization via Generative Diffusion Models}\label{section2}
This section presents an overview of GDMs, their applications, principles, and extensions to facilitate network optimization. A step-by-step tutorial is provided, using a simple, yet representative, sum rate maximization problem as a demonstrative example, to illustrate the applications of GDMs in wireless environments.

\subsection{Applications of Generative Diffusion Models}
GDMs are known for their unique capabilities, theoretical robustness, and recent improvements in training and sampling efficiency, leading to their adoption in various domains~\cite{cao2022survey,yang2022diffusion}.
\subsubsection{Computer Vision}
The evolution and applications of GDMs in the field of vision have been marked by a series of interconnected advancements. Beginning with the DDPM~\cite{ho2020denoising} and DDIM~\cite{song2020denoising}, the field has shifted towards dynamic and flexible frameworks that can generate high-quality images from noise. Building on this foundation, the reflected diffusion models~\cite{lou2023reflected} integrated constraints into the generative process, leading to more faithful samples and expanding the potential applications of GDMs. This concept of flexibility and adaptability was further extended by the DiffCollage model~\cite{zhang2023diffcollage}, which demonstrated the ability of GDMs to generate large-scale content in parallel. The latent flow diffusion models~\cite{ni2023conditional} then bridged the gap between image and video generation, synthesizing optical flow sequences~\cite{enkelmann1988investigations} in the latent space to create videos with realistic spatial details and temporal motion. Furthermore, the video diffusion models~\cite{ho2022video} marked a significant milestone in generative modeling research, showcasing the potential of GDMs in generating temporally coherent, high-fidelity videos.
\subsubsection{Text}
Unlike Transformer-based models such as GPT, which focus primarily on sequence data, GDMs offer a unique advantage in their ability to model complex data distributions, making them more versatile for various tasks. Integrating language models into the diffusion process by Diffusion-LM~\cite{li2022diffusion} has enhanced the controllability and coherence of the generated text, demonstrating the adaptability of GDMs to different text generation tasks. This adaptability was further evidenced by the latent diffusion energy-based model~\cite{yu2022latent}, which introduced an energy-based model into the diffusion process, thereby improving the interpretability and quality of text modeling. The versatility of GDMs was showcased by the DiffuSeq~\cite{gong2022diffuseq} and DiffuSum~\cite{zhang2023diffusum} models, which applied GDMs to diverse tasks such as sequence-to-sequence generation and extractive summarization. Lastly, the innovative approach of the DiffusER model~\cite{reid2023diffuser} in formulating text editing as a diffusion process further expanded the scope of GDM applications, demonstrating their potential in complex text editing tasks.
\subsubsection{Audio}
GDMs have been leveraged to create a transformative shift in audio generation. The symbolic music generation model~\cite{mittal2021symbolic} demonstrated the potential of GDMs in generating complex symbolic music. The ProDiff model~\cite{huang2022prodiff} further showcases the ability of GDMs to generate high-quality text-to-speech outputs rapidly. The MM-Diffusion model~\cite{ruan2023mm} further extended the versatility of GDMs, demonstrating their capability to generate joint audio and video content. The DiffWave model~\cite{kongdiffwave} and the DiffSinger model~\cite{liu2022diffsinger} enhanced audio synthesis by generating high-fidelity waveforms and expressive singing voices, respectively. Moreover, the CRASH model~\cite{rouard2021crash} used the GDM in raw audio synthesis, demonstrating GDMs' ability to generate high-resolution percussive sounds, offering a more flexible generation capability compared to traditional methods.

\subsubsection{Others}
GDMs were also applied widely to other application domains. 
In cyber security, GDMs are both robust defense mechanisms and potential attack tools. On the defense side, GDMs offer a novel approach to safeguard against adversarial attacks and enhance privacy through differential privacy techniques~\cite{ankile2023denoising,ghalebikesabi2023differentially}. Conversely, GDMs can be manipulated for adversarial example generation and deception attacks, threatening the integrity of systems~\cite{maungmaung2023generative,blasingame2023diffusion}.
In graph generation, GDMs have been utilized to generate intricate graph structures, as demonstrated by the works in~\cite{niu2020permutation, vignac2022digress,chen2023efficient}. These models have effectively harnessed the power of GDMs to handle discrete data types, showcasing their adaptability in representing complex relationships and structures inherent in graph data. 
This adaptability extends to the field of molecular and material generation, where models like MolDiff~\cite{peng2023moldiff}, DiffDock-PP~\cite{ketata2023diffdock}, and MDM~\cite{huang2022mdm} demonstrated how GDMs can be utilized to generate intricate molecular structures, such as proteins in the field of molecular biology and material science. GDMs have shown great potential in handling heterogeneous features and synthesizing diverse tabular and time-series data types. The models presented in CoDi~\cite{lee2023codi}, TabDDPM~\cite{kotelnikov2022tabddpm}, and DiffECG~\cite{neifar2023diffecg} have demonstrated the versatility of GDMs in tasks ranging from synthesizing tabular data to ECG signal synthesis.

The exceptional performance and broad applicability of GDMs can be attributed to their unique design. This has garnered significant attention, particularly in generating diverse high-resolution images, with large-scale models such as GLIDE~\cite{nichol2022glide}, DALLE-2~\cite{dalle2}, Imagen~\cite{Imagen}, and the fully open-source Stable Diffusion~\cite{stabdiff} being developed by leading organizations like OpenAI, Nvidia, and Google. Given the widespread use and success of GDMs in the CV domain, we introduce the principles and theory of GDMs in this context in Section~\ref{2daf}. This is a foundation for our subsequent discussion on how GDMs can be extended to facilitate network optimization in Section~\ref{sfafe}.

\subsection{Principles of the GDMs}\label{2daf}
Unlike GANs that generate samples from a latent vector in a single forward pass through the Generator network~\cite{gui2021review}, GDMs utilize a denoising network to iteratively converge to an approximation of a real sample $x \sim q(x)$ over a series of estimation steps \cite{10172151}, where $q(x)$ is the data distribution. This unique design has made GDMs emerge as a powerful tool in the field of generative modeling~\cite{zhang2023text}.

\begin{figure}[t]
\centering
\includegraphics[width=0.45\textwidth]{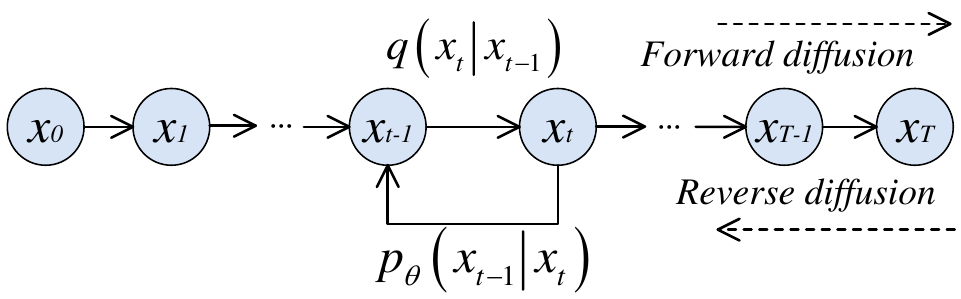}
\caption{Illustration of the forward and reverse diffusion processes. The forward diffusion process involves the addition of noise, typically Gaussian noise, to the existing training data. Subsequently, the reverse diffusion process, also referred to as ``denoising,'' aims to recover the original data from the noise-added version.}
\label{visio}
\end{figure}

As shown in Fig.~\ref{visio}, the underlying principle of GDMs is simple. With an initial input, GDMs progressively introduce Gaussian noise through a series of steps, i.e., the forward diffusion process, which generates the targets for the denoising neural network. Subsequently, the neural network is trained to reverse the noising process and recover the data and content~\cite{ho2020denoising}. The reverse diffusion process allows for the generation of new data. In the following, we show the mechanisms of forward diffusion and reverse denoising processes, utilizing an original data point ${\bm{x}}_0$, e.g., network solution or signal matrices, as our exemplar.

\subsubsection{Forward Diffusion Process}
The forward diffusion process can be modeled as a Markov chain with $T$ steps. Let ${\bf{x}}_0$ denote the original data. At each step, i.e., $t$, in the Markov chain, a Gaussian noise with a variance of $\beta_t$ is added to ${\bf{x}}_{t-1}$ to yield ${\bf{x}}_t$ with the distribution $q\left({\bf{x}}_{t}|{\bf{x}}_{t-1}\right)$. This process is represented as
\begin{equation}
q\left( {\left. {{{\bf{x}}_t}} \right|{{\bf{x}}_{t - 1}}} \right) = {\cal N}\left( {{{\bf{x}}_t};{{\bm{\mu}}_t} = \sqrt {1 - {\beta _t}} {{\bf{x}}_{t - 1}},{{\bf{\Sigma }}_t} = {\beta _t}{\bf{I}}} \right),
\end{equation}
where $q\left( {\left. {{{\bf{x}}_t}} \right|{{\bf{x}}_{t - 1}}} \right)$ is a normal distribution, characterized by the mean ${{\bm{\mu }}_t}$ and the variance ${\bf{\Sigma }}$, and ${\bf{I}}$ is the identity matrix indicating that each dimension has the same standard deviation ${\beta _t}$.

Then, from the original data ${\bf{x}}_0$ to the final ${\bf{x}}_T$, the posterior probability can be expressed in a tractable form as
\begin{equation}\label{qterni}
q\left( {\left. {{{\bf{x}}_{1:T}}} \right|{{\bf{x}}_0}} \right) = \prod\limits_{t = 1}^T q \left( {\left. {{{\bf{x}}_t}} \right|{{\bf{x}}_{t - 1}}} \right)
\end{equation}

However, according to \eqref{qterni}, sampling ${\bm x}_t$ $\left( t \in \{ 0, 1, \ldots, T \} \right)$ necessitates $t$ times of calculation, which becomes computationally intensive when $t$ is large. To avoid this, we define $ {\alpha _t} = 1 - {\beta _t} $ and $ {{\bar \alpha }_t} = \prod\limits_{j = 0}^t {{\alpha_j}} $, enabling us to express ${{{\bf{x}}_t}}$ as
\begin{align}
{{\bf{x}}_t} &= \sqrt {1 - {\beta _t}} {{\bf{x}}_{t - 1}} + \sqrt {{\beta _t}} {\bm{ \epsilon}}_{t - 1}  = \sqrt {{\alpha _t}} {{\bf{x}}_{t - 2}} + \sqrt {1 - {\alpha _t}} {{\bm{ \epsilon}}_{t - 2}}\notag\\
&=\cdots = \sqrt {{{\bar \alpha }_t}} {{\bf{x}}_0} + \sqrt {1 - {{\bar \alpha }_t}} {{\bm{ \epsilon}}_{\bf{0}}},
\end{align}
where ${{\bm{ \epsilon}}_{\bf{0}}},\ldots,{{\bm{ \epsilon}}_{\bf{t-1}}}\sim {\cal N}\left( {{\bf{0}},{\bf{I}}} \right)$. Consequently, ${\bf{x}}_t$ can be obtained using the following distribution:
\begin{equation}
{{\bf{x}}_t} \sim q\left( {{{\bf{x}}_t}\mid {{\bf{x}}_0}} \right) = {\cal N}\left( {{{\bf{x}}_t};\sqrt {{{\bar \alpha }_t}} {{\bf{x}}_0},\left( {1 - {{\bar \alpha }_t}} \right){\bf{I}}} \right).
\end{equation}
Given that $\beta_t$ is a hyperparameter, we can precompute $ {\alpha _t} $ and $ {{\bar \alpha }_t} $ for all timesteps. This allows us to sample noise at any timestep $t$ and obtain ${{\bf{x}}_t}$. Therefore, we can sample our latent variable ${{\bf{x}}_t}$ at any arbitrary timestep. The variance parameter $\beta_t$ can be fixed to a constant or chosen under a $\beta_t$-schedule~\cite{ho2020denoising} over $T$ timesteps.

\subsubsection{Reverse Diffusion Process}
When $T$ is large, $x_T$ approximates an isotropic Gaussian distribution~\cite{ho2020denoising}. If we can learn the reverse distribution $ q\left( {\left. {{{\bf{x}}_{t - 1}}} \right|{{\bf{x}}_{t}}} \right) $, we can sample ${{\bf{x}}_T}$ from $ {\cal N}\left( {{\bf{0}},{\bf{I}}} \right) $, execute the reverse process, and obtain a sample from $q\left(x_0\right)$.

However, statistical estimates of $ q\left( {\left. {{{\bf{x}}_{t - 1}}} \right|{{\bf{x}}_{t}}}\right) $ require computations involving the data distribution, which is practically intractable. Therefore, our aim is to estimate $ q\left( {\left. {{{\bf{x}}_{t - 1}}} \right|{{\bf{x}}_{t}}} \right) $ with a parameterized model $ {p_\theta } $ as follows:
\begin{equation}
p_\theta({\bf{x}}_{t-1}|{\bf{x}}_t) = \mathcal{N}\left({\bf{x}}_{t-1}; {\bm \mu}_\theta({\bf{x}}_t,t),{\bm \Sigma}_\theta({\bf{x}}_t,t)\right).
\end{equation}
Subsequently, we can obtain the trajectory from ${{\bf{x}}_T}$ to ${{\bf{x}}_0}$ as
\begin{equation}
{p_\theta }\left( {{{\bf{x}}_{0:T}}} \right) = {p_\theta }\left( {{{\bf{x}}_T}} \right)\prod\limits_{t = 1}^T {{p_\theta }} \left( {{{\bf{x}}_{t - 1}}\mid {{\bf{x}}_t}} \right).
\end{equation}
By conditioning the model on timestep $t$, it can learn to predict the Gaussian parameters, i.e., the mean ${\bm\mu}_\theta({\bf{x}}_t,t)$ and the covariance matrix ${\bm\Sigma}_\theta({\bf{x}}_t,t)$ for each timestep.

The training of the GDM involves an optimization of the negative log-likelihood of the training data. According to~\cite{ho2020denoising}, adding the condition information, e.g., ${\bm{g}}$, in the denoising process, ${p_\theta({\bf{x}}_{t-1}|{\bf{x}}_t,{\bm{g}})}$ can be modeled as a noise prediction model with the covariance matrix fixed as
\begin{equation}
\bf{\Sigma}_\theta\left(\bf{x}_t, \bm{g}, t\right)=\beta_t \bf{I},
\end{equation}
and the mean is constructed as
\begin{equation}
{{\bm{\mu }}_\theta }\left( {{{\bm{x}}_t},{\bm{g}},t} \right) = \frac{1}{{\sqrt {{\alpha_t}} }}\left( {{{\bm{x}}_t} - \frac{{{\beta_t}}}{{\sqrt {1 - {{\bar \alpha }_t}} }}{{\bm{\epsilon}} _\theta }\left( {{{\bm{x}}_t},{\bm{g}},t} \right)} \right).
\end{equation}
We first sample $ {{\bf{x}}^T}\sim\mathcal{N}({\bm{0}},{\bm{I}}) $ and then from the reverse diffusion chain parameterized by $ \theta  $ as
\begin{equation}\label{denoise}
{{\bm{x}}_{t - 1}}\mid {{\bm{x}}_t} = \frac{{{{\bm{x}}_t}}}{{\sqrt {{\alpha _t}} }} - \frac{{{\beta _t}}}{{\sqrt {{\alpha _t}\left( {1 - {{\bar \alpha }_t}} \right)} }}{{\bm{\epsilon}} _\theta }\left( {{{\bm{x}}_t},{\bm{g}},t} \right) + \sqrt {{\beta _t}} {\bm{\epsilon}},
\end{equation}
where ${\bm{ \epsilon}} \sim\mathcal{N}({\bm{0}},{\bm{I}}) $ and $t = 1,\ldots,T$. Furthermore, the authors in~\cite{ho2020denoising} introduced simplifications to the original loss function by disregarding a specific weighting term:
\begin{equation}\label{faef}
{{\cal L}_t} = {{\mathbb{E}}_{{{\bf{x}}_0},t,{\bm{ \epsilon}}}}\left[ {{{\left\| {{\bm{ \epsilon}} - {{\bm{ \epsilon}}_\theta }\left( {\sqrt {{{\bar a}_t}} {{\bf{x}}_0} + \sqrt {1 - {{\bar a}_t}} {\bm{ \epsilon}},t} \right)} \right\|}^2}} \right].
\end{equation}
This effectively shows that instead of predicting the mean of the distribution, the model predicts the noise ${\bm{\epsilon}}$ at each timestep $t$.

\subsection{Motivations of using GDMs in Network Optimization}\label{sfafe}
\renewcommand{\arraystretch}{1.1}
\begin{table*}[ht]
\centering
\begin{tabular}{m{1.2cm}|m{7cm}|m{7.5cm}}
\toprule[1pt]
\hline
\textbf{Paper} & \textbf{Key Contributions} & \textbf{Role of Diffusion Model} \\
\hline
\cite{krishnamoorthy2023diffusion} & Introduces a framework for applying diffusion models in black-box optimization scenarios. & Employs diffusion processes to generate high-quality solutions iteratively. \\
\hline
\cite{li2024diffusion} & Proposes a novel diffusion model approach for enhancing data-driven black-box optimization. & Applies diffusion models to refine solutions using data-driven insights iteratively. \\
\hline
\cite{sun2024difusco} & Develops a graph-based diffusion solver specifically tailored for combinatorial optimization problems. & Uses diffusion techniques on graphs to solve complex combinatorial tasks more efficiently. \\
\hline
\cite{liu2024graph} & Focuses on optimizing reinforcement learning policies using graph-based diffusion methods. & Integrates diffusion processes into policy learning to improve decision making in complex environments. \\
\hline
\cite{zhang2024enhancing} & Aims to improve the robustness of models against adversarial attacks. & Utilizes diffusion models to optimize and enhance model robustness. \\
\hline
\cite{giannone2024aligning} & Focuses on generating designs under constraints using diffusion models. & Aids in aligning design generation processes with optimization trajectories. \\
\hline
\cite{luo2022antigen} & Proposes a method for designing antigen-specific antibodies. & Employs diffusion models for the optimization and design of specific antibodies. \\
\hline
\cite{chen2023score} & Enhances policy optimization in reinforcement learning through diffusion behavior. & Leverages diffusion models for regularization and improvement of policy optimization. \\
\hline
\cite{zhou2024adaptive} & Presents a framework for adaptive online replanning using diffusion models. & Facilitates real-time optimization and replanning in dynamic environments. \\
\hline
\cite{xu2024stage} & Introduces a wavelet-based optimization technique for enhancing CT image reconstruction from sparse views. & Facilitates progressive image enhancement and noise reduction through iterative refinement. \\
\hline
\cite{wu2023data} & Develops a model for reconstructing high-quality CT images from ultra-sparse data. & Utilized in iterative reconstruction processes to improve image stability and quality. \\
\hline
\cite{jiang2024back} & Presents a zero-shot approach for 3D human pose estimation using diffusion models. & Enables effective optimization of pose estimation in zero-shot scenarios by leveraging generative capabilities. \\
\hline
\cite{huang2023diffusion} & Proposes methods for 3D scene generation, optimization, and planning. & Plays crucial in generating and optimizing 3D scenes for planning tasks. \\
\hline
\cite{huang2023dreamtime} & Offers an optimization strategy for converting text descriptions into 3D content. & Improves the text-to-3D conversion process by enhancing content creation and optimization. \\
\hline
\cite{urain2023se} & Introduces DiffusionFields for optimizing robotic grasp and motion planning. & Assists in learning cost functions for effective optimization of robotic tasks. \\
\hline
\cite{maze2023diffusion} & Demonstrates the superiority of diffusion models over GANs in topology optimization tasks. & Achieves more effective and efficient topology optimization. \\
\hline
\cite{park2021neural} & Introduces a framework for stochastic optimization based on controlled SDEs. & Applies diffusion models for optimizing processes in continuous-time datasets. \\
\hline
\cite{liu2023dipper} & Develops a diffusion-based path planning method for legged robots. & Utilizes diffusion models for optimizing 2D path planning tasks. \\
\hline
\bottomrule[1pt]
\end{tabular}
\caption{Summary of Papers on Diffusion Models in Optimization}
\label{fjlae}
\end{table*}

We acknowledge that diffusion models, as a type of generative learning technology, were not initially designed for optimization problems. Originally conceived for tasks such as image and audio generation, where their ability to model complex data distributions and generate high-quality samples was paramount, diffusion models have seen their potential for broader applications, as shown in Table~\ref{fjlae}. Specifically, the motivation for using GDMs in network optimization, particularly in intelligent networks, stems from their unique characteristics and capabilities.

{\textit{First, GDMs possess a robust generative capability, which is suitable in dynamic network optimization with or without expert datasets, i.e., labeled optimal solutions.}} Unlike conventional applications of GDMs, such as in image or text domains, network optimization does not typically have access to large datasets suitable for offline training~\cite{zappone2019model}. The lack of an expert dataset presents challenges when applying GDMs to facilitate network optimization. Fortunately, in addressing this challenge, the reverse diffusion process of GDMs, involving a denoising network, can be effectively utilized. Specifically, instead of relying on the standard loss function as illustrated in~\eqref{faef}, the denoising network can be trained to maximize the {\textit{value}} of the final generated solution output~\cite{du2023ai}. Here, the {\textit{value}} is related to the optimization objective function, which is designed to either maximize or minimize a specific outcome based on the given application. In network optimization, the {\textit{value}} can be a performance metric like sum rate, latency, or energy efficiency.
This training process can be achieved by executing the generated solution within the network environment, followed by network parameter adjustments based on the received feedback. Thus, the obstacle presented by the absence of a suitable dataset transmutes into an opportunity for dynamic online learning and optimization~\cite{du2023generative}. 
Notably, when expert datasets are accessible, adjustments can be made to minimize the loss between the expert and the generated solutions. These adjustments enable the GDM to continuously refine its output based on loss, leading to progressively more optimized network solutions with higher objective values.

{\textit{Second, GDMs can easily incorporate conditioning information into the denoising process.}} In intelligent networks, optimal solutions, e.g., power allocation schemes and incentive mechanism designs, typically change with the dynamic wireless environment~\cite{lin2006tutorial}. Therefore, the wireless environment information, such as path loss and small-scale fading channel parameters, can be used as the conditioning information in the denoising process~\cite{liu2023deep}. After sufficient training, the denoising network should be able to generate the optimal solution given any dynamic wireless environment condition~\cite{du2023ai}. This ability to adapt to dynamic environments and generate optimal solutions is valuable in wireless network optimization.

{\textit{Furthermore, the relationship between GDMs and DRL in intelligent network optimization is not just the substitution or competition but rather a compliment and/or supplement of each other that allows for mutual enhancement and learning.}} Specifically, training the denoising network in GDMs, which is guided by feedback from the external environment, embodies a reinforcement learning paradigm~\cite{du2023ai}. Thus, techniques such as Q-networks can facilitate more effective training of the denoising network~\cite{osband2016deep}. Moreover, GDMs can be leveraged to enhance the performance of various DRL algorithms~\cite{du2023generative}. For instance, the robust generative capabilities of GDMs can be harnessed in imitation learning, thereby augmenting the performance of offline DRL~\cite{reuss2023goal,wang2023diffusion}. In addition, GDMs can substitute the action network in DRL algorithms, where actions are treated as the output of the denoising process~\cite{wang2022diffusion}.

\subsection{Tutorial with an Example}\label{tutoriale}
\begin{figure}[!t]
\centering
\subfloat[The three orthogonal channel gains are $1$, $0.5$, and $2.5$, respectively.]{
\includegraphics[width=0.45\textwidth]{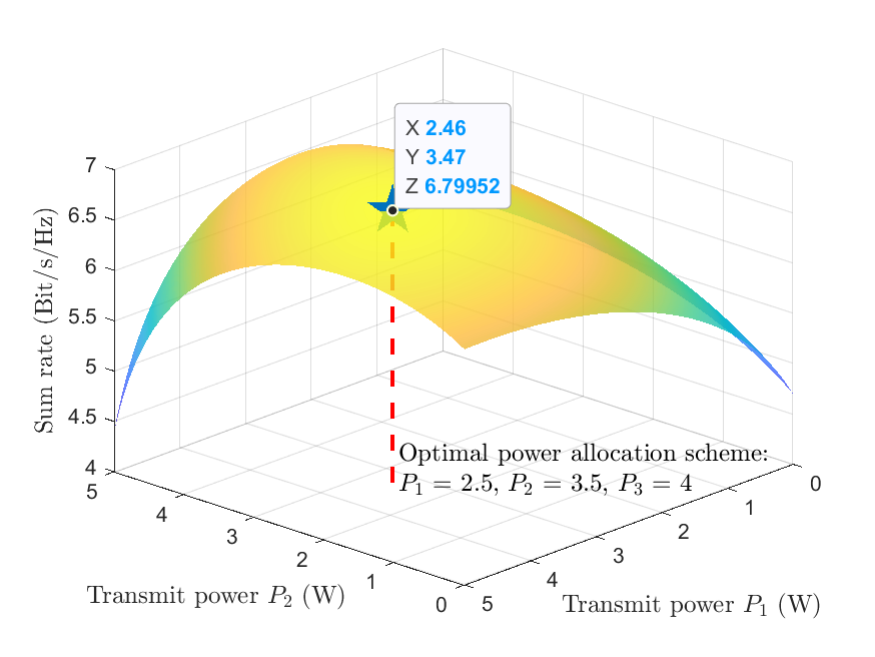}
\label{fig:figure1}
}
\hfill
\subfloat[The three orthogonal channel gains are $3$, $1$, and $3$, respectively.]{
\includegraphics[width=0.45\textwidth]{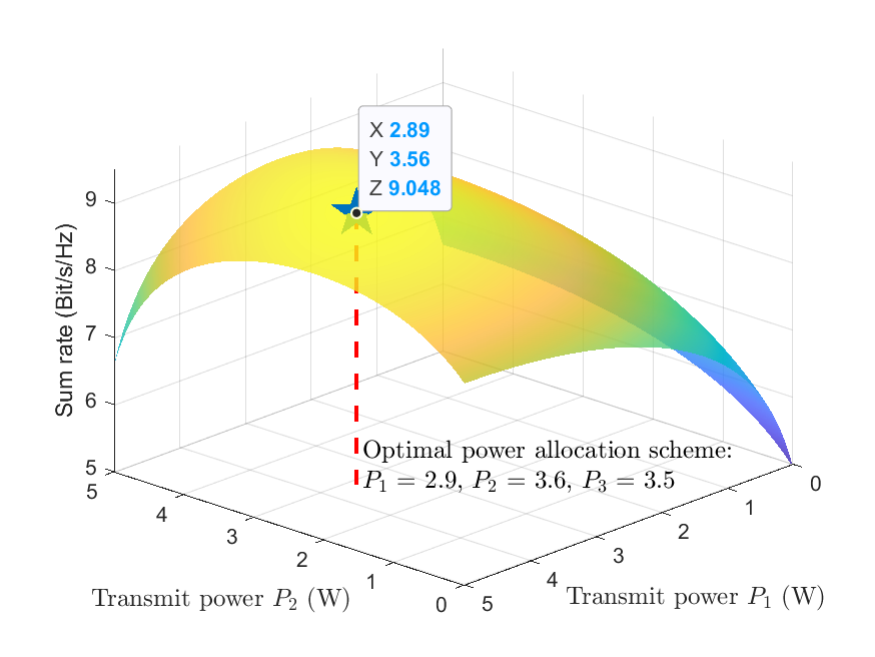}
\label{fig:figure2}
}
\hfill
\subfloat[The three orthogonal channel gains are $1$, $3$, and $1$, respectively.]{
\includegraphics[width=0.45\textwidth]{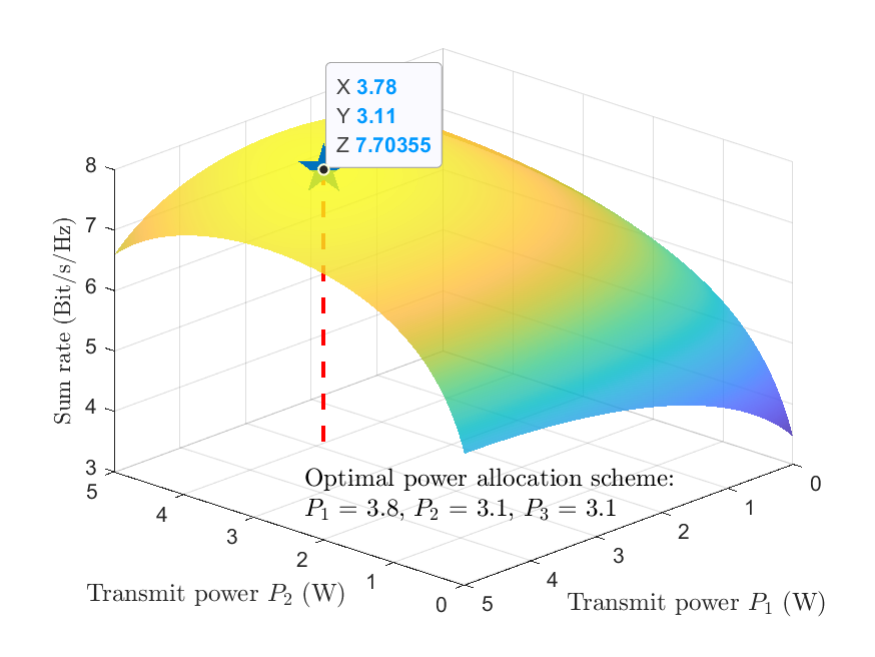}
\label{fig:figure3}
}
\caption{The sum rate values for different power allocation schemes and different channel gains with $M = 3$ and total power is 10 $\rm W$. We con observe that the optimal power allocation scheme and the corresponding peak sum rate values keep changing because of the dynamic wireless environment.}
\label{fig:three_figures}
\end{figure}
In this part, we representatively formulate an optimization problem in a wireless network and show a step-by-step tutorial to solve it by using GDMs. We compare the solutions generated by GDMs with the traditional DRL methods, such as Soft Actor-Critic (SAC)~\cite{haarnoja2018soft} and Proximal Policy Optimization (PPO)~\cite{schulman2017proximal}. The code is available at \url{https://github.com/HongyangDu/GDMOPT}.

\subsubsection{Problem Formulation}
Consider a wireless communication network where a base station with total power $P_T$ serves a set of users over multiple orthogonal channels. The objective is to maximize the sum rate of all channels by optimally allocating power among the channels. Let $g_n$ denote the channel gain for the $n^{\rm th}$ channel and $p_n$ denote the power allocated to that channel. The sum rate of all $M$ orthogonal channels is given by the sum of their individual rates~\cite{goldsmith2005wireless}, which can be expressed as
\begin{equation}
\sum_{m=1}^M \text{log}_2 \left(1 + g_m p_m / N_0\right),
\end{equation}
where $N_0$ is the noise level that can be set as $1$ without loss of generality for the analysis.  The problem is to find the power allocation scheme $\left\{ {{p_1}, \ldots,{p_M}} \right\}$ that maximizes the capacity $C$ under the power budget and the non-negativity constraints as
\begin{equation}\label{optimization}
\begin{array}{*{20}{l}}
{\mathop {\max }\limits_{\left\{ {{p_1}, \ldots ,{p_M}} \right\}} }&{C = \sum\limits_{m = 1}^M {{\rm{lo}}{{\rm{g}}_2}} \left( {1 + {g_m}{p_m}} \right)}\\
{\:\:\quad{\rm{s.t.}},}&{\left\{ \begin{array}{l}
{p_m} \ge 0,\forall m,\\
\sum\limits_{m = 1}^M {{p_m}}  \le {P_T}.
\end{array} \right.}
\end{array}
\end{equation}
The dynamic nature of the wireless environment presents a significant challenge, as the values of the channel gains, denoted as $\left\{ {{g_1}, \ldots,{g_M}} \right\}$, can fluctuate within a range. This variability is illustrated in Fig.~\ref{fig:three_figures}, which depicts the sum rate values for different power allocation schemes and channel gains when $M = 3$. It is evident that changes in channel conditions can significantly impact the optimal power allocation scheme. While various solutions have been proposed to address this issue, the following problems exist:
\begin{itemize}
\item Traditional mathematical solutions depend on accurate channel estimation~\cite{zheng2019intelligent}. However, even with precise estimation, the resources and energy consumed by pilot signals and the algorithm to perform the estimation are considerable and also introduce latency.
\item Heuristic algorithms~\cite{desale2015heuristic} can achieve near-optimal solutions; but they involve multiple iterations in the solution process, leading to increased energy consumption and additional delays.
\item The water-filling algorithm~\cite{yu2004iterative}, which can optimally solve this problem and provide an upper bound on the achievable sum rate, involves an iterative process to determine the correct number of channels for power allocation. The iteration stems from the fact that power is added to channels until the marginal increase in capacity is equal across all channels, or the power budget is consumed~\cite{yu2004iterative}. This process can be computationally intensive, particularly when dealing with a large number of channels.
\end{itemize}
Given these challenges, AI-based solutions have been proposed. For example, despite requiring a certain overhead, DRL allows for direct model deployment once training is complete. The delay in inferring an optimal solution for a given wireless environment is minimal. However, as the performance of the DRL algorithms continues to improve, the model design becomes more complex. For example, the SAC~\cite{haarnoja2018soft}, a state-of-the-art DRL method, involves five networks, including two Q-networks and their target networks and a policy network, which increases the complexity of the model.

As discussed in Section~\ref{sfafe}, GDMs are characterized by their simplicity, directness, and robustness. Furthermore, GDMs can easily incorporate the wireless environment as the condition in the denoising process, leveraging their strong generative capacity to generate optimal solutions. For example, the environmental factors such as channel gains and noise, that can influence the optimal solution can be modeled as a vector ${\bm{g}}$ in~\eqref{denoise}.


\subsubsection{GDM as the solution}
\begin{figure*}[!t]
\centering
\includegraphics[width=0.95\textwidth]{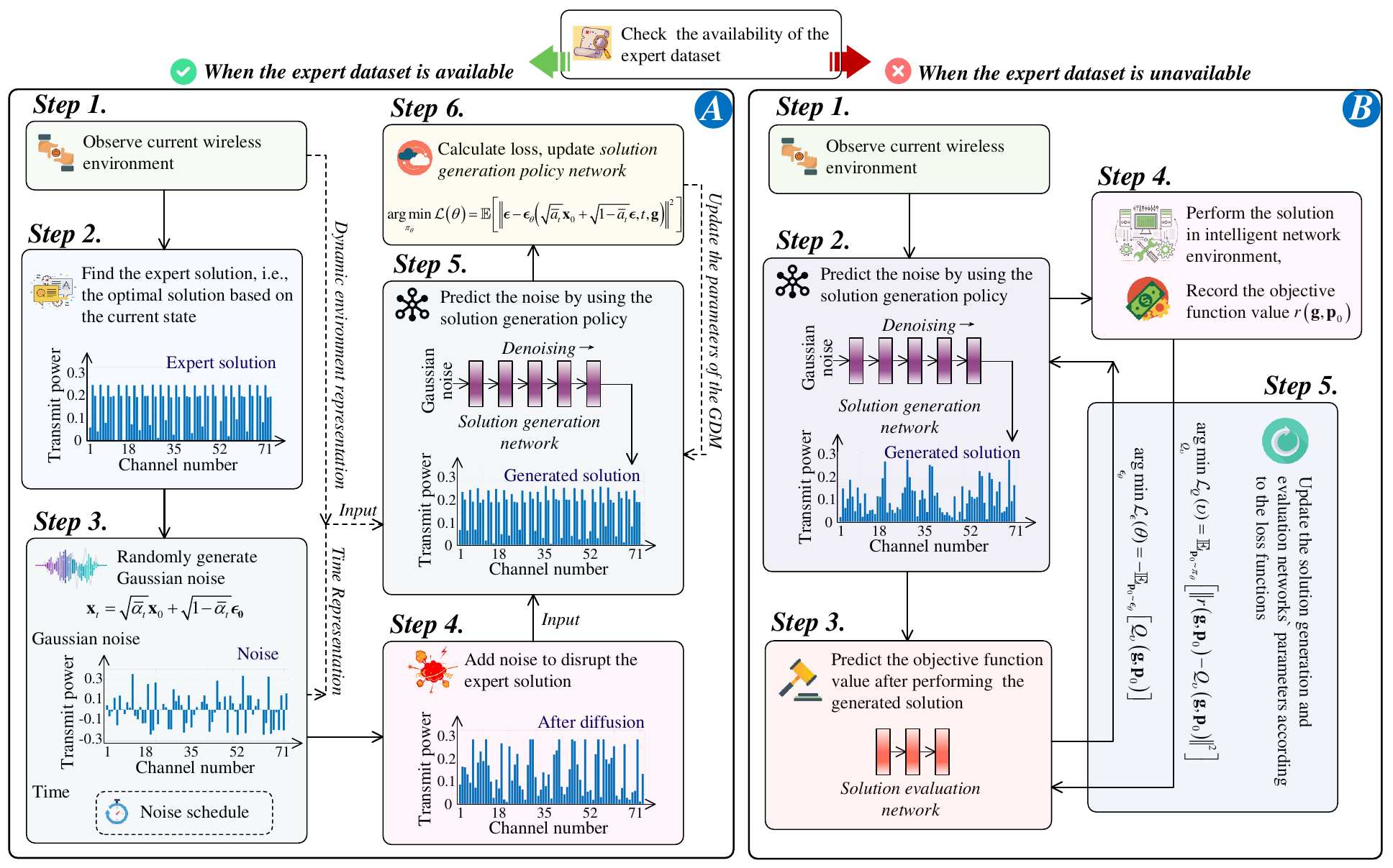}
\caption{GDM training approaches with and without an expert dataset. {\textbf{Part A}} illustrates the GDM training scenario when an expert database is accessible. The process learns from the GDM applications in the image domain: the optimal solution is retrieved from the expert database upon observing an environmental condition, followed by the GDM learning to replicate this optimal solution through forward diffusion and reverse denoising process. {\textbf{Part B}} presents the scenario where no expert database exists. In this case, GDM, with the assistance of a jointly trained solution evaluation network, learns to generate the optimal solution for a given environmental condition by actively exploring the unknown environment.}
\label{falwiou}
\end{figure*}
Next, we demonstrate how to solve the problem using GDMs. The GDM is trained to generate a power allocation scheme that maximizes the sum rate. The steps to solve the problem using diffusion models are as follows:
\begin{enumerate}
\item {\textbf{Solution Space Definition:}} The first step in wireless network optimization is to define the solution space. The AI-generated solution represents the optimal power allocation scheme that maximizes the sum rate. This scheme is generated by the GDM through a series of denoising steps applied to Gaussian noise. As shown in Algorithm~\ref{Algorithm1} line 2, in the considered problem, the dimension of the solution vector should be the number of channels, i.e., $M$. Then, it should be performed in the wireless environment, as shown in Algorithm~\ref{Algorithm1} lines 3-7.

\item {\textbf{Objective Function Definition:}} The next step is to define the objective function to be maximized or minimized. In this context, the training objective of the diffusion model is to maximize the sum rate achieved by the GDM-generated power allocation, as shown in Algorithm~\ref{Algorithm1} line 8. The upper bound can be provided by the water-filling algorithm~\cite{yu2004iterative}.

\begin{algorithm}[t]
\caption{Objective function and solution space definitions}
\label{Algorithm1}
\begin{algorithmic}[1]
\Procedure{ComputeObjective}{$env\_state, solutions$}
\State \# $solutions.dimension = M$
\State $total\_power \gets P_T$, e.g., $10$
\State $weights \gets solutions / \text{sum}(solutions)$
\State $a \gets weights * total\_power$
\State $snr \gets g\_n * a $
\State $rate \gets \text{np.log}_2(1 + snr)$
\State $value \gets \text{np.sum}(rate)$
\State \# $\text{upper}\:\text{bound:}$ $\text{water}(g\_n, total\_power)$
\State \Return $value$
\EndProcedure
\end{algorithmic}
\end{algorithm}

\item {\textbf{Dynamic Environment Definition:}} In wireless networks, the channel conditions can vary among different users, resulting in a dynamic and diverse environment. To accommodate this variability, GDM is designed to generate the optimal power allocation scheme corresponding to a given set of channel conditions. Thus, we consider a general case that each channel gains, e.g., $g_m$ $\left(m = 1,\ldots,M \right)$, change randomly over a range, e.g., $(0.5, 2.5)$, as shown in Algorithm~\ref{Algorithm2}. Note that here we consider the general case. In practice, the uniform distribution can also be replaced with a specific channel fading distribution, e.g., Rayleigh, Rician, or Nakagami-$m$. The upper and lower bounds of the channel gains can be chosen correspondingly as needed.

\begin{algorithm}[t]
\caption{Dynamic Environment Definition}
\label{Algorithm2}
\begin{algorithmic}[1]
\Procedure{GenerateState}{}
\State $env\_state \gets \text{np.zeros}(M)$
\State $env\_state[0] \gets \text{np.random.uniform}(min, max)$
\State $\cdots$
\State $env\_state[M-1] \gets \text{np.random.uniform}(min, max)$
\State \Return $env\_state$
\EndProcedure
\end{algorithmic}
\end{algorithm}

\item {\textbf{Training and Inference:}} The conditional GDM is proposed to generate the power allocation scheme. This approach diverges from back-propagation algorithms in neural networks or DRL techniques that directly optimize model parameters. Instead, GDMs strive to generate the optimal power allocation scheme by denoising the initial distribution. The power allocation scheme designed in the given environment is denoted as ${\bm{p}}$. The GDM that maps environment states to power allocation schemes is referred to as the {\textit{solution generation network}}, i.e., ${{\bm{\epsilon}}_\theta }\left( {\left. {\bm{p}} \right|{\bm{g}}} \right)$ with neural network parameters $\theta$. The objective of ${{\bm{\epsilon}}_\theta }\left( {\left. {\bm{p}} \right|{\bm{g}}} \right)$ is to output a deterministic power allocation scheme that maximizes the expected objective function values as defined in Algorithm~\ref{Algorithm1}. The {\textit{solution generation network}} is represented via the reverse process of a conditional GDM, according to~\eqref{denoise}. The end sample of the reverse chain is the final chosen power allocation scheme. According to whether the expert dataset, i.e., the optimal ${\bm{p}}$ under given ${\bm{g}}$, is available, there are two ways to train the ${\bm{\epsilon}}_\theta$:
\item[4.1)] {\textit{When there is no expert dataset:}} A {\textit{solution evaluation network}} $Q_\upsilon$ is introduced, which can assign a Q-value that represents the expected objective function to an environment-power allocation pair, i.e., ${\bm g}$ and ${\bm p}$. Here, the $Q_\upsilon$ network acts as a guidance tool for the training of the GDM network, i.e., {\textit{solution generation network}} ${\bm{\epsilon}}_\theta$. The optimal ${\bm{\epsilon}}_\theta$ is the network that generates the power allocation scheme ${\bm p}_0$ according to~\eqref{denoise} that has the highest expected Q-value. Thus, the optimal {\textit{solution generation network}} can be computed by
\begin{equation}\label{actortrain}
\mathop {\arg \min }\limits_{{{\bm{\epsilon}}_\theta }} \mathcal{L}_{\bm{\epsilon}}(\theta) = - {\mathbb{E}_{{{\bm{p}}_0}\sim{{\bm{\epsilon}}_\theta }}}\left[ {{Q_\upsilon }\left( {{\bm{g}},{{\bm{p}}_0}} \right)} \right].
\end{equation}
The training goal of the {\textit{solution evaluation network}} $Q_\upsilon$ is to minimize the difference between the predicted Q-value by the current network and the real Q-value. Thus, the optimization of $Q_\upsilon$ is
\begin{equation}\label{qualitytrain}
\mathop {\arg \min }\limits_{{Q_\upsilon }} \mathcal{L}_Q(\upsilon) = {\mathbb{E}_{{\bm{p}}_0\sim{\pi _{{\theta }}}}} \left[ {{{\left\| { {r}\!\left( {{\bm{g}},{{\bm{p}}_0}} \right)  -  {Q_{{\upsilon}}}\left( {{\bm{g}},{{\bm{p}}_0}} \right)} \right\|}^2}} \right],
\end{equation}
where $r$ denotes the objective function value when the generated power allocation scheme ${{\bm{p}}_0}$ is performed in the environment ${\bm{g}}$. Then, the network structure for training is shown in Part B of Fig~\ref{falwiou}, and the overall algorithm of GDM in sum rate maximization is given in Algorithm~\ref{sddfaeg}.
\item[4.2)] {\textit{When an expert database is available:}} In some instances of intelligent network optimization, a dataset of expert solutions might already be available. For example, applying traditional optimization schemes over time makes it feasible to obtain the optimal power allocation schemes corresponding to various channel conditions. Utilizing this expert dataset, the loss function can be designed to minimize the gap between the generated power allocation and the expert schemes as follows:
\begin{equation}
\mathop {\arg \min }\limits_{{\pi_\theta }} \mathcal{L}(\theta) = {\mathbb{E}_{{\bm{p}}_0\sim{\pi _{{\theta }}}}} \left[ {{{\left\| { {r} \left( {{\bm{g}},{{\bm{p}}_0}} \right)  - r_{\rm{exp}} \left( {{\bm{g}}} \right)} \right\|}^2}} \right],
\end{equation}
where {\small $r_{\rm{exp}} \left( {{\bm{g}}} \right)$} is the objective function value under the given ${{\bm{g}}}$. 

To achieve efficient training, we can use a similar process to that used for GDM in the image domain. Let $ {{{\bf{x}}_0}} $ denote the expert solution $r_{\rm{exp}}$. As shown in Part A of Fig~\ref{falwiou}, to train GDM by forward diffusion and inverse denoising processes, the optimization of the loss function of the GDM network can be expressed as
\begin{equation}
\mathop {\arg \min }\limits_{{\pi_\theta }} \mathcal{L}(\theta) \!=\!{{\mathbb{E}}}\!\left[ {{{\left\| { {\bm{\epsilon}} \! -\! {{\bm{\epsilon}}_\theta }\!\left( \!{\sqrt {{{\bar a}_t}} {{\bf{x}}_0} +\! \sqrt {1 \!- \!{{\bar a}_t}}{\bm{\epsilon}},t,{\bm{g}}} \right)} \right\|}^2}} \right]\!,
\end{equation}
where ${\bm{\epsilon}}$ is the added Gaussian noise, ${\sqrt {{{\bar a}_t}} {{\bf{x}}_0} +\! \sqrt {1 \!- \!{{\bar a}_t}}{\bm{\epsilon}}}$ denotes the expert solution after the forward diffusion process, and the network ${\bm{\epsilon}}_\theta$ can accurately predict the added noise with the inputs including the disrupted expert solution, the timestep information $t$, and the environment information condition $\bm{g}$.

After training, when the channel conditions change again, the GDM network ${\bm{\epsilon}}_\theta$ is capable of efficiently generating the corresponding optimal solution according to~\eqref{denoise}.
\end{enumerate}

\begin{rem}
The Algorithm~\ref{sddfaeg} is designed for scenarios where an optimal solution needs to be obtained under specific environmental conditions. However, in intelligent networking, there are many situations where the value of the objective function is not immediately obtained after executing a solution in the environment~\cite{feriani2021single,yu2019deep}. A typical example of this is the service provider selection problem, where tasks from users are allocated across various servers, each of which is with unique computing capability~\cite{du2023enabling,zhou2010novel,du2023generative}. The total utility of all users, which is designed as the objective function to be maximized, can only be calculated after a long period of the allocation process. As a result, a decision-making process, such as allocating user tasks to desired servers, has to be modeled by forming a Markov chain~\cite{ching2006markov}. In such cases, our proposed Algorithm~\ref{sddfaeg} remains useful with minor adjustments. Specifically, the reward part in Algorithm~\ref{sddfaeg} (lines 7-13) needs to be adjusted to take into account the dynamics of the Markov chain and add the discount factor in the loss function model. More details on how to do this, along with examples, are discussed in Section~\ref{section3}.
\end{rem}

\begin{rem}
In situations where expert strategies are unavailable for guidance, GDM utilizes a solution evaluation network during the training phase. This is inspired by the Q-network commonly used in DRL~\cite{xu2020service,ohira2021novel,iqbal2021double}. The solution evaluation network estimates the quality of a given solution, e.g., the power allocation scheme in the discussed example, under specific environmental conditions. This quality assessment guides the GDM during its iterative denoising process.
    Moreover, other advanced techniques from the DRL field can be adopted to make GDM training even more efficient. For example, the double Q-learning technique~\cite{hasselt2010double}, which aims at reducing over-estimation in Q-learning, can be adopted. 
    This approach maintains two Q-networks, using the smaller Q-value for updates, thus offering a conservative estimate and mitigating over-optimistic solution assessments~\cite{hasselt2010double,vimal2020energy}. Incorporating such methods can augment GDM training, promoting robustness and efficiency.
\end{rem}

\begin{algorithm}[t]
{\small \caption{GDM in Network Optimization}}
\hspace*{0.02in} {\bf{\textit{Training Phase:}}}
\begin{algorithmic}[1]
\State Input hyper-parameters: denoising step $N$, exploration noise $\epsilon$
\vspace{0.1cm}
\State \#\#{\textit{ \quad Initialize Neural Networks}}
\State Initialize solution generation network ${\bm{\varepsilon}}_\theta$ with weights $\theta$, solution evaluation network $Q_\upsilon$ with weights $\upsilon$
\vspace{0.1cm}
\State \#\#{\textit{\quad Begin Learning Process}}
\State Initialize a random process ${\mathcal{N}}$ for power allocation exploration
\While{not converge}
\State At the $j^{\rm th}$ time moment, observe the current environment ${\bm{g}}^{(j)}$, which can be simulated by using Algorithm~\ref{Algorithm2}
\State {\textbf{Set ${\bm{p}}_N$ as Gaussian noise. Generate power allocation ${\bm{p}}_0^{(j)}$ by denoising ${\bm{p}}_N$ using ${\bm{\varepsilon}}_{\theta}$}}, according to~\eqref{denoise}
\State Add the exploration noise to ${\bm{p}}_0^{(j)}$
\State Apply the generated power allocation scheme ${\bm{p}}_0^{(j)}$ to the environment and observe the objective function value by using Algorithm~\ref{Algorithm1}.
\State Record the real objective function value $r^{(j)}\left({\bm{g}}^{(j)},{\bm{p}_0^{(j)}}\right)$
\State Update the $Q_\upsilon$ according to~\eqref{qualitytrain}
\State Update the ${\bm{\varepsilon}}_\theta$ according to~\eqref{actortrain}
\EndWhile
\State \Return The trained solution generation network ${\bm{\varepsilon}}_\theta$
\end{algorithmic}
\hspace*{0.02in} {\bf{\textit{Inference Phase:}}}
\begin{algorithmic}[1]
\State Observe the environment vector ${\bm g}$
\State Generate the optimal power allocation ${\bm{p}}_0$ by denoising Gaussian noise using ${\bm{\varepsilon}}_\theta$
\State \Return The optimal power allocation ${\bm p}_0$
\label{sddfaeg}
\end{algorithmic}
\end{algorithm}

\subsubsection{Insights}
To better understand the proposed GDM method, we implemented Algorithm~\ref{sddfaeg} to solve the optimization problem in \eqref{optimization} and observed the results. We denote the sum rate obtained by performing the power allocation scheme generated by the GDM in the training process as the {\textit{test sum rate}} and use the water-filling algorithm~\cite{yu2004iterative} to obtain the upper bound, i.e., the {\textit{achievable sum rate}}. The experimental platform for running our proposed algorithms was built on a generic Ubuntu 20.04 system with an AMD Ryzen Threadripper PRO 3975WX 32-Cores CPU and an NVIDIA RTX A5000 GPU.

First, we considered a scenario with $M = 3$ channels. The channel gain values were randomly selected from $0.5$ to $2.5$. Note that the upper and lower channel gain limits here can be changed accordingly depending on the actual channel conditions. The number of denoising steps, denoted by $T$, was set to $9$. We then investigated the impact of different learning rates and $\beta$ schedulers on the algorithm's performance. 
\begin{figure}[!t]
\centering
\includegraphics[width=0.48\textwidth]{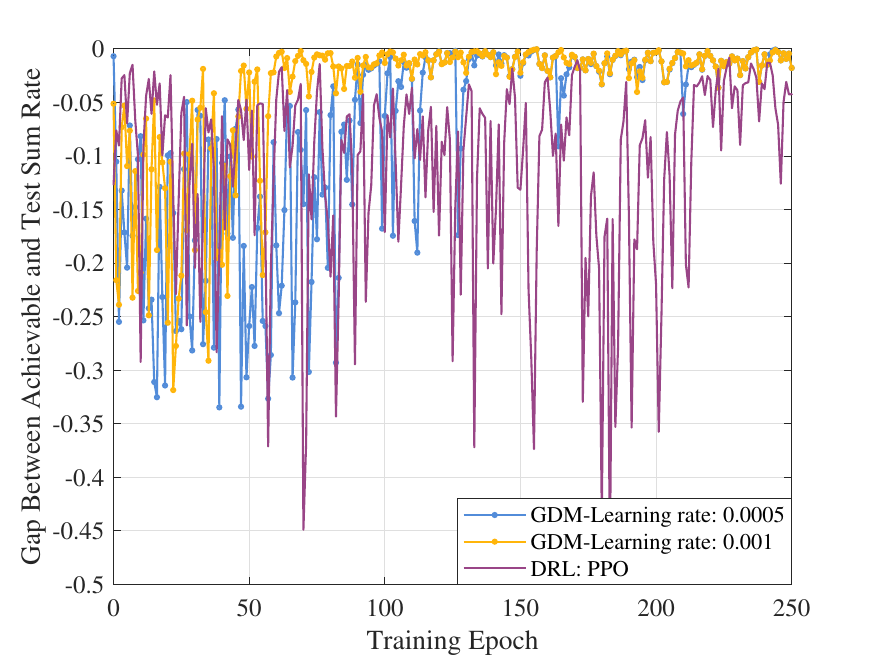}
\caption{Test reward curves of GDM-aided and DRL-aided optimization methods under different learning rate values, with the number of channels $M = 3$, and the channel gains vary within $0.5$ and $2.5$.}
\label{fig:11}
\end{figure}
\begin{figure}[!t]
\centering
\includegraphics[width=0.48\textwidth]{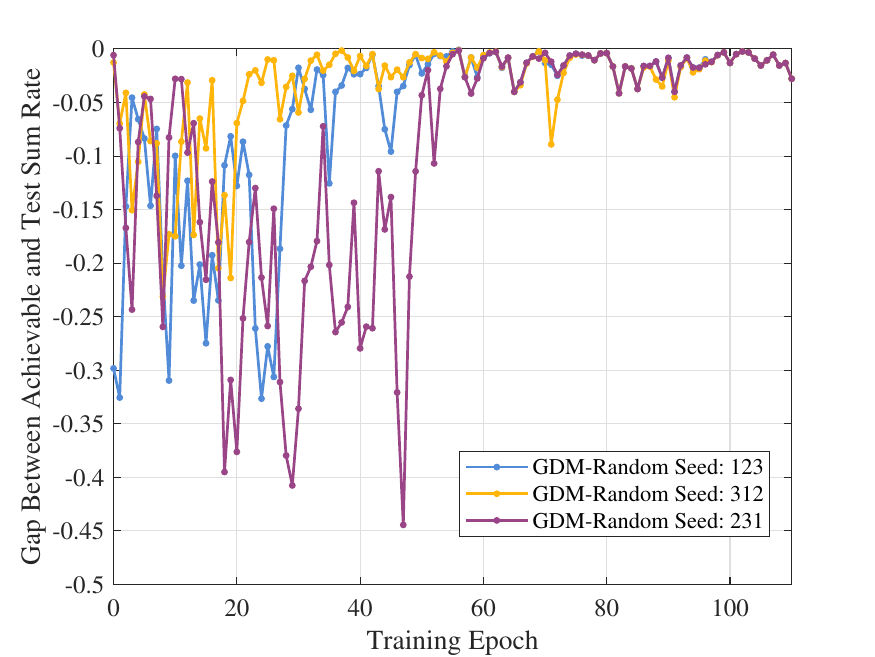}
\caption{Test reward curves of GDM-aided optimization methods under different random seed values, with the number of channels $M = 3$, and the channel gains vary within $0.5$ and $2.5$.}
\label{fig:12}
\end{figure}

Figure~\ref{fig:11} illustrates the gap between achievable and test sum rates against the training epoch. We observe that the conventional DRL method, i.e., PPO, exhibits more significant fluctuations and less effective convergence. The challenges are from the problem's inherent complexity, the environmental variability, or the influence of specific hyperparameters. However, despite these challenges, both GDM methods outperform the PPO method, irrespective of their learning rates. In the first case, GDM with a learning rate of 0.001 achieves rapid convergence to zero, taking approximately 48 seconds across 60 epochs, underscoring the method's efficiency. Conversely, with a learning rate of 0.0005, GDM converges more slowly yet effectively reaches zero, requiring about 104 seconds over 130 epochs, reflecting a steadier learning trajectory due to smaller adjustments per iteration. These variations in learning times directly depend on the chosen learning rates, with faster rates enabling quicker learning at the potential cost of overshooting minima. Furthermore, it is pertinent to note the correlation between dataset size and learning dynamics. While not explicitly analyzed in this context, the number of epochs typically reflects the dataset's size, with more extensive datasets requiring {\color{blue}more} epochs to achieve thorough learning.
This superior performance manifests the GDM's ability to capture complex patterns and relationships between observations, leading to more accurate action decisions. This ability is advantageous in network optimization problems requiring high-performance, time-efficient, fast-converging solutions.

Fig.~\ref{fig:12} further shows the robustness of the GDM methods, examining how varying random seeds influence the training performance. The figure delineates three distinct curves, each corresponding to a different random seed. While the random seed is known to significantly sway outcomes in image-related GDM applications such as Stable Diffusion~\cite{stabdiff}, our findings reveal a contrasting scenario. After about 50 timesteps, all three cases stabilize, maintaining a gap to zero (where zero signifies the theoretical upper bound) within a negligible margin of 0.05. This observation shows that, unlike in image-related applications where identical text prompts can yield vastly different images based on the seed, the random seed's impact on performance in this context is minimal. This insight highlights the GDM's resilience against varying initial conditions, suggesting its consistent ability to learn the power allocation scheme and achieve near-optimal performance, especially in similar network optimization problems.

Then we consider a more complex case that the number of channels is $5$ and the channel gains of these 5 channels vary within $0.5$ and $5$. We compare the performance of GDM and DRL algorithms and study the impact of denoising steps.
\begin{figure}[!t]
\centering
\includegraphics[width=0.48\textwidth]{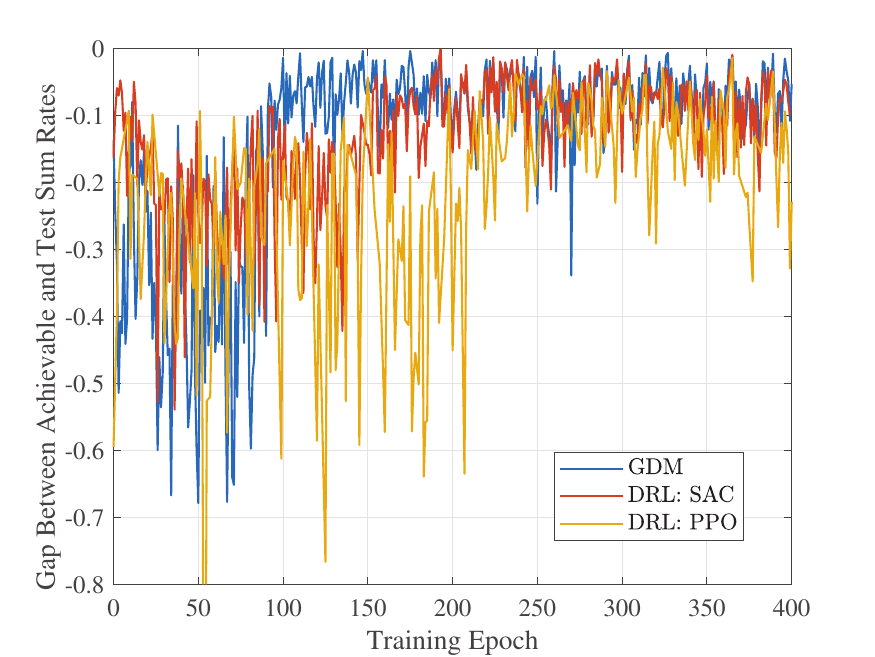}
\caption{Test reward curves of GDM-aided and DRL-aided optimization methods, with the number of channels $M = 5$, and the channel gains vary within $0.5$ and $5$.}
\label{fig:21}
\end{figure}

\begin{figure}[!t]
\centering
\includegraphics[width=0.48\textwidth]{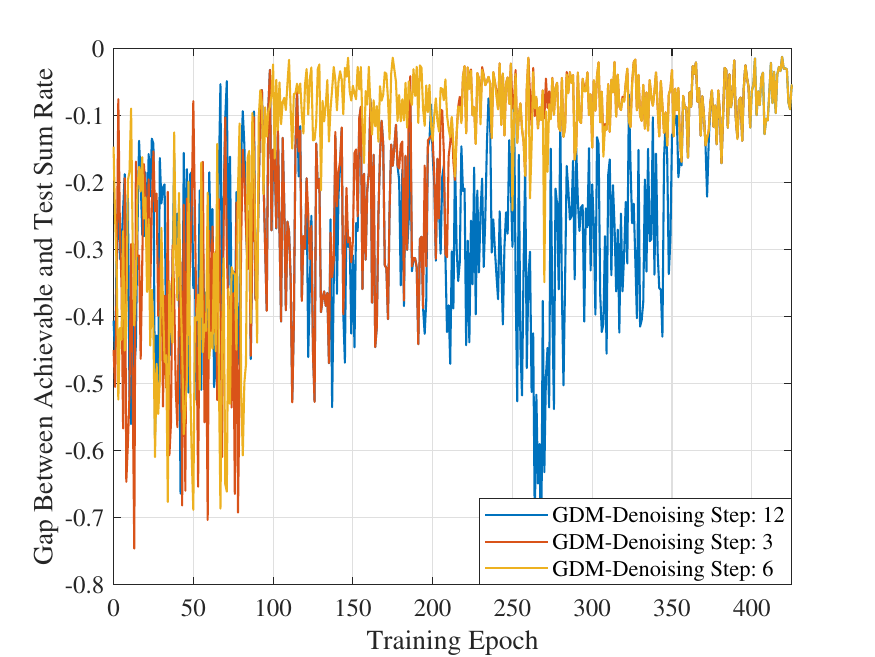}
\caption{Test reward curves of GDM-aided optimization methods under different denoising steps, with the number of channels $M = 5$, and the channel gains vary within $0.5$ and $5$.}
\label{fig:22}
\end{figure}

In Fig.~\ref{fig:21}, we examine the performance of the GDM method compared to two DRL methods, i.e., SAC and PPO. All three methods demonstrate convergence, while the final gap values for GDM and SAC are closer to zero, indicating a better power allocation scheme. In contrast, PPO exhibits larger fluctuations and slower convergence. While the final results of GDM and SAC are similar, GDM converges faster, which is attributed to its ability to capture complex patterns and relationships more efficiently. This faster convergence of GDM is particularly beneficial in scenarios where time efficiency is crucial. 

Furthermore, we study the impact of different denoising steps on the performance of the GDM in Fig.~\ref{fig:22}. The figure presents three curves, each corresponding to a different number of denoising steps. The first curve, representing $6$ denoising steps, exhibits the fastest convergence. The second curve, corresponding to $3$ denoising steps, converges slower. This slower convergence rate could be attributed to insufficient denoising when the number of steps is small, leading to greater uncertainty in generated power allocation schemes. However, when the number of steps is too larger, as in the third curve where the number of denoising steps is $12$, the convergence is slowest. This could be due to the model losing its ability to explore the environment effectively, as excessive denoising might lead to overfitting the training data. This analysis underscores the importance of carefully selecting the number of denoising steps in the GDM, striking a balance between sufficient denoising and maintaining the GDM's ability to explore the environment.

\begin{figure}[!t]
\centering
\includegraphics[width=0.48\textwidth]{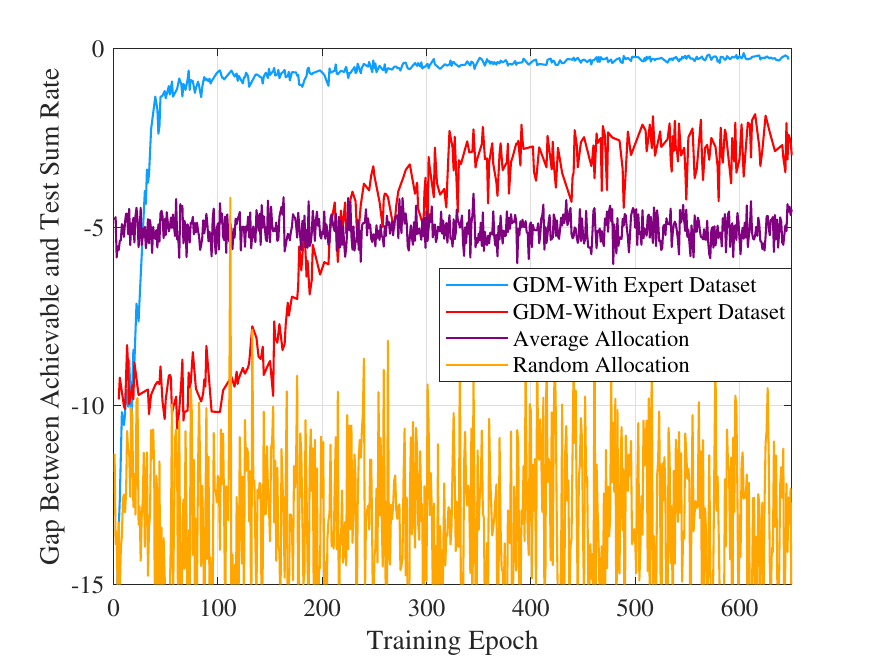}
\caption{Test reward curves of GDM-aided optimization methods with and without expert dataset, with the number of channels is $71$, i.e., $M = 71$, and the channel gains vary within $2$ and $25$.}
\label{fig:3}
\end{figure}

\begin{figure*}[!t]
\centering
\includegraphics[width=1\textwidth]{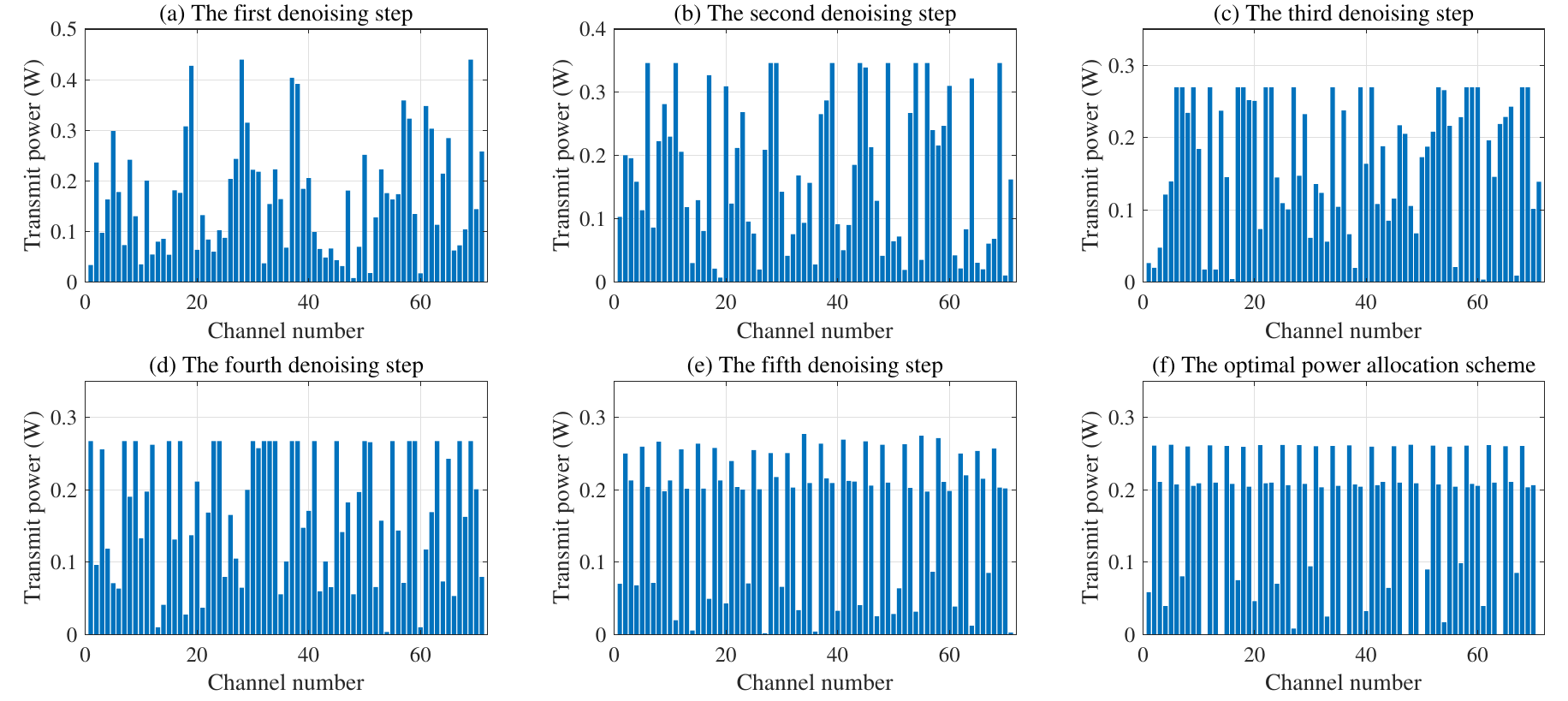}
\caption{{\textbf{Sub-figures (a) to (e)}} illustrate the process of $5$-step denoising Gaussian noise into the transmit power allocation schemes using a well-trained GDM. Here, we consider $71$ channels with the total transmission power of $12$ ${\rm W}$. In these $71$ channels, the channel gains differ randomly. Some channels fall within the range of $2$ to $5$, others between $10$ to $15$, and the remaining channels exhibit gains varying from $20$ to $25$. We simulate using a set of observations obtained by random sampling. {\textbf{Sub-figure (f)}} is the optimal power allocation scheme obtained by the water-filling algorithm~\cite{yu2004iterative}.}
\label{fig:4}
\end{figure*}
Fig.~\ref{fig:3} shows the test reward curves for GDM-aided optimization methods, both with and without access to an expert dataset, in a scenario with $71$ channels, i.e., $M=71$, and channel gains varying between $2$ and $25$. The figure further validates the efficacy of the GDM approaches, irrespective of the availability of the expert dataset. Using an expert dataset in GDM training significantly accelerates the convergence process. However, even without an expert dataset, the GDM approach can independently decrease the gap between the achieved sum rate and the upper bound. Furthermore, two straightforward power allocation schemes, namely average and random allocation, are also presented for comparison. Average allocation, which evenly distributes power among the channels, outperforms random allocation, which arbitrarily assigns power. However, GDM, with its advanced learning capability, outperforms both strategies.

Fig.~\ref{fig:4} visualizes the process of the well-trained GDM generating the power allocation scheme from the Gaussian noise. We consider $71$ channels with a total transmission power of $12$ $\rm W$, where the specific channel gains of the $71$ channels randomly vary between $(2, 5)$, $(10, 15)$, or $(20, 25)$. Figs. \ref{fig:4} (a)-(e) show the progressive refinement of the power allocation scheme through the denoising process. Fig. \ref{fig:4} (f) presents the optimal power allocation scheme obtained by the water-filling algorithm~\cite{yu2004iterative}. This series of figures demonstrates the capability of GDM to generate near-optimal power allocation schemes through iterative denoising, even when confronted with complex and variable channel conditions. It also highlights the close agreement between the GDM-generated and water-filling algorithm-generated power allocation schemes, emphasizing the effectiveness of GDM in learning and imitating expert solutions. The gap between the sum rate under the power allocation scheme shown in Fig. \ref{fig:4} (e) and the upper bound is $0.11$ ${\rm{bit/s/Hz}}$.

{\textbf{Lesson Learned:}} From the above showcase discussions, we glean several insights into the application of GDMs in network optimization. Firstly, the superior performance of GDMs over traditional DRL methods underscores the transformative potential of GDMs in complex optimization tasks. This is particularly notable in scenarios where rapid convergence and high performance are paramount. Secondly, the learning-related parameters in GDM, such as learning rates and denoising steps, facilitate a novel balance between exploration and exploitation. Notably, the denoising process, acting as a pivotal mechanism in GDMs, introduces a fresh perspective to this classic trade-off in RL as we discussed in Fig.~\ref{fig:22}. Thirdly, the resilience of GDMs to varying initial conditions and their consistent near-optimal performance, even in the absence of an expert dataset, show the robustness and adaptability. This robustness is particularly crucial in real-world applications where conditions can be unpredictable and data may be imperfect or incomplete. Lastly, the ability of GDMs to generate near-optimal power allocation schemes that are closely aligned with expert solutions underscores their capacity for sophisticated pattern recognition and imitation. This suggests that GDMs can be used as a powerful tool for learning from and leveraging expert knowledge in complex domains in network optimization tasks.

\section{Deep Reinforcement Learning}\label{section3}
This section first discusses DRL algorithms and their applications in network optimization~\cite{luong2019applications,xu2019load}, followed by examining the integration of GDMs within DRL frameworks~\cite{zhang2022lad,wang2022diffusion,chen2022offline,janner2022planning,lu2023contrastive,reuss2023goal,wang2023pdpp,brehmer2023edgi,cao2023multi,wang2023diffusion,liang2023adaptdiffuser,pearce2023imitating,ajay2022conditional}. We then present a case study on AIGC service provider selection in edge networks~\cite{du2023generative}.

\subsection{Fundamentals of DRL}\label{aelfjl}
DRL is a powerful approach that combines the strengths of both deep learning and reinforcement learning, enabling the development of algorithms capable of learning to make optimal decisions through interactions with their environment~\cite{luong2019applications,xu2019load}. The DRL framework comprises two main components: the agent and the environment \cite{9195488}. The agent, a decision-making entity, learns to interact optimally with the environment to maximize a cumulative reward \cite{10172220}. The environment provides feedback to the agent in the form of rewards based on the actions taken by the agent \cite{9575181}. This interaction forms the basis of the learning process in DRL. We summarize several representative DRL algorithms as
\begin{itemize}
\item {\textbf{Deep Q-Network (DQN):}} DQN uses a deep neural network for approximating the Q-value function, enabling it to handle high-dimensional state spaces. However, it struggles with high-dimensional or continuous action spaces~\cite{mnih2015human}.
\item {\textbf{Prioritized DQN:}} This variant of DQN prioritizes experiences with high temporal-difference error, leading to faster learning but introducing additional complexity~\cite{schaul2015prioritized}.
\item {\textbf{Deep Recurrent Q-Network (DRQN):}} DRQN extends DQN with recurrent neural networks for tasks requiring memory of past information, which is however challenging to train~\cite{hausknecht2015deep}.
\item {\textbf{PPO:}} PPO is a stable policy gradient method that keeps policy updates close to zero, which however may require more samples to learn effectively~\cite{schulman2017proximal, 10032267}.
\item {\textbf{REINFORCE:}} REINFORCE directly optimizes the policy function, making it widely applicable but suffering from high variance~\cite{williams1992simple}.
\item {\textbf{SAC:}} SAC maximizes both the expected return and the policy's entropy, leading to better performance in complex environments at the cost of computational complexity~\cite{haarnoja2018soft}.
\item {\textbf{Rainbow:}} Rainbow combines seven DQN improvements, enhancing performance but increasing implementation complexity~\cite{hessel2018rainbow}.
\end{itemize}

In the context of wireless communications, DRL offers several advantages. First, DRL is adept at handling complex network optimization problems, enabling network controllers to find optimal solutions even without complete and precise network information~\cite{luong2019applications,wang2021incorporating}. This strength is further complemented by DRL's capacity to enable network entities to learn and accumulate knowledge about the communication and networking environment. This facilitates learning optimal policies without knowing the channel model and mobility pattern~\cite{li2019deep,luong2019applications}. Furthermore, DRL supports autonomous decision-making, reducing communication overheads and boosting network security and robustness~\cite{tang2022constructing,du2023generative}.

Given these advantages, DRL has found extensive applications in network optimizations \cite{10107766}. However, it is important to note that DRL also has its limitations, which, however, may be mitigated by the introduction of GDMs:
\begin{itemize}
\item {\textbf{Sample Inefficiency:}} DRL often requires a large number of interactions with the environment to learn effectively, which can be computationally expensive and time-consuming~\cite{luong2019applications}. GDMs, with the strong ability to model complex data distributions, could reduce the number of samples required.
\item {\textbf{Hyperparameter Sensitivity:}} The performance of DRL algorithms can be significantly influenced by hyperparameters, demanding meticulous tuning for diverse tasks \cite{ashraf2021optimizing}. GDMs, with their flexible structure and adaptability to various data distributions, could provide a more robust solution.
\item {\textbf{Difficulty in Modeling Complex Environments:}} DRL algorithms may struggle with environments characterized by complex and high-dimensional state and action spaces. By accurately capturing the underlying data distributions, GDMs could provide a more efficient representation of the environment.
\item {\textbf{Instability and Slow Convergence:}} DRL algorithms may suffer from instability and slow convergence. The unique structure of GDMs involves a diffusion process, potentially offering a more stable and efficient learning process.
\end{itemize}

\subsection{Applications of GDM in DRL}
The distinctive characteristics of GDMs have been effectively utilized to enhance DRL. These advantages include high expressiveness, the ability to capture multi-modal action distributions, and the potential to integrate with other RL strategies seamlessly.  One notable application of GDMs in DRL is presented in \cite{wang2022diffusion}, where the authors introduced Diffusion Q-learning (Diffusion-QL). This innovative method utilized a GDM as the policy representation, more specifically, a DDPM~\cite{ho2020denoising} based on a Multilayer Perceptron (MLP). The authors incorporated the Q-learning guidance into the reverse diffusion chain, facilitating optimal action selection. Through this integration, they demonstrated the expressiveness of GDMs in capturing multi-modal action distributions and showcased their effectiveness in enhancing behavior cloning and policy improvement processes. As a result, Diffusion-QL surpassed previous methods across several D4RL benchmark tasks~\cite{fu2020d4rl} for offline RL. Complementarily, the work in \cite{chen2022offline} improves offline RL further by addressing the limitations of distributional expressivity in policy models.
In contrast to the approach in \cite{wang2022diffusion}, the authors in \cite{chen2022offline} decoupled the learned policy into a generative behavior model and an action evaluation model. This separation facilitated the introduction of a diffusion-based generative behavior model capable of modeling diverse behaviors such as agent's trajectories. The optimal selection of actions from this behavior model was achieved through importance sampling in concert with an action evaluation model. They also incorporated an in-sample planning technique to mitigate extrapolation error and enhance computational efficiency. The resulting methodology outperformed traditional offline RL methods on D4RL datasets~\cite{fu2020d4rl} and showed proficiency in learning from heterogeneous datasets. These highlighted studies represent just a subset of the burgeoning body of work on GDMs in DRL. For an extended discussion, Table~\ref{xxasfa} reviews various papers about GDM and DRL, summarizing their contributions and impacts. The distinctive ability of GDMs to accurately model complex distributions significantly enhances DRL algorithms, particularly in network settings where decision-making processes frequently require navigating through intricate solution spaces. This capability facilitates more effective and efficient optimization of network configurations and resource allocations compared to traditional models, offering advanced solutions that can dynamically adapt to the complexities inherent in network management.

\renewcommand{\arraystretch}{1.1}
\begin{table*}[!t]
\centering
\begin{tabular}{m{1cm}|m{7.8cm}|m{7.8cm}}
\toprule[1pt]
\hline
\textbf{Paper} & \textbf{Key Contributions} & \textbf{Results} \\
\hline
\cite{zhang2022lad} & Leverage Language Augmented Diffusion (LAD) models for language-based skills in RL & Achieve an average success rate of 72\% on the CALVIN language robotics benchmark \\
\hline
\cite{wang2022diffusion} & Propose Diffusion Q-learning (Diffusion-QL) for offline RL and represent the policy as a GDM & Achieve state-of-the-art performance on the majority of D4RL benchmark tasks \\
\hline
\cite{chen2022offline} & Decouple policy learning into behavior learning and action evaluation and introduce a generative approach for offline RL & Achieve superior performance on complex tasks such as AntMaze on D4RL \\
\hline
\cite{janner2022planning} & Develop a diffusion probabilistic model for trajectory optimization and introduce a model directly amenable to trajectory optimization & Demonstrate effectiveness in control settings emphasizing long-horizon decision-making and test-time flexibility \\
\hline
\cite{lu2023contrastive} & Introduce Contrastive Energy Prediction (CEP) for learning the exact guidance in diffusion sampling & Demonstrate effectiveness in offline RL and image synthesis, outperforming existing state-of-the-art algorithms on D4RL benchmarks \\
\hline
\cite{reuss2023goal} & Propose a robust version of the Diffusion Implicit Models (DIMs) for better generalization to unseen states in RL & Show the new approach provides more stable policy improvement and outperforms the baseline DIM methods on various complex tasks \\
\hline
\cite{wang2023pdpp} & Treat procedure planning as a distribution fitting problem, remove the expensive intermediate supervision and use task labels instead & Achieve state-of-the-art performance on three instructional video datasets across different prediction time horizons without task supervision \\
\hline
\cite{brehmer2023edgi} & Introduce the Equivariant Diffuser for Generating Interactions (EDGI), an algorithm for MBRL and planning & Improve sample efficiency and generalization in 3D navigation and robotic object manipulation environments \\
\hline
\cite{cao2023multi} & Propose a general adversarial training framework for multi-agent systems using diffusion learning, enhancing robustness to adversarial attacks & Demonstrate enhanced robustness to adversarial attacks in simulations with FGM and DeepFool perturbations \\
\hline
\cite{wang2023diffusion} & Introduce a new imitation learning framework that leverages both conditional and joint probability of the expert distribution, and explore the use of different generative models in the framework & Outperform baselines in various continuous control tasks including navigation, robot arm manipulation, dexterous manipulation, and locomotion \\
\hline
\cite{liang2023adaptdiffuser} & Introduce a self-evolving method for diffusion-based planners in offline reinforcement learning, demonstrating an ability to improve planning performance for both known and unseen tasks & Outperform the previous state-of-the-art Diffuser by 20.8\% on Maze2D and 7.5\% on MuJoCo locomotion, and show better adaptation to new tasks, e.g., KUKA pick-and-place, by 27.9\%\\
\hline
\cite{pearce2023imitating} & Introduce innovations for diffusion models in sequential environments & Accurately model complex action distributions, outperform state-of-the-art methods on a simulated robotic benchmark, and scale to model human gameplay in complex 3D environments\\
\hline
\cite{ajay2022conditional} & Apply conditional generative modeling to the problem of sequential decision-making and investigate conditioning on constraints and skills & Outperform existing offline RL approaches and demonstrate the flexible combination of constraints and composition of skills at test time \\
\hline
\bottomrule[1pt]
\end{tabular}
\caption{Extended summary of papers on GDM in DRL}
\label{xxasfa}
\end{table*}

In summary, the integration of GDMs into DRL, as demonstrated by these representative studies and further summarized in Table~\ref{xxasfa}, leverages several key advantages offered by GDMs. The key advantages that GDMs offer to address the disadvantages of DRL as we discussed in Section~\ref{aelfjl} are listed below:
\begin{itemize}
\item \textbf{Expressiveness:} GDMs are capable of modeling complex data distributions, making them well-suited for representing policies in DRL~\cite{vargas2023expressiveness}. For instance, in a dynamic traffic routing scenario, the policy needs to adapt to various traffic conditions, road structures, and vehicle behaviors~\cite{liu2016balanced}. GDMs can effectively model such a policy.
\item \textbf{Sample Quality:} GDMs are known for generating high-quality samples~\cite{watson2021learning,huang2022prodiff}. In the context of DRL, this translates into the generation of high-quality actions or strategies~\cite{hong2022improving}. For example, in a network resource allocation task, the quality of the generated allocation decisions directly impacts the network performance. GDMs can generate high-quality decisions, leading to improved network performance.
\item \textbf{Flexibility:} The ability of GDMs to model diverse behaviors is particularly useful in DRL, where the agent needs to adapt to a variety of situations and tasks~\cite{lyu2022accelerating}. In a network management task, for instance, the network may need to adapt to various traffic conditions and user demands. GDMs can model a wide range of behaviors, enabling the network to adapt to these diverse conditions.
\item \textbf{Planning Capability:} GDMs can be used for planning by iteratively denoising trajectories, providing a novel perspective on the decision-making processes in DRL~\cite{wang2023diffusion}. For example, a DRL agent could use a GDM to plan the network operations, iteratively refining the plan to optimize the network efficiency~\cite{liang2023adaptdiffuser,pearce2023imitating}.
\end{itemize}

While GDMs offer promising advantages in DRL, they also present certain challenges. The iterative nature of GDMs can lead to increased computational complexity, which could be a hurdle in large-scale DRL tasks such as optimizing city-wide communication networks~\cite{wang2023diffusion}. Additionally, GDMs may struggle to accurately model certain data distributions, especially those with high noise levels or irregularities. This could pose challenges in DRL tasks involving real-world network traffic data, which may contain stronhg noise and outliers~\cite{yang2022diffusion}. 
While these challenges underline the limitations of GDMs, they also present opportunities for innovative approaches that can effectively harness the benefits of GDMs while mitigating their shortcomings. Leveraging GDMs within advanced DRL algorithms offers a promising solution to both computational complexity and modeling limitations. An example could be found in combining GDMs with SAC~\cite{du2023generative}, a state-of-the-art DRL method known for its efficient learning and robustness. This combination capitalizes on the strength of GDMs in modeling complex action distributions while utilizing the optimization capabilities of SAC, yielding a hybrid model with the potential for enhanced performance and efficiency in complex network optimization tasks. To illustrate this, we delve into a case study, introducing an innovative combination of GDM and SAC.

\subsection{Case Study: AIGC Service Provider Selection}
\subsubsection{System Model}
\begin{figure}[!t]
\centering
\includegraphics[width=0.4\textwidth]{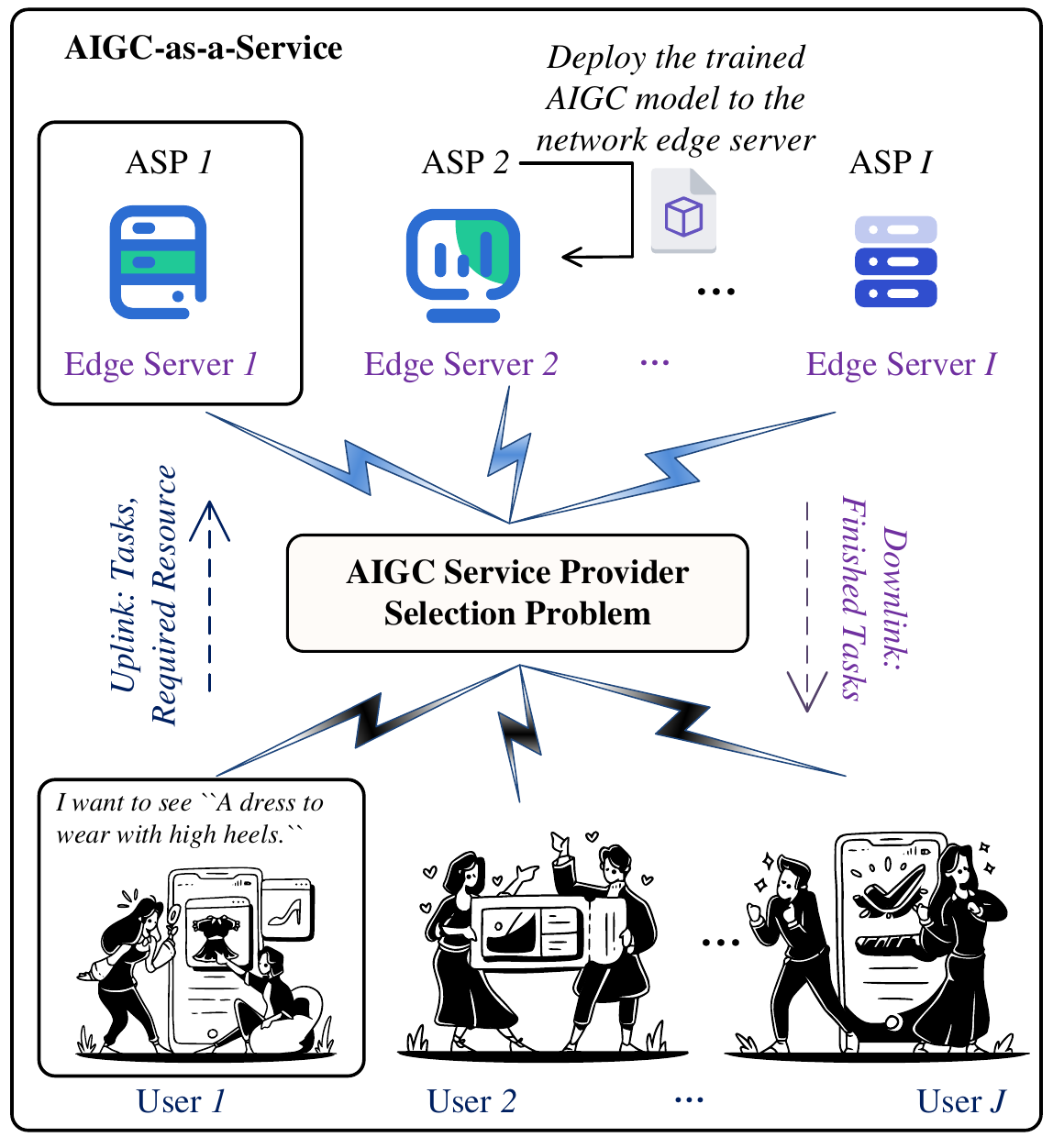}
\caption{AIGC service provider selection problem. Following the paradigm of ``AIGC-as-a-Service'', various ASPs deploy their AIGC models onto network edge servers. With user requests arriving, an optimal task scheduler should be designed for real-time user task allocation. The goal is to maximize total user QoE, considering the unique capabilities of each AIGC model and the computing resource constraints of edge servers \cite{du2023generative}.}
\label{drlmodel}
\end{figure}
The AIGC service provider selection problem depicted in Fig.~\ref{drlmodel} and detailed in \cite{du2023generative}, can be regarded as an extension of the resource-constrained task assignment problem. This is a well-known challenge in wireless networks where resources are scarce and their efficient utilization is critical to achieving the desired performance \cite{dai2022psaccf}. Specifically, we consider a set of sequential tasks and available ASPs, each of which possesses a unique utility function. The objective is to assign users' AIGC tasks to ASPs in a way that maximizes the overall user utility. This user utility is a function of the required computing resource for each task and it is related to the AIGC model that performs the task. In addition, we acknowledge that the computing resources of each ASP is limited.

From a mathematical perspective, the ASP selection problem can be modeled as an integer programming problem, with the decision variables representing the sequence of task assignments to available ASPs. The formulation also incorporates constraints that capture the limitations on available resources. Failing to meet these constraints can have severe consequences, such as the crash of an ASP and the subsequent termination and restart of its running tasks.

\subsubsection{GDM-based Optimal Decision Generation}
\begin{figure}[!t]
\centering
\includegraphics[width=0.48\textwidth]{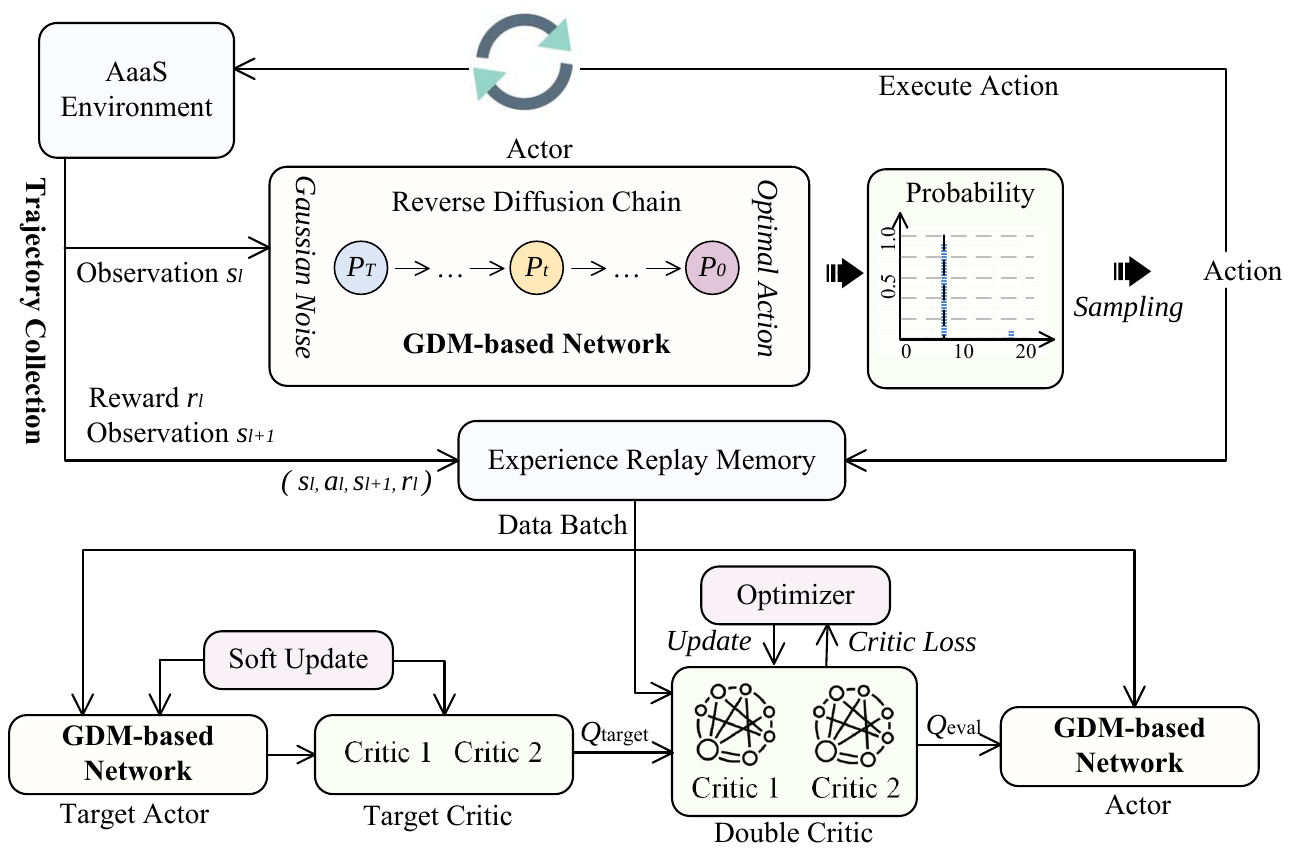}
\caption{The overall architecture of the D2SAC~algorithm \cite{du2023generative}.}
\label{asfaef2}
\end{figure}
The authors in~\cite{du2023generative} applied GDM to the actor-critic architecture-based DRL paradigm and proposed the Deep Diffusion Soft Actor-Critic (D2SAC) as a deep diffusion reinforcement learning algorithm.
As shown in Fig.~\ref{asfaef2}, the D2SAC algorithm incorporates several key components to optimize the policy, including an actor network, a double critic network, a target actor, a target critic, an experience replay memory, and the environment. Here's a summary and explanation of these components and their roles:
\begin{itemize}
    \item {\textbf{Trajectory Collection:}} The agent observes the environment and collects transitions of state by executing actions in the environment. These transitions are regarded as experiences and are added to the experience replay memory. The actor network generates an action distribution over all possible actions given an environment observation and samples an action from this distribution. This action is performed, transitioning to a new state and returning an immediate reward as feedback.
    \item {\textbf{GDM as the Policy:}} The core of the actor network is the GDM, which effectively encodes the observation's representation. It captures the dependencies between the observation and the action space.
    \item {\textbf{Experience Replay Memory:}} This is a method to handle the delay in receiving reward feedback. Experiences are stored and the missing reward is filled in later before updating the GDM-based network. Off-policy training is used to improve the handling of delayed feedback~\cite{gu2017deep}.
    \item {\textbf{Double Critic Network:}} During the policy improvement process, the actor network is optimized by sampling mini-batches of transitions from the experience replay memory. The double critic network, composed of two separate critic networks, is used to reduce the overestimation bias by providing a conservative estimate of the Q-value function~\cite{hasselt2010double}.
    \item {\textbf{Policy Improvement:}} The actor learns to maximize the expected cumulative reward for each action at the current state. The maximization problem is solved using the gradient ascent algorithm~\cite{khaneja2005optimal}. Specifically, gradients are calculated over a mini-batch of transitions sampled from the experience replay memory, and the actor network is updated by performing gradient descent on these gradients.
    \item {\textbf{Action Entropy Regularization:}} An entropy regularization term is introduced to prevent the policy from becoming overly confident in certain actions and converging prematurely to a suboptimal solution~\cite{zhao2020domain}. This encourages exploration.
    \item {\textbf{Q-function Improvement:}} The Q-function, used for estimating the future rewards of actions, must be accurately estimated for successful optimization. To achieve this, the Temporal Difference (TD) error between two Q networks is minimized during training~\cite{tesauro1995temporal}.
\end{itemize}
Next, we discuss the performance of D2SAC and compare it with seven DRL algorithms as discussed in Section~\ref{aelfjl}. Furthermore, we demonstrate the efficacy of D2SAC across various benchmark tasks within the DRL domain.

\subsubsection{Numerical Results}
The authors in \cite{du2023generative} compared D2SAC with benchmark reinforcement learning algorithms: DQN, DRQN, Prioritized-DQN, Rainbow, REINFORCE, PPO, and SAC. As shown in Fig.~6 in \cite{du2023generative}, D2SAC's reward acquisition over time demonstrates its superior ability to balance exploration and exploitation, resulting in more optimal policy decisions.


\begin{table*}
\centering
\caption{Performance Comparisons on General Benchmark Tasks.}
\label{tab:comparison-on-general}
\footnotesize
\renewcommand\arraystretch{1.25}
\begin{tabular}{c|c|cccc}
\hline
\multicolumn{2}{c|}{\textbf{Policy}} & \textbf{Acrobot-v1} & \textbf{CartPole-v1} & \textbf{CoinRun-v0} & \textbf{Maze-v0} \\
\hline\hline
\multirow{7}{*}{DRL} & DQN & -81.81 $\pm$ 17.19 & 499.80 $\pm$ 0.14 & 6.00 $\pm$ 4.90 & 3.00 $\pm$ 4.58 \\
& Prioritized-DQN & -105.20 $\pm$ 14.74 & 498.70 $\pm$ 1.43 & 5.00 $\pm$ 5.00 & 2.00 $\pm$ 4.00 \\ 
& DRQN & -82.26 $\pm$ 14.34 & 132.50 $\pm$ 69.79 & $-$ & $-$ \\ 
& REINFORCE & -104.80 $\pm$ 14.51 & 500.00 $\pm$ 0.00 & 0.00 $\pm$ 0.00 & 0.00 $\pm$ 0.00 \\ 
& PPO & -77.22 $\pm$ 8.45 & 499.90 $\pm$ 0.33 & 0.00 $\pm$ 0.00 & 2.00 $\pm$ 4.00 \\ 
& Rainbow & -158.10 $\pm$ 55.48 & 478.30 $\pm$ 29.28 & 5.00 $\pm$ 5.00 & 2.00 $\pm$ 4.00 \\ 
& SAC & -121.00 $\pm$ 35.31 & 500.00 $\pm$ 0.00 & 10.00 $\pm$ 0.00 & 3.00 $\pm$ 4.58 \\ 
\hline\hline
\multirow{8}{*}{Online\cite{pmlr-v119-cobbe20a,rl-zoo}} & A2C & -86.62 $\pm$ 25.10 & 499.90 $\pm$ 1.67 & $-$ & $-$ \\
& ACER & -90.85 $\pm$ 32.80 & 498.62 $\pm$ 23.86 & $-$ & $-$ \\
& ACKTR & -91.28 $\pm$ 32.52 & 487.57 $\pm$ 63.87 & $-$ & $-$ \\
& PPO2 & -85.14 $\pm$ 26.27 & 500.00 $\pm$ 0.00 & $-$ & $-$ \\
& DQN & -88.10 $\pm$ 33.04 & 500.00 $\pm$ 0.00 & $-$ & $-$ \\
& TRPO & $-$ & 485.39 $\pm$ 70.51 & $-$ & $-$ \\
& PPO + IMPALA & $-$ & $-$ & 8.95 & \textbf{9.88} \\
& Rainbow + IMPALA & $-$ & $-$ & 5.50 & 4.24 \\
\hline\hline
\textbf{Ours} & \textbf{D2SAC} & \textbf{-70.77} $\pm$ \textbf{4.12} & \textbf{500.00} $\pm$ \textbf{0.00} & \textbf{10.00} $\pm$ \textbf{0.00} & 7.00 $\pm$ 4.58 \\
\hline
\end{tabular}
\end{table*}

Table \ref{tab:comparison-on-general} presents comparative performance metrics of various control tasks in the Gym environment~\cite{rl-zoo}
\begin{itemize}
\item {\textbf{Acrobot-v1:}} A two-link pendulum simulation, with the goal of maintaining an upright position. The reward system is designed to favor lesser negative values.
\item {\textbf{CartPole-v1:}} A cart-pole system model, where the objective is to prevent a pole from falling. The performance measure here is the average reward, with higher values being desirable.
\item {\textbf{CoinRun-v0:}} A platform game task where the agent's goal is to collect a coin while avoiding obstacles. The performance is gauged through the average reward per episode, aiming for higher values.
\item {\textbf{Maze-v0:}} A maze navigation task, where reaching the goal while taking fewer steps is rewarded. Similar to the previous tasks, higher average reward values indicate better performance.
\end{itemize}
These benchmarks cover a diverse range of problems, including physics-based control (Acrobot-v1, CartPole-v1), strategy (CoinRun-v0), and pathfinding (Maze-v0). A closer examination of the table reveals that D2SAC significantly outperforms most of the compared policies on these tasks. Specifically, for the Acrobot-v1 task, D2SAC achieves the least negative reward, implying superior performance in the complex task of manipulating the two-link pendulum. In the CartPole-v1 and CoinRun-v0 tasks, D2SAC matches the top-performing algorithms with perfect average rewards of 500 and 10, respectively, indicating a consistent ability to keep the pole upright and successfully collect coins in the platform game. The performance on Maze-v0, although not the highest, is competitive and within the performance range of top-performing policies.

\section{Incentive Mechanism Design}\label{section4}
In this section, we investigate the applicability of GDM for shaping robust and efficient incentive mechanisms in network designs~\cite{du2023generative,liu2023blockchainempowered, liu2023deep}.

\subsection{Fundamentals of Incentive Mechanisms}
Incentive mechanism \cite{liu2023deep, du2023generative} plays an important role in network optimization for maintaining the network operationality and long-term economic sustainability.
Specifically, the mechanism rewards the network participants who share computing, communication, and information resources and services.
Take CrowdOut~\cite{crowdout}, a mobile crowdsourcing system for road safety, as an example.
Drivers (using smartphones or vehicular sensors) can report road safety situations that they experience in their urban environments, e.g., speeding, illegal parking, and damaged roads, to the central management center.
However, the drivers consume their computing and communication resources, e.g., battery power, CPU, and wireless bandwidth, to sense and report issues.
They might be discouraged from actively joining such cooperations without appropriate rewards, especially in the long term.
Accordingly, the incentive mechanisms aim at answering the following series of questions: 1) how to encourage the network entities to behave in a certain way that is beneficial to the network, e.g., through the use of rewards, reputation, or credit \cite{8664132}, 2) how to motivate the contribution of resources, 3) how to discourage and prevent the malicious behavior, and 4) how to ensure the fairness.
To do so, the incentive mechanisms should be designed to satisfy several properties, including but not limited to Individual Rationality (IR), Incentive Compatibility (IC), fairness, Pareto Efficiency (PE), Collusion Resistance (CR), and Budget Balance (BB) \cite{zeng2021comprehensive}.
With years of research, various incentive mechanisms have been presented and widely adopted in network optimization.
We consider the following representative techniques for developing incentive mechanisms, including the Stackelberg game, auction, contract theory, and Shapley value.

\subsubsection{Stackelberg Game}
In game theory, the Stackelberg game refers to an iterative process, in which a leader makes the first move and the remaining followers move sequentially, until reaching the equilibrium \cite{yang2013coping}.
In the network context, the leader, typically a network operator, first determines the resource prices or service charges. 
Network users, i.e., followers, then determine their resource demands based on the given prices, with the goal of balancing their utility against the cost that they paid for the resources. 
At the Stackelberg equilibrium, the followers cannot increase their utility by changing their demands, and the leader cannot increase its profit by altering the price. 
In this way, the network efficiency and the participants' utilities can be balanced, thereby promoting efficient cooperation. 
With wide adoption, the Stackelberg game provides a robust foundation for designing network incentive mechanisms.

\subsubsection{Auction}
An auction mechanism is widely adopted for incentivizing resource trading \cite{9773059}. 
Specifically, an auctioneer conducts an auction for trading network resources, e.g., bandwidth or computing power, that are subject to allocation among bidders. 
The auction process begins with the auctioneer announcing the resources to be traded and soliciting bids. 
Each bidder evaluates its demand and willingness to pay, submitting a bid accordingly. 
The auctioneer then chooses a subset of bidders as the winners based on the bid amount or more complex rules. 
Finally, the auctioneer calculates the payment from each winner, which could be the bid amount or another value depending on the auction type, and performs the resource allocation. 
Auctions can foster competition among bidders, aiming to maximize social welfare in terms of network utilities while satisfying certain constraints like budget balance, i.e., the auctioneer's revenue should be positive.

\subsubsection{Contract Theory}
Contract-theoretic incentive mechanisms can effectively address network information asymmetry~\cite{contractkang}. 
In this setup, an employer (typically the network operator or service provider) and an employee (the network user) engage in a contractual agreement. 
The employer designs contracts specifying service charges, Quality of Service (QoS) levels, and resource allocations. 
However, it may not have complete information about the employees' preferences and behaviors, which is called information asymmetry~\cite{contractkang}. 
With contract theory, the employers can launch a series of contracts, which ensures the IR, i.e., the utility of the employee is higher than the threshold and IC, i.e., the employees can acquire the highest utility by faithfully following the contracts that they signed properties of the employees.
Hence, the employees behave honestly, driven by utilities, circumventing the undesirable effects, such as selfish strategies, caused by the information asymmetry.
Contract-theoretic incentive mechanisms have been widely adopted in various network scenarios and have many variants to support high-dimension resource allocation, heterogeneous employees, etc.

\subsubsection{Shapley Value}
The Shapley Value (SV) is a solution from cooperative game theory, quantifying a player's marginal contribution across potential coalitions.
In the incentive mechanism design, the players contribute to the network and are subject to being rewarded. 
Hence, SV for each player, denoted by $i$, can be defined as
\begin{equation}
SV(i) = \sum_{\mathbb{S} \subseteq \mathbb{N} \backslash {i}} \frac{|\mathbb{S}|!(|\mathbb{N}|-|\mathbb{S}|-1)!}{|\mathbb{N}|!} [v(\mathbb{S} \cup {i}) - v(\mathbb{S})],
\end{equation}
where $\mathbb{S}$ represents a coalition without $i$, $v$ represents the value function, $n$ is the total number of players.
SV can be used to allocate rewards, reputation, or credits, in which the player contributing more resources to the network will have higher SVs, thereby encouraging cooperation and resource contribution to the network.

\subsection{Applications of GDM in Incentive Mechanism Design}
From the above description, we observe that the overall procedure of incentive mechanism design is to model the participants' utility and thus formulate an optimization problem under constraints.
Hence, the problem becomes solving an optimization and finding the optimal incentive mechanism strategies that can maximize the utility.
Traditionally, researchers find the optimal solutions following the optimization principle.
Nonetheless, this method requires complete and accurate information about the network and, more importantly, is not applicable to complex network scenarios with complicated utility functions. 
Thanks to the strong ability to model complex environments, GDMs provide new possibilities for solving optimization problems.
A typical process of adopting GDMs to design incentive mechanisms contains the following steps.
\begin{itemize}
\item \textbf{Model the network states}: The first step is to model the network states. To do so, we typically use a vector, say ${\textbf{e}}$, which contains many factors, e.g., the upstream and downstream bandwidth, number of participants, bit error rate, and other scenario-specific factors, to depict the given network environment. 
\item \textbf{Formulate the utilities of participants}: Based on the factors in ${\textbf{e}}$ and other hyperparameters, e.g., the weights of these factors, we can formulate the utility function, as well as the associated constraints. Generally, the incentive mechanism design problem is to maximize the utility while satisfying all the constraints.
\item \textbf{Customize the GDM settings}: Thirdly, we customize the GDM settings according to the incentive mechanism design task. The \textit{solution space} is the universe of all the possible incentive mechanism strategies. For instance, the action space contains all the possible contracts in the contract-theoretic incentive mechanism. The \textit{objective function} takes the value of the utility function acquired in Step 2 if all the constraints are satisfied. Otherwise, it takes a large negative value as the constraint violation punishment. The \textit{dynamic environment} is the vector ${\textbf{e}}$.
\item \textbf{Train GDM and perform inference}: Finally, we can perform GDM training. The well-trained GDM can then be used for finding the optimal incentive mechanism design in any given network state ${\textbf{e}}$. The details of the training process are elaborated in Section~\ref{tutoriale}.
\end{itemize}

\subsection{Case Study: GDM-based Contract-Theoretic Incentive Mechanism}
\subsubsection{Background}
In this part, we conduct a case study to illustrate how to apply GDMs in a practical incentive mechanism design problem.
Specifically, we consider an emerging network scenario, namely mobile AIGC \cite{liu2023blockchainempowered, liu2023deep}.
Currently, the success of ChatGPT ignited the boom of AIGC, while the substantial resource costs of large AIGC models prevent numerous end users from enjoying the easy-accessible AIGC services.
To this end, researchers recently presented the concept of mobile AIGC, employing Mobile AIGC Service Providers (MASPs) to provide low-latency and customized AIGC inferences, leveraging mobile communications and edge computing capabilities.
Hence, the mobile AIGC network is composed of users and MASPs.
The former requests AIGC services from MASPs, and the latter operates the local AIGC models to perform inferences.
Given that AIGC inferences are resource-intensive, we utilize contract theory to design an incentive mechanism that rewards the MASPs according to their contributed resources.

\subsubsection{System Model}
Considering the diversity and heterogeneity of the current AIGC models, we divide all MASPs into $\mathcal{Z}$ levels according to the complexity of their local models, i.e., from level-$1$ to level-$\mathcal{Z}$.
The model complexity of each level of MASPs (denoted by $\theta_1$, $\dots$, $\theta_\mathcal{Z}$) can be quantified from different aspects, such as the number of model parameters \cite{complexity}.
Typically, the higher the model complexity, the more powerful the model is, and simultaneously, the more computing resources are required during the inference \cite{performance}.
In our system, we let the index of level follow the ascending order of model complexity, i.e., the higher the model complexity, the higher the index. 
Finally, we use $p_z$ to denote the proportion of level-$z$ ($z \in \{1, 2, \ldots, Z \}$) MASPs in the entire mobile AIGC network.

\subsubsection{Utility Formulation} 
For simplicity, we assume users evaluate the AIGC services using the most fundamental metric, i.e., the service latency.
Considering the heterogeneity of MASPs, the expected service quality and the required service fees for different levels of MASPs are different.
Hence, the utility of users towards level-$z$ ($z \in \{1, 2, \ldots, Z \}$) MASPs can be defined as \cite{contractkang}
\begin{equation}
U_\mathrm{U}^z = \big [\alpha_1(\theta_z)^{\beta_1} - \alpha_2(\mathcal{L}_z/\mathcal{L}_{max})^{\beta_2}] - \mathcal{R}_z,
\end{equation}
where $\big [\alpha_1(\theta_z)^{\beta_1} - \alpha_2(\mathcal{L}_z/\mathcal{L}_{max})^{\beta_2}]$ is a complexity-latency metric \cite{contractkang}, indicating the revenue that the client can gain.
$\mathcal{L}_z$ is the latency requirement of users for level-$z$ MASPs, while $\mathcal{L}_{max}$ is the maximum expected latency.
$\alpha_1$, $\alpha_2$, $\beta_1$, and $\beta_2$ are weighting factors.
$\mathcal{R}_z$ represents the rewards that users need to pay for level-$z$ MASPs.

For MASPs, they sell the computational resources by performing AIGC inferences for users.
Therefore, the utility of level-$z$ MASPs can be defined as
\begin{equation}
U_\mathrm{SP}^z = R_z - \Big[ \frac{(\mathcal{L}_{max}-\mathcal{L}_z)}{\mathcal{L}_z}\cdot\theta_z\Big],
\end{equation}
where $\Big[ \frac{(\mathcal{L}_{max}-\mathcal{L}_z)}{\mathcal{L}_z}\cdot\theta_z\Big]$ represents the costs of level-$z$ MASPs, which is determined by two factors, the model complexity $\theta_z$ and the latency $\mathcal{L}_z$.
Firstly, with $\theta_z$ fixed, the higher the $\mathcal{L}_z$, i.e., the longer latency can be tolerated by the users, the smaller the costs.
Meanwhile, the larger the $\theta_z$, the larger the costs of MASPs, since we have mentioned that complex models typically consume more resources for inference.

\subsubsection{GDM-based Optimal Contract Generation}
Based on the above descriptions, we design the following contract-theoretic incentive mechanism. 
Specifically, the users produce a specific contract, formed by \{$\mathcal{L}_z, \mathcal{R}_z$\} ($z \in \{1, 2, \ldots, Z \}$), for each level of MASPs, which then decide whether to sign.
The contract design should be optimal, maximizing $U_\mathrm{C}$ while satisfying the IR and IC constraints, i.e.,
\begin{equation}
\begin{array}{*{20}{l}}
{\mathop {\max }\limits_{{{\cal L}_z},{{\cal R}_z}} }&{\sum\limits_{z = 1}^{\cal Z} {{p_z}} U_{\rm{U}}^z\left( {{{\cal L}_z},{{\cal R}_z},{\theta _z}} \right),}\\
{\:\:\:{\rm{s}}{\rm{.t}}{\rm{.}}}&{\begin{array}{*{20}{l}}
{({\rm{IR}}):\;U_{{\rm{SP}}}^z({{\cal L}_z},{{\cal R}_z},{\theta _z}) \ge {U_{th}},}\\
{ z \in \{ 1, \ldots ,{\cal Z}\} ,}\\
{({\rm{IC}}):\;U_{{\rm{SP}}}^z({{\cal L}_z},{{\cal R}_z},{\theta _z}) \ge U_{{\rm{SP}}}^z({{\cal L}_j},{{\cal R}_j},{\theta _z}),}\\
{ z,j \in \{ 1, \ldots ,{\cal Z}\} ,z \ne j,}
\end{array}}
\end{array}
\end{equation}
where $U_{th}$ is the utility lower bound for MASPs.
Finally, we apply the aforementioned four-step procedure to formulate the GDM training paradigm and find the optimal contract design.
\begin{itemize}
\item \textbf{Model the network state}: For simplicity, we consider two types of MASPs in the mobile AIGC network. Hence, the network state vector in our case is defined as [$n$, $L_{max}$, $p_1$, $p_2$, $\theta_1$, $\theta_2$].
\item \textbf{Formulate the utility of participants}: There are two utility functions in our case, i.e., $U_\mathrm{U}$ and $U_\mathrm{SP}$. The former is the major utility that we intend to maximize. The latter is used in calculating the constraints, i.e., IR and IC. 
\item \textbf{Customize the GDM settings}: The space is formed as the universe of the contract design. Each bundle is formed as \{$\mathcal{L}_1$, $\mathcal{R}_1$, $\mathcal{L}_2$, $\mathcal{R}_2$\}. The hyperparameters $\alpha_1$, $\alpha_2$, $\beta_1$, and $\beta_2$ are set as 30, 5, 1 and 1, respectively.
\item \textbf{Train GDM and perform inference}: We train the GDM for more than 50000 epochs. The numerical results are discussed below.
\end{itemize}
\subsubsection{Numerical Results}
Our experiments validate the GDM's effectiveness in designing incentive mechanisms. Echoing the observations from Fig. 4 in \cite{liu2023blockchainempowered}, we found that GDM performs comparably to PPO in terms of coverage speed. GDM notably excels in achieving significantly higher rewards than PPO. This superior performance is attributed to two key factors: 1) the GDM keeps denoising and testing new samples in the training process, which fine-tunes the parameters of the \textit{solution generation network}, and 2) the randomness and dynamics in the wireless environment can be overcome due to the higher sample quality.
Additionally, our analysis extends to contract design under three heterogeneous network states, examining the utility function $U_\mathrm{U}$.
Our findings indicate that GDM consistently ensures high $U_\mathrm{U}$ values, maintaining stability and meeting the IC and IR constraints across various network conditions.

\section{Semantic Communications}\label{section6}
In this section, we consider the SemCom technique and explore the involvement of GDM within the SemCom framework~\cite{yang2022semantic,lin2023semantic,lin2023unified}.
\subsection{Fundamentals of Semantic Communications}
SemCom~\cite{yang2022semantic} refers to extracting and transmitting the most relevant semantic information from raw data to the receivers using AI technology. It aims to lower network loads by selectively transmitting meaningful and contextually relevant information instead of transmitting the entire raw data~\cite{du2023semantic}. SemCom consists of three main components: the semantic encoder, the wireless channel, and the semantic decoder~\cite{liang2023generative}. 

\subsubsection{Semantic Encoder}
It is responsible for extracting and transmitting relevant semantic information from the raw data provided by the transmitting users~\cite{du2023generativeica}. This is typically achieved by utilizing neural networks, which encode the raw data into meaningful semantic representations. The semantic encoder employs various techniques such as feature extraction and dimensionality reduction to capture the essential semantic information~\cite{du2022rethinking}.

\subsubsection{Wireless Channels}
However, during transmission, the semantic information is subject to physical noise introduced by the wireless channel ~\cite{kang2022personalized}. Physical noise refers to external factors that interfere with the transmission of the message. It can result in noise-corrupted semantic information, which is then transmitted to the receivers for further processing. The channel component of SemCom handles the transmission of this noise-corrupted semantic information, taking into account the wireless channel characteristics and the potential effects of noise and interference~\cite{van2023generative}.

\subsubsection{Semantic Decoder}
The receivers employ a semantic decoder, e.g., implemented by neural networks, to decode the received noise-corrupted semantic information and reconstruct the distorted data. The semantic decoder utilizes its learning capabilities to reverse the encoding process and extract the intended semantic meaning from the received information \cite{lin2023commag}. Semantic noise arises from the use of symbols that are ambiguous to the receivers. It can also occur when there is a mismatch in understanding between the sender and receiver. By employing sophisticated neural network architectures, the semantic decoder aims to minimize the effects of semantic noise and accurately obtain the original semantic data. 

The ultimate objective of SemCom is to effectively convey the intended meaning of the transmitted symbols, rather than transmitting the raw bits directly, thereby reducing communication overhead and enhancing communication effectiveness \cite{lin2023semantic}.

\subsection{Case Study: GDM-based Resource allocation for SemCom-aided AIGC services}
\begin{figure}[!t]
\centering
\includegraphics[width=0.45\textwidth]{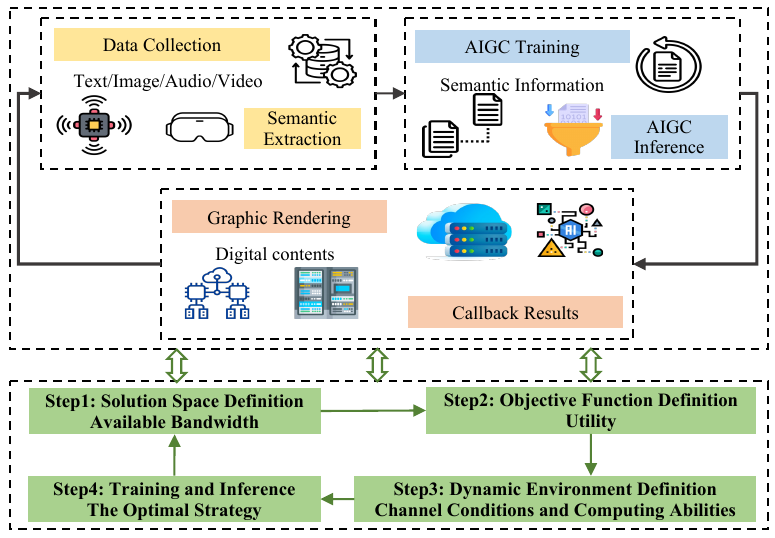}
\caption{Resource allocation problem in a SemCom-aided AIGC service scenario. First, the edge devices collect raw data, e.g., photos, and extract semantic information. Then, the AIGC service providers use the received semantic information to perform the AIGC inference using GenAI models to obtain meaningful content, e.g., animated style photos. These contents are further used by the multimedia service provider, e.g., Metaverse service provider, to render digital content for the users, e.g., animated style avatars \cite{lin2023unified}.}
\label{fig:semantic_framework}
\end{figure}

\subsubsection{Motivation}

There are several examples of integrating GenAI technologies in SemCom \cite{lin2023blockchain}. For instance, GANs have been employed to develop semantic decoders that tackle the out-of-distribution problem of SemCom \cite{zhang2022deep}. GANs are used to generate realistic and meaningful semantic information based on the available data. Additionally, a variational autoencoder (VAE) is utilized to calculate the lower bound of semantic distortion and derive the corresponding loss function \cite{alemi2016deep}. By incorporating GANs and VAEs, SemCom can enhance the accuracy and fidelity of semantic decoding, thereby improving the overall communication performance~\cite{liang2023generative}. 

To elucidate the role of GDMs in SemCom, we consider their application in an AIGC service process, illustrated in Fig. \ref{fig:semantic_framework}. 
The process begins with edge devices collecting primary data, such as photographs. These edge devices then extract semantic information from the data, focusing on meaningful content rather than raw data transmission. The extracted semantic information is significant for AIGC Service Providers (ASPs). Then, ASPs employ GenAI models, inclusive of GDMs, to conduct AIGC inference, transforming semantic information into enriched content, such as stylized animations~\cite{wang2023unified}. 
The final stage involves multimedia service providers, like Metaverse platforms, leveraging this semantically-enriched content to craft digital offerings for end-users, such as animated avatars~\cite{lin2023unified}.
We formulate a unified resource allocation problem for this workflow, considering the limited computing and communication resources allocated to the semantic extraction, AIGC inference, and graphic rendering modules. The objective is to maximize the overall utility by efficiently allocating these resources.

\subsubsection{Problem Formulation}
The integration gain includes the computing time for semantic extraction ($T_s^{comp}$), AIGC inference ($T_a^{comp}$), and graphic rendering ($T_m^{comp}$). These times are influenced by the available computing resources and the current computing resource congestion, introducing uncertainty to the utility optimization problem. Concurrently, the transmission time is associated with the transfer of semantic information ($T_a^{comm}$), AIGC content ($T_{m,u}^{comm}$), and rendering results ($T_{m,d}^{comm}$). These times are affected by the allocated communication resources to each part. Specifically, we consider the allocation of bandwidth resources with $W_a^m$, $W_m^s$, and $W_s^a$ denoting the bandwidths for semantic information, AIGC content, and rendering results transmissions, respectively. The objective function is given by ${\ln \left( { R_s^a} \right) + \ln \left( { R_a^m} \right)} + \ln \left( {R_m^s} \right)$, where $R_s^a$, $R_s^a$ and $R_s^a$ are the data rates for the transmissions of semantic information, AIGC content, and rendering results, respectively. The logarithmic form is used as we assume that the subjective user experience follows a logarithmic law to the objective performance metrics~\cite{du2023attention}. The objective function is considered as the reward in the GDM-based resource allocation scheme to find a near-optimal strategy. Following \cite{liu2019deep,yan2022qoe}, we construct the bandwidth allocation problem as follows:

\begin{equation}
\begin{array}{*{20}{l}}
{}\\
{\begin{array}{*{20}{l}}
{\mathop {\max }\limits_{W_a^m,W_m^s,W_s^a} }&{\ln \left( { R_s^a} \right) + \ln \left( { R_a^m} \right)} + \ln \left( {R_m^s} \right),
\\
{\quad\:\:{\rm{s}}{\rm{.t}}{\rm{.}}}&{T_s^{comp} + T_a^{comm} + T_a^{comp}}\\
{}&{\quad + T_{m,u}^{comm} + T_{m,d}^{comm} + T_m^{comp} \le {T_{\max }},}
\vspace{0.1cm}
\\
{}&{W_a^m + W_m^s + W_s^a \le {W_{\max }}}.
\end{array}}
\end{array}
\end{equation}

\subsubsection{GDM-based Resource Allocation Scheme Generation}
The optimal bandwidth resource allocation scheme can be generated according to the following steps
\begin{itemize}
\item \textbf{Step 1: Solution Space Definition:}  The solution space in the proposed problem encompasses allocating available bandwidth for transmission among the semantic extraction, AIGC inference, and rendering modules. The goal is to optimize the utilization of bandwidth resources to ensure efficient communication and collaboration between these modules.

\item \textbf{Step 2: Objective Function Definition:} The training objective of the proposed problem is to maximize the utility of the system, which is served as rewards that are obtained by dynamic resource allocation strategies. It should consider the total tolerable transmission time and available resources among these modules.  

\item \textbf{Step 3: Dynamic Environment Definition:} GDMs are utilized to generate an optimal bandwidth allocation scheme based on a given set of wireless channel conditions and computing capabilities involved in the three modules, such as the semantic entropy and the transmit power. Semantic entropy is defined as the minimum expected number of semantic symbols about the data that is sufficient to predict the task \cite{yan2022qoe}. The semantic entropy and the transmit power are randomly varied within a specific range associated with a given task.   

\item \textbf{Step 4: Training and Inference:} The conditional GDM generates the optimal bandwidth allocation strategy by mapping different environments to bandwidth allocation designs. The optimal strategy is achieved through the reverse process, where the GDM trains and infers the corresponding allocation policies to maximize the expected cumulative utility.
\end{itemize}

\subsubsection{Numerical Results}
As studied in \cite{lin2023unified}, the proposed method is implemented on a system running Ubuntu 20.04, equipped with a 32-core CPU and an NVIDIA RTX A5000 GPU. The dynamic environment parameters are sampled using uniform distributions, while the additive Gaussian noise is applied by sampling from normal distributions within the AIGC and rendering modules. {\color{blue}Fig. 5 in \cite{lin2023unified}} presents the test reward results for GDM and DRL, i.e., PPO in the bandwidth allocation task. This comparison is conducted over 400 training epochs with learning rates set at $3\times10^{-7}$ and $3\times10^{-6}$, buffer size 1,000,000, and an exploration noise of 0.01 according to \cite{lin2023unified}. As depicted in {\color{blue}Fig. 5 in \cite{lin2023unified}}, the curve for DRL exhibits greater volatility compared to that of GDM. Besides, the reward values for GDM are more compact, indicating more stable performance. As the number of training epochs increases, neither exhibits a clear upward or downward trend, which confirms both GDM and DRL converge. Therefore, GDM outperforms DRL in the bandwidth allocation task. To compare the utilities generated by various bandwidth allocation strategies, characterized by the parameters $[W_s^a, W_a^m, W_m^s]$, GDM and PPO select two distinct network states under dynamic network conditions. These are designated as $\mathsf{GDM}_1$, $\mathsf{GDM}_2$, $\mathsf{PPO}_1$, and $\mathsf{PPO}_2$. The definition of network states {\color{blue}follows} that presented in \cite{lin2023unified}. As shown in {\color{blue}Fig. 6 in \cite{lin2023unified}}, the strategies exhibit close alignment in the allocation of bandwidth for $W_a^m$, yet there are considerable differences in the other two parameters, $W_s^a$ and $W_m^s$. There is a significant variation in allocating different types of bandwidth across various network states. Additionally, the strategies generated by GDM demonstrate higher utilities than PPO across different network states. Therefore, GDM outperforms PPO in terms of generated strategies in dynamic environments. This superiority can be attributed to the optimal bandwidth allocation mechanism inferred by GDMs, which enables fine-tuning output through denoising steps and facilitates exploration. Consequently, the proposed mechanism exhibits enhanced flexibility, mitigating the effects of uncertainty and noise encountered during the transmission and computing among semantic extraction, AIGC inference, and graphic rendering modules.



\section{Internet of Vehicles Networks}\label{section7}
In this section, we introduce the concept of IoV networks, discuss the role of GDM in IoV networks, and provide a case study~\cite{zhou2020evolutionary,zhang2023generative}.

\subsection{Fundamentals of IoV Networks}
Drawing inspiration from the Internet of Things (IoT), the IoV network turns moving vehicles into information-gathering nodes \cite{zhou2020evolutionary,9136587}. Harnessing emerging information and communication technologies facilitates network connectivity between vehicles and other elements, i.e., other vehicles, users, infrastructure, and service platforms. For the IoV network, the goal is to enhance the overall intelligence of the vehicle, as well as improve the safety, fuel efficiency, and driving experience \cite{8967260}.

In the IoV network, vehicles are regarded as data agents for collecting and disseminating data such as traffic patterns, road conditions, and navigation guidance \cite{7913583}. Managing large amounts of data in the IoV network is a very complex task. As a remedy, GenAI is proposed. In particular,  GenAI performs the critical functions of organizing and restoring the data collected within the IoV. Additionally, it can generate synthetic data, enhancing the efficacy of machine learning model training within the network. Furthermore, the contributions of GenAI go beyond simple data management. It utilizes the collected data to inform the real-time decision-making process. This includes predicting traffic conditions, identifying potential hazards, and determining the best route for the driver.

\subsection{Applications of GDM in IoV Networks}
The field of GenAI is composed of several models, and each model brings unique capabilities to various applications. The GDM has attracted much attention among these models due to its unique advantages.  Applying the GDM model within IoV networks yields promising results. In particular there are two specific applications as follow:

\subsubsection{Recovery of Images sent by vehicles}
In IoV networks, vehicles usually transmit images to communicate information about their environment for safe driving. However, these images may be distorted or lose quality due to transmission errors, noise, or interference. The GDM, with its ability to generate high-quality images, can be employed to recover the original quality of these transmitted images. In particular, the vehicles adopt semantic technology to extract information from images, i.e., as a prompt at the transmitter, and recover it using GDM at the receiver. By doing so, the transmitted data and communication delays can be reduced in IoV.

\subsubsection{Optimization Based on GDM}
The GDM iterative framework suits the IoV network optimization tasks, including path planning and resource allocation \cite{Huang_2023_CVPR}. Using stochastic differential equations (SDEs), the model refines solutions progressively via a diffusion process. For example, in path planning, GDM begins with a random path, making iterative refinements based on performance criteria such as travel time and energy consumption. The model uses gradients of these metrics to guide the path updates toward an optimal or near-optimal solution, stopping iterations when updates become negligible.

Therefore, thanks to the ability to recover high-quality images from transmitted data and iteratively optimize solutions, the GDM provides a powerful tool for enhancing the efficiency and robustness of IoV networks. 

\begin{figure}[!t]
\centering
\includegraphics[width=0.45\textwidth]{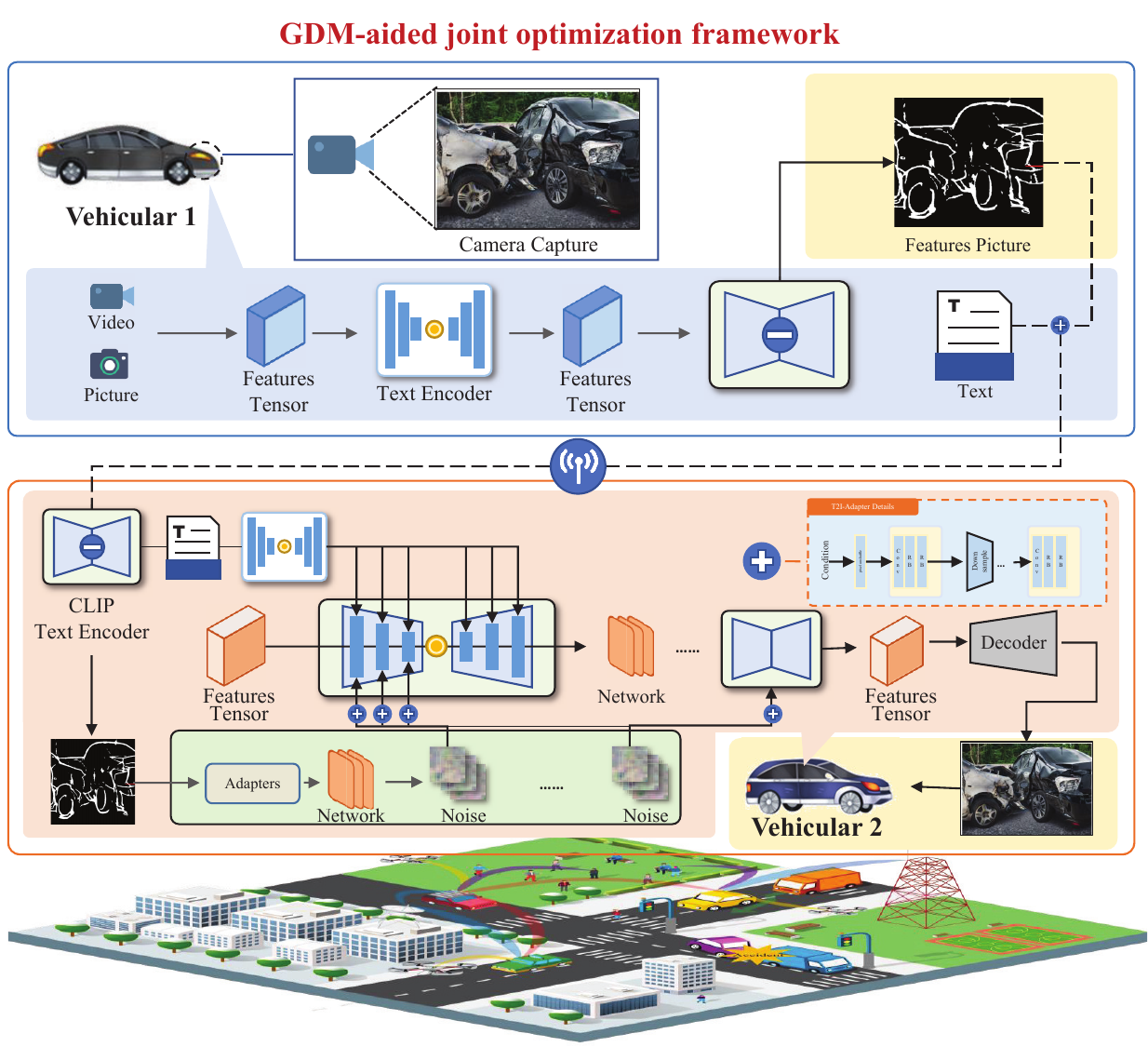}
\caption{GenAI-enabled IoV network, where the semantic information extraction step, image skeleton extraction step, wireless transmission step,  GenAI-enabled 
image generation step and image reconstruction step are involved \cite{zhang2023generative}.}
\label{fig:IoV}
\end{figure}
\subsection{Case Study: A GenAI-driven IoV network}
In this part, we conduct a case study to illustrate how to apply GDMs in  IoV design.

\subsubsection{System Model}
Under the 3GPP V2X standard \cite{8792382}, we consider a GenAI-driven IoV network with multiple V2V links as shown in Fig.~\ref{fig:IoV}. We aim to ensure reliable, real-time information transmission in our considered network. The orthogonal frequency division multiplexing technology is adopted, where each V2V link can achieve dynamic transmission rates on different sub-channels. Moreover, a successful image transmission rate is introduced as a constraint. This rate is affected by different parameters such as achievable transmission rate, image similarity measure, channel coherence time, and generated image payload.

\subsubsection{Problem Formulation}
In our considered work, We consider transmission rate and image similarity as the performance indicators, and hence they are combined into a unified QoE indicator and used as the optimization goal. As described in (\ref{eq_IOV}), an optimization problem is formulated to maximize the system QoE under the constraints of the transmission power budget and the probability of successful transmission for each vehicle, where the channel selection strategy, the transmission power for each vehicle, and the diffusion steps for inserting the skeleton are jointly optimized.
\begin{subequations}
\label{eq_IOV}
\begin{flalign}
\mathop{\rm max}\limits_{\{P_{v}, d_{v}, c_{v}\}}{\kern 1pt} {\kern 1pt} &\sum_{v \in V} {\rm QoE}(v)  \\
 {{\rm s.t.}\quad\:} &\sum_{v \in V} p_v \leq P_{\text{max}}, \text{(Power Budget)}\\
&{\rm Pr}(v) \geq {\rm Pr}_{\text{min}},  \text{(Transmission Constraint)} \\
&c_v \in C, \text{(Channel Selection Constraint)}\\
&d_v \in \mathbb{N}^+, \text{(Diffusion Steps Constraint)}\\
&\forall v \in V.  \nonumber
\end{flalign}
\end{subequations}

\subsubsection{GDM-based Joint Channel Selection and Power Allocation}
For the formulated problem, a GDM-based DDPG approach is proposed, where the corresponding three tuples of MDP and the network design are as follows.
\begin{itemize}
\item \textbf{MDP design:}  The state space consists of the current information and previously selected actions, where the current information includes the channel information of each V2V link, the transmission rate of each V2V link, and the generated image payload. The action space consists of the selectable channel, the transmit power, and the diffusion steps for inserting the skeleton. The reward function consists of an instant reward term and a penalty term. The design principle follows that a larger penalty will be given when the constraints are not met, while an instant reward will be given when the constraints are met or the goal becomes higher. Accordingly, the agent can achieve high QoE while satisfying the corresponding constraints.
\item \textbf{GDM-based IoV network design:}  
In our proposed approach, we adopt the GDM-based network. Specifically, the GDM-based network design employs GDMs in two distinct roles. Firstly, GDMs reconstruct received images at the receivers in vehicular networks. Leveraging the multi-modal technique, we utilize the contrastive language-image pre-training (CLIP) framework to incorporate both text and image information in the diffusion process for image reconstruction, which is a task that incorporates denoising steps for image generation and transmits power values. 
Secondly, another GDM is tasked with optimizing the number of denoising steps, the channel selection strategies, and the transmit power values. In particular, the IoV network uses a diffusion process to map environmental states to resource allocation strategies, incorporating a crucial denoising step to eliminate less important information and enhance signal clarity during training.
The corresponding network operates through a chain mechanism, where each step incrementally refines the solution, ensuring it adapts to temporal dependencies and dynamic environments. This approach can be fine-tuned to generate samples over multiple time steps, enhancing its ability to handle tasks with long-term dependencies.
\end{itemize}


\subsubsection{Numerical Results} 
We conduct experiments to prove the validity of our proposed method.  In our simulation setup, the GDM-based approach utilizes a learning rate 3e-7 for both the actor and the critic network. The exploration noise is set at 0.01, and the time step of the diffusion chain is 1. We employ a tanh activation function with a hidden layer of 256 units. The output layer is designed as the cardinality of the action space, while the input layer corresponds to the cardinality of the state space. The discount factor $\gamma$ is set to 0.95. It is shown that the average cumulative rewards obtained by different types of schemes versus the number of training episodes, where the curves have been smoothened to show the trend more clearly. Our proposed GDM-based approach always outperforms other baselines (i.e., DRL-DDPG, DRL-DQN, greedy, and random schemes) under the same parameter settings when all schemes converge.  Although the proposed GDM-based DDPG approach and DDPG-based obtain roughly similar rewards during the training phase, the proposed GDM-based DDPG approach outperforms DDPG after convergence. The reason is that traditional DRL methods may not be able to effectively filter out noise (i.e., useless information in the buffer) in environments. In contrast, the diffusion model in the GDM-based method enhances environment exploration, and its denoising process helps distinguish signal from noise, thereby improving learning results to find reasonable actions.

\section{Miscellaneous Issues}~\label{section77}
In this section, we discuss the applications of GDM to several other network issues, including channel estimation, error correction coding, and channel denoising.
\subsection{Channel Estimation}
\subsubsection{Motivations}
In wireless communication systems, the wireless channel depends on various factors such as fading, interference, and noise, which can lead to distortions in the received signal. Consequently, researchers introduce channel estimation techniques to estimate the channel response, which can be used to mitigate the impacts caused by the aforementioned factors, thereby enhancing the quality of the received signal. As such, accurate channel estimation is crucial for reliable communication and efficient use of the available bandwidth~\cite{dovelos2021channel}.

So far, several kinds of channel estimation techniques have been proposed, including pilot-based, compressed sensing-based, etc. The pilot-based methods use known pilot symbols inserted in the transmitted signal to estimate the channel response. For instance, the minimum mean square error (MMSE) based method achieves channel estimation by multiplying the received signal with the conjugate of the transmitted signal, followed by division by the sum of the power of the transmitted signal and the noise variance. This method not only minimizes the mean square error between the received signal and the estimated signal but also considers the noise variance, which is important for determining the reliability of the estimated channel coefficients~\cite{liu2014channel}. The compressed sensing-based methods exploit the sparsity of the channel response to estimate it from a small number of measurements. For example, the authors in~\cite{nawaz2012superimposed} create a training signal using a random sequence with a known pilot sequence.  At the receiver, first-order statistics and the compressed sensing method are applied to estimate the wireless channels with sparse impulse response. Unlike these two methods, data-driven methods employ machine learning algorithms to learn the channel response from the received signal without relying on any prior knowledge of the channel during the offline training phase. After trained, the data-driven methods can estimate the channel in an online phase. For instance, the authors in~\cite{liao2019chanestnet} first use the convolutional neural network (CNN) to extract channel response feature vectors, and then employs recurrent neural network (RNN) for channel estimation. Besides, there are some other techniques, such as optimization-based methods, which use mathematical optimization, such as convex optimization, to estimate the channel response, and hybrid methods that combine different techniques to improve the accuracy and efficiency of channel estimation.

While effective, existing methods still faces several challenges. One of the main challenges is the dynamic nature of the channel, which means that the channel can change rapidly due to various factors such as mobility and interference. This requires channel estimation to be robust to test-time distributional shifts~\cite{arvinte2022mimo}. These shifts naturally occur when the test environment no longer matches the algorithm design conditions, especially at the user side (could be transmitter or receiver), where the propagation conditions may change from indoor to outdoor, whenever the user is moving. An effective solution to this challenge is to use GenAI for robust channel estimation, because of the following main reasons.
\begin{itemize}
\item The GenAI model can extract complex patterns from large amount of data and learn in a changing environment. This not only enhances the model's generalization ability but also enables it to adapt to the dynamic characteristics of the channel, thereby improving the robustness of the estimation.
\item The GenAI model can directly learn the distribution of channel responses from the received signals and use the structure captured by the deep generative model as a prior for inference, eliminating the need for prior knowledge of the sparsifying basis.
\end{itemize}
Next, we further illustrate applications of GenAI in channel estimation, using MIMO channel estimation via the GenAI model as a case study~\cite{arvinte2022mimo}.

\subsubsection{Case Study: MIMO Channel Estimation Utilizing Diffusion Model}
\begin{figure}[!t]
\centering
\includegraphics[width=0.4\textwidth]{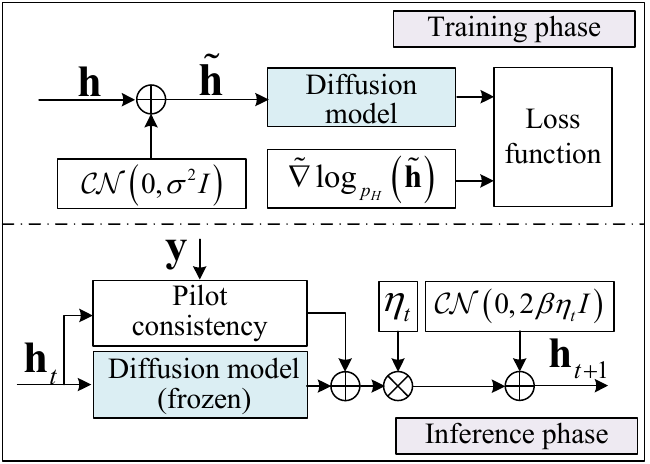}
\caption{During training, the noise is first added to ${\bf{h}}$ to obtain ${\bf{\tilde h}}$. Then a regression target for the gradient of ${\log _{{p_H}}}\left( {{\bf{\tilde h}}} \right)$ is produced. After that, the ${l_2}$-loss is used to train the parameters of the deep neural network via back-propagation. After training, the current channel estimate is updated by a pilot consistency term, a diffusion update, and added noise to achieve inference.}
\label{fig:GAICHN}
\end{figure}

Channel estimation using diffusion model~\cite{arvinte2022mimo} primarily involves training and inference phases, as shown in Fig.~\ref{fig:GAICHN}. The training phase involves using a deep neural network to learn the underlying structure of the channel from a set of noisy channel estimates. The main steps include the following:
\begin{itemize}
\item \textbf{Step 1}: Using the received pilot symbols to calculate the noisy channel estimation ${\bf{h}}$.
\item \textbf{Step 2}: Adding the noise to the training channel ${\bf{h}}$ to produce a perturbed channel ${\bf{\tilde h}}$.
\item \textbf{Step 3}: Computing the gradient of ${\log _{{p_H}}}\left( {{\bf{\tilde h}}} \right)$.
\item \textbf{Step 4}: Producing a regression target for the gradient using the diffusion model.
\item \textbf{Step 5}: Training the parameters of the deep neural network using back-propagation and the ${l_2}$-loss.
\end{itemize}

The inference stage involves utilizing the trained model to estimate the channel based on a set of received pilot symbols. The primary steps are as follows:
\begin{itemize}
\item \textbf{Step 1}: Updating the current channel estimation via the pilot consistency term, which enforces consistency between the received pilot symbols and the estimated channel.
\item \textbf{Step 2}: The diffusion update is applied to the channel estimate, which smooths out the estimate and helps to reduce noise.
\item \textbf{Step 3}: To prevent the model from converging to a sub-optimal solution, noise is added to the updated channel estimate at each step.
\item \textbf{Step 4}: The process is repeated until convergence, at which point the final estimate of the channel is produced.
\end{itemize}
It is noteworthy that the iterative algorithm operates independently of the training phase and can accommodate other impairments such as interference scenarios or few-bit quantization of the received pilots.

The proposed model is evaluated by training an NCSNv2 model~\cite{song2020improved} on complex-valued channel matrices. The model architecture, RefineNet~\cite{lin2017refinenet}, comprises eight layers and approximately 5.2 million parameters. To accommodate complex-valued inputs, the real and imaginary components of the matrix are processed as two separate input channels. Training is performed on a dataset of $20,000$ channel realizations, derived from the clustered delay line (CDL) channel model, with an equal distribution between two antenna spacings~\cite{arvinte2022mimo}.

Fig.~2 in~\cite{arvinte2022mimo} presents the test results for in-distribution CDL channels in a blind SNR configuration with $\alpha = 0.4$. The top plot reveals that the comparison algorithm, WGAN~\cite{balevi2020high}, captures some aspects of the channel structure for very low antenna spacing. However, its performance peaks, about -26 dB, rapidly in high SNR conditions. Another comparative algorithm, i.e., Lasso~\cite{schniter2014channel}, similarly exhibits a trend, with its peak value approximately at -22 dB. This effect is more pronounced with an antenna spacing of half wavelength and fewer structural components, indicating that neither baseline employs a suitable prior knowledge. In contrast, the diffusion-based approach exhibits a near-linear reduction in the normalized mean square error (NMSE), aligning with the theoretical findings in~\cite{jalal2021robust}, without explicit learning of a prior. At an SNR level of 15 dB, the NMSE of the diffusion-based approach is over 12 dB lower than both baseline methods, underscoring the superiority of the diffusion-based approach.

\subsection{Error Correction Coding}
\subsubsection{Motivations}
Developing codes that can be decoded effectively in noisy environments is imperative in wireless communications. Decoding methods fall into two categories: hard and soft decoding~\cite{biglieri2005coding}. Hard decoding strictly uses the most probable value of the received signal, ignoring any signal quality metrics. In contrast, soft decoding incorporates the most probable signal value and additional signal quality information, thus improving decoding accuracy. 
While these strategies offer some level of efficacy, decoding complexity escalates with advanced encoding systems, such as algebraic block codes, presenting significant challenges~\cite{biglieri2005coding}. Decoding these systems optimally often involves adhering to the maximum-likelihood principle—identifying the codeword that maximizes the likelihood of the received signal. However, this approach is identified as NP-hard, implying that an exhaustive search is generally required for the optimal solution, rendering it impractical for real-world applications.

Recent studies, notably those employing model-free machine learning approaches, have aimed at addressing this challenge~\cite{choukroun2022error}. Specifically, a transformer-based decoder, which integrates the encoder within its architecture, demonstrated superior performance over traditional methods with significantly reduced time complexity, as detailed in~\cite{choukroun2022error}. Despite these advancements, the model-free paradigm faces critical limitations. Firstly, it demands substantial storage and memory capacity, posing issues for resource-limited devices. Secondly, its non-iterative nature mandates a uniform, computationally demanding neural decoding procedure, irrespective of the extent of codeword corruption.

To this end, GDMs have been explored for decoding tasks, as evidenced by recent works~\cite{choukroun2022denoising, bao2024improving}. GDMs employ an iterative decoding approach while efficiently adapting to varying degrees of codeword corruption and reducing computational complexity. 
Specifically, the authors in ~\cite{choukroun2022denoising} consider the corruption of channel codewords as a forward diffusion process of GDM. This perspective allows the corruption to be methodically reversed using an adaptive denoising diffusion probabilistic model, presenting a sophisticated yet efficient solution for error correction and signal restoration in communication systems. Moreover, the authors in~\cite{bao2024improving} proposed a diffusion-based image restoration method, Diffusion-based Error Contraction and Correction (DiffECC), leveraging an Ordinary Differential Equation (ODE)-based sampler to formulate a detailed update equation for conditional diffusion. The application of the Adam optimizer enhances neural estimations. The objective is for backward diffusion to reach a stable point at each timestep, aiming for the error term $\epsilon_\theta\left(x_t, t\right)$ to meet a set error benchmark $\epsilon$, especially in generating clean images. For image restoration, where unknown factors initially distort inputs, DiffECC innovatively adjusts neural predictions by amalgamating outputs from consecutive denoising stages as a regularization factor.

\subsubsection{Case Study: Denoising Diffusion Error Correction Codes}

\begin{figure}[!t]
\centering
\includegraphics[width=0.4\textwidth]{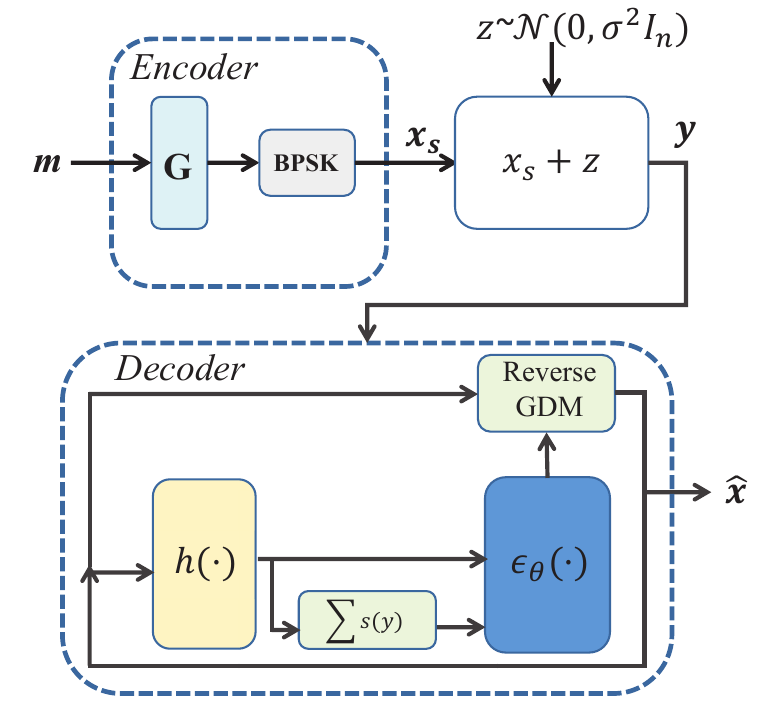}
\caption{The denoising diffusion error correction codes architecture, where the decoding is performed via the reverse diffusion process \cite{choukroun2022denoising}.}
\label{fig:code_error}
\end{figure}

As shown in Fig.~\ref{fig:code_error}, the elements of the denoising diffusion used for decoding and the proposed architecture are summarized, where the training process is as follows.
\begin{itemize}
\item \textbf{Decoding as a Reverse Diffusion Process.} In this stage, a process of ``forward diffusion'' is used to process codewords sampled from a particular encoding distribution. Specifically, the process gradually transmits codewords by gradually adding a small amount of Gaussian noise, with the size of each step controlled by a specific variance table. Next, data transmission over a noisy communication channel is regarded as a modified iterative diffusion process that requires inversion at the receiving end to decode the original data. Finally, decoding is regarded as a reverse diffusion process, transforming the posterior probability into a Gaussian distribution as per the Bayesian theorem~\cite{choukroun2022error}. The goal of the decoder can be defined to predict the channel's noise.

\item \textbf{Denoising via Parity Check Conditioning.} In the decoding process, it is regarded as the reverse denoising process of the GDM, which relies on time steps and can reverse the entire diffusion process by sampling Gaussian noise corresponding to the final step. During training, a time step is randomly sampled, generating noise and a syndrome requiring correction. Owing to its invariance to the transmitted codeword, diffusion decoding can be trained using a single codeword. During inference, the denoising model predicts multiplicative noise, converts it into additive noise, and performs the gradient step in the original additive diffusion process. 
\end{itemize}


Fig.~4 in~\cite{choukroun2022denoising} shows BER obtained by three schemes in terms of the normalized SNR values, i.e., $E_b/N_0$ (EbNo), over the Rayleigh fading channel environment. It shows that with the increment of the value of EbNo, the GDM-based scheme is superior to other benchmarks. In particular, when the EbNo is 4 {\rm dB}, the BER obtained by GDM scheme is 50\% of that obtained by Binary Phase (BP) scheme, and 11\% of that obtained by error correction code transformer (ECCT) scheme~\cite{choukroun2022error}. The reason is that the GDM is able to learn to decode, even under some serious noisy fading channels.

\subsection{Channel Denoising}
\subsubsection{Motivations}
GDM-based models are characterized by the ability to add Gaussian noise to the training data gradually and then learn to restore the original data from the noise through a back sampling process. The process is similar to that of a receiver in a wireless communication system, which is required to recover the transmitted signal from the noisy received signal.

Thus, in \cite{wu2024cddm}, a GDM-based approach for denoising wireless communication channels is introduced to predict and mitigate channel noise for post-channel equalization and enhance overall system performance. Distinctively, this GDM-based model proposed in~\cite{wu2024cddm} operates solely on the principles of forward diffusion, independent of any received signal. When integrated into semantic communication systems utilizing Joint Source-Channel Coding (JSCC), the GDM-based model significantly minimizes the disparity between transmitted and received signals across both Rayleigh fading and additive white Gaussian noise (AWGN) channels. Furthermore, the authors in \cite{kim2023learning} assess the diffusion model's capabilities in channel generation and its performance in End-to-End (E2E) communication scenarios, subject to AWGN and authentic Rayleigh fading channels. Their findings validate the diffusion model's ability to learn the channel distribution accurately. Additionally, it is shown that the E2E framework, facilitated by the diffusion model, achieves a symbol error rate remarkably comparable to that obtained with a channel-aware framework, applicable to both AWGN and Rayleigh fading environments.

\subsubsection{Case Study: GDM-based Channel Denoising Model}

\begin{figure}[!t]
\centering
\includegraphics[width=0.46\textwidth]{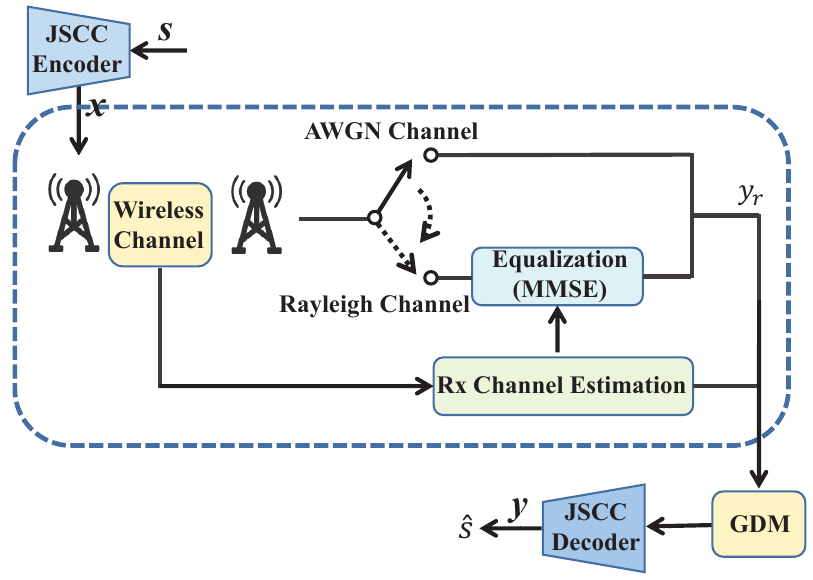}
\caption{The joint GDM and JSCC system architecture, where GDM is trained using a specialized noise schedule \cite{wu2024cddm}.}
\label{fig:Channel_Denoising}
\end{figure}
As shown in Fig.~\ref{fig:Channel_Denoising}, the joint GDM and JSCC architecture is summarized, where the training process is as follows.
\begin{itemize}
\item \textbf{Conditional Distribution of The Received Signals:} Real-valued and complex-valued symbols are transformed and transmitted in the wireless channel, where the transformation combines the effects of Rayleigh fading gain and additive white Gaussian noise. The received signal is then processed through an MMSE equalizer to produce an equalized complex signal. Study conditional distributions of real-valued vectors using known signal and channel state information. Based on the noise impact and channel state, the signal is reparameterized and a GDM-based channel denoising model is trained to obtain noise estimates.

\item \textbf{Training Algorithm of GDM:}
In the training process of GDM, the original source signal is first represented in a new parameterized form. At the beginning of training, the Kullback-Leibler divergence~\cite{complexity} is mainly used to optimize the variational upper bound of the negative log-likelihood. During training, the optimal value of a key hyper-parameter is required to be determined. 
Next, the optimization objective for a series of loss functions is simplified by re-parameterization and re-weighting methods. Finally, the overall loss function is minimized, effectively recovering the original source signal.
\end{itemize}

Figs.~5 and 6 in~\cite{wu2024cddm} show PSNR obtained by three schemes regarding the SNR over the AWGN channel and Rayleigh fading channel environments. To achieve optimal performance, both GDM-based JSCC scheme and JSCC scheme must be retrained for a given SNR. It shows that for different values of SNR, the GDM-based JSCC scheme is superior to others. For example, over Rayleigh fading channel with SNR of 20 {\rm dB}, compared with the JSCC scheme, the GDM-based JSCC scheme can obtain about 1.06 {\rm dB} gain.

\section{Future Directions}\label{section8}
This section elucidates potential research avenues warranting further examination.
\subsection{Space-air-ground Integrated Network}
The Space-Air-Ground Integrated Network (SAGIN) is a promising paradigm for future wireless networks, characterized by its three-dimensional coverage, high capacity, and reliable communications~\cite{cheng2018air,cao2021hap,du2022performance}. However, the optimization of SAGIN is a complex task due to the high dimensionality of the network configuration, the heterogeneity of the network elements, and the dynamic nature of the network environment~\cite{li2022age,jia2021toward}. GDMs, with their ability in complex data distribution modeling, could be a powerful tool for optimizing SAGIN~\cite{zhang2023generativeSAGIN}. 
\begin{itemize}
\item {\textbf{Dynamic Network Environment Modeling and Prediction:}} The dynamic nature of the SAGIN environment poses a significant challenge for its optimization~\cite{cao2021hap,cui2022space}. GDMs can be used to model and predict these dynamic network environments. This would allow for more efficient resource allocation, network scheduling, and routing strategies, as the predictions could provide valuable insights into future network states~\cite{cheng2020comprehensive}.
\item {\textbf{Synthetic Network Scenario Generation:}} Testing and validating network optimization algorithms require a variety of network scenarios~\cite{liu2018space}. GDMs can generate synthetic network scenarios that closely mimic real-world conditions, providing a robust platform for testing and validating these algorithms.
\item {\textbf{Network Scheduling and Routing:}} SAGIN involves a variety of network elements, each with its unique characteristics and requirements~\cite{ye2020space, 10058144}. GDMs can capture these unique characteristics and model the complex interactions between different network elements, facilitating more efficient network scheduling and routing strategies. 
\end{itemize}

\subsection{Extremely Large-Scale MIMO}
Extremely Large-Scale MIMO (XL-MIMO) is an emerging technology that is expected to play a pivotal role in the 6G of wireless mobile networks~\cite{wang2024tutorial,wang2023extremely,du2021millimeter}. XL-MIMO offers vast spatial degrees of freedom by deploying an extremely large number of antennas, leading to significant enhancements in spectral efficiency and spatial degrees of freedom. However, implementing XL-MIMO introduces new challenges, including the need for more flexible hardware designs, a much larger number of antennas, smaller antenna spacing, new electromagnetic characteristics, and near-field-based signal processing schemes~\cite{wang2022uplink,du2022performancethz}. GDMs can be instrumental in addressing these challenges and optimizing the performance of XL-MIMO systems. Here are some potential research directions:
\begin{itemize}
\item {\textbf{Hybrid Channel Estimation and Modeling:}} XL-MIMO systems involve a large number of antennas, leading to high-dimensional data~\cite{wang2023uplink}, and also the co-existence of near-field and far-field channels within the coverage of cellular networks. Especially, in the near-field channel, the channel response vectors depend on both the distance and direction between the transceiver of each antenna element, unlike the far-field channel. 
Therefore, the increased ``huge'' complexity for near-field channel estimation may not be resolved with the conventional approaches. GDMs can be used to model and estimate such hybrid channel state information efficiently. They can exploit the inherent graph structure in the spatial domain, where antennas can be considered as nodes and the spatial correlation between antennas as edges. This can lead to more accurate and efficient channel estimation methods.
\item {\textbf{Signal Processing:}} The signal processing in XL-MIMO systems can be complex due to the large number of antennas and the near-field communication characteristics. Especially, in the latter case, the interference caused by multi-user transmissions can be effectively mitigated by utilizing the higher degree of freedom existing in the distance and direction of near-field channel response vectors. GDMs can be used to develop efficient signal processing algorithms that can handle high-dimensional data and exploit the spatial correlation in the antenna array. This can lead to improved performance in terms of data rate and reliability.
\item {\textbf{Hardware Design and Implementation:}} XL-MIMO systems involve different hardware designs, such as uniform linear array (ULA)-based, uniform planar array (UPA)-based, and continuous aperture phased (CAP)-based XL-MIMO. GDMs can be used to model and analyze these different designs, helping to understand their characteristics and interrelationships. This can guide the design and implementation of XL-MIMO systems.
\end{itemize}


\subsection{Integrated Sensing and Communications}
The ISAC unifies wireless sensing and communication systems to efficiently employ limited resources for mutual benefits~\cite{cui2021integrating}. It is a key element in future wireless systems, supporting various applications like autonomous driving and indoor localization~\cite{cheng2022integrated,wang2020twpalo}. The GDM can be utilized in ISAC systems for both data processing and generation. As a processing technique, it can classify and recover ISAC-related data. Moreover, it can generate synthetic ISAC data, a vital function for boosting the training efficiency of neural networks within the ISAC systems. Specifically, GDM has applications in various aspects of the ISAC system.

\begin{itemize}
\item {\textbf{ISAC Data Generation:}} The GDM can be used to generate samples for ISAC network training. For example, in indoor localization based on received signal strength indication (RSSI), the authors in~\cite{njima2021indoor} proposed a GAN for RSSI data augmentation. This network generates fake RSSI based on a small set of real collected labeled data. Using these data, the experimental results show that overall localization accuracy of the system has improved by 15.36\%. Compared to GAN, GDM has stronger inference capabilities, which enable it to generate better fake data, thereby further enhancing system performance.
\item {\textbf{ISAC Data Processing:}} Apart from data generation, GenAI models are also commonly used to process ISAC data~\cite{chen2020fido}. For instance, given that the GAN-based semi-supervised learning can handle unlabeled and labeled data, the authors in~\cite{xiao2019csigan} introduced a complement generator that uses a limited amount of unlabeled data to generate samples for training the discriminator. Building on this, they further adjust the number of probability outputs and utilize manifold regularization to stabilize the learning process, enhancing the human activity recognition performance in both semi-supervised and supervised scenarios.
\end{itemize}  

\subsection{Movable Antenna System}
The future of wireless communication networks is expected to be shaped significantly by the integration of movable antennas~\cite{new2023fluid,khammassi2023new}. Movable or fluid antennas, unlike conventional fixed-position antennas, have the capability of flexible movement and can be deployed at positions with more favorable channel conditions to achieve higher spatial diversity gains~\cite{shojaeifard2022mimo}. This flexibility enables better coverage and adaptability to changing environmental conditions. By strategically relocating the antenna, it becomes possible to mitigate signal blockage or interference caused by various obstacles, including buildings and vegetation. Therefore, the movable antennas can reap the full diversity in the given spatial region~\cite{shojaeifard2022mimo}. The complex and dynamic nature of wireless environments, characterized by high-dimensional configurations and non-linear relationships, necessitates sophisticated models like GDMs that can capture such high-dimensional and complex structures.
\begin{itemize}
\item {\textbf{Optimization of Antenna Positioning:}} GDMs can be used to optimize the positioning of movable antennas in real time. By modeling the wireless environment and the effects of different antenna positions, GDMs can generate optimal antenna positions that maximize signal strength and minimize interference.
\item {\textbf{Dynamic Resource Allocation:}} GDMs can be applied to the dynamic resource allocation problem in movable antennas. By modeling the resource demands and availability in the network, GDMs can generate optimal resource allocation strategies that balance the needs of different network users and maximize network efficiency~\cite{tlebaldiyeva2022enhancing}.
\item {\textbf{Predictive Maintenance:}} Based on historical data, GDMs can be used to predict potential failures in movable antennas. By modeling antenna performance and failure patterns, GDMs can generate predictions about future failures, allowing for proactive maintenance and minimizing network downtime.
\item {\textbf{Integration with Reinforcement Learning:}} As demonstrated in Section~\ref{section3}, the integration of GDMs with reinforcement learning techniques can be further explored in the context of movable antennas. This can lead to more robust and efficient resource slicing and scheduling strategies, enhancing the performance of 5G networks \cite{10038824} and autonomous vehicles \cite{lin2023drl}.
\end{itemize}               

\section{Conclusions}\label{section9}
In this tutorial, the transformative potential of GDMs in intelligent network optimization has been thoroughly explored. The unique strengths of GDMs, including their broad applicability and capability to model complex data distributions, were studied. 
We highlighted their potential in enhancing the DRL algorithms and providing solutions in key intelligent network scenarios, such as incentive mechanism design, SemCom, IoV networks, channel estimation, error correction coding, and channel denoising.
These explorations demonstrated the practicality and efficacy of GDMs in real-world applications.
The tutorial concluded by emphasizing the research directions of GDMs in shaping the future of intelligent network optimization and encouraging further exploration in this promising field.

\bibliographystyle{IEEEtran}
\bibliography{citations}

\begin{thebibliography}{100}
\providecommand{\url}[1]{#1}
\csname url@samestyle\endcsname
\providecommand{\newblock}{\relax}
\providecommand{\bibinfo}[2]{#2}
\providecommand{\BIBentrySTDinterwordspacing}{\spaceskip=0pt\relax}
\providecommand{\BIBentryALTinterwordstretchfactor}{4}
\providecommand{\BIBentryALTinterwordspacing}{\spaceskip=\fontdimen2\font plus
\BIBentryALTinterwordstretchfactor\fontdimen3\font minus
  \fontdimen4\font\relax}
\providecommand{\BIBforeignlanguage}[2]{{%
\expandafter\ifx\csname l@#1\endcsname\relax
\typeout{** WARNING: IEEEtran.bst: No hyphenation pattern has been}%
\typeout{** loaded for the language `#1'. Using the pattern for}%
\typeout{** the default language instead.}%
\else
\language=\csname l@#1\endcsname
\fi
#2}}
\providecommand{\BIBdecl}{\relax}
\BIBdecl

\bibitem{jovanovic2022generative}
M.~Jovanovic and M.~Campbell, ``Generative artificial intelligence: {T}rends
  and prospects,'' \emph{Computer}, vol.~55, no.~10, pp. 107--112, Oct. 2022.

\bibitem{korzynski2023generative}
P.~Korzynski, G.~Mazurek, A.~Altmann, J.~Ejdys, R.~Kazlauskaite,
  J.~Paliszkiewicz, K.~Wach, and E.~Ziemba, ``Generative artificial
  intelligence as a new context for management theories: {A}nalysis of
  {ChatGPT},'' \emph{Central Eur. Manag. J.}, 2023.

\bibitem{peres2023chatgpt}
R.~Peres, M.~Schreier, D.~Schweidel, and A.~Sorescu, ``On {ChatGPT} and beyond:
  {H}ow generative artificial intelligence may affect research, teaching, and
  practice,'' \emph{Int. J. Res. Mark.}, 2023.

\bibitem{van2023processgan}
C.~van Dun, L.~Moder, W.~Kratsch, and M.~R{\"o}glinger, ``Process{GAN}:
  {S}upporting the creation of business process improvement ideas through
  generative machine learning,'' \emph{Decis. Support Syst.}, vol. 165, p.
  113880, 2023.

\bibitem{Accenture}
Accenture, ``2023 technology vision report,''
  \url{https://www.accenture.com/us-en/insights/technology/technology-trends-2023}.

\bibitem{ni2023generative}
B.~Ni, D.~L. Kaplan, and M.~J. Buehler, ``Generative design of de novo proteins
  based on secondary-structure constraints using an attention-based diffusion
  model,'' \emph{Chem}, 2023.

\bibitem{srinivasan2021biases}
R.~Srinivasan and K.~Uchino, ``Biases in generative art: {A} causal look from
  the lens of art history,'' in \emph{Proc. ACM Conf. Fair. Account. Transp.},
  2021, pp. 41--51.

\bibitem{vaswani2017attention}
A.~Vaswani, N.~Shazeer, N.~Parmar, J.~Uszkoreit, L.~Jones, A.~N. Gomez,
  {\L}.~Kaiser, and I.~Polosukhin, ``Attention is all you need,'' \emph{Adv.
  Neural Inf. Process. Syst.}, vol.~30, 2017.

\bibitem{GPTReport}
O.~AI, ``Gpt-4 technical report,'' \emph{arXiv preprint arXiv:2303.08774},
  2023.

\bibitem{goodfellow2020generative}
I.~Goodfellow, J.~Pouget-Abadie, M.~Mirza, B.~Xu, D.~Warde-Farley, S.~Ozair,
  A.~Courville, and Y.~Bengio, ``Generative adversarial networks,''
  \emph{Commun. ACM}, vol.~63, no.~11, pp. 139--144, Nov. 2020.

\bibitem{kingma2019introduction}
D.~P. Kingma, M.~Welling \emph{et~al.}, ``An introduction to variational
  autoencoders,'' \emph{Found. Trends Mach. Learn.}, vol.~12, no.~4, pp.
  307--392, Apr. 2019.

\bibitem{rezende2015variational}
D.~Rezende and S.~Mohamed, ``Variational inference with normalizing flows,'' in
  \emph{Proc. Int. Conf. Mach. Learn.}, Lille, France, Jul. 2015, pp.
  1530--1538.

\bibitem{zhao2017energy}
J.~Zhao, M.~Mathieu, and Y.~LeCun, ``Energy-based generative adversarial
  networks,'' in \emph{Proc. Int. Conf. Mach. Learn.}, 2017.

\bibitem{sohl2015deep}
J.~Sohl-Dickstein, E.~Weiss, N.~Maheswaranathan, and S.~Ganguli, ``Deep
  unsupervised learning using nonequilibrium thermodynamics,'' in \emph{Proc.
  Int. Conf. Mach. Learn.}, 2015, pp. 2256--2265.

\bibitem{songscore}
Y.~Song, J.~Sohl-Dickstein, D.~P. Kingma, A.~Kumar, S.~Ermon, and B.~Poole,
  ``Score-based generative modeling through stochastic differential
  equations,'' in \emph{Proc. Int. Conf. Learn. Represent.}, 2021.

\bibitem{peng2019stochastic}
S.~Peng and S.~Peng, \emph{Stochastic differential equations}.\hskip 1em plus
  0.5em minus 0.4em\relax Springer, 2019.

\bibitem{cao2022survey}
H.~Cao, C.~Tan, Z.~Gao, G.~Chen, P.-A. Heng, and S.~Z. Li, ``A survey on
  generative diffusion model,'' \emph{IEEE Trans. Knowledge Data Eng.}, to
  appear, 2024.

\bibitem{stabdiff}
S.~AI, ``Stable diffusion,'' \url{https://stability.ai/}.

\bibitem{ho2020denoising}
J.~Ho, A.~Jain, and P.~Abbeel, ``Denoising diffusion probabilistic models,''
  \emph{Adv. Neural Inf. Process. Syst.}, vol.~33, pp. 6840--6851, 2020.

\bibitem{song2020denoising}
J.~Song, C.~Meng, and S.~Ermon, ``Denoising diffusion implicit models,''
  \emph{Proc. Int. Conf. Learn. Represent.}, 2020.

\bibitem{li2022diffusion}
X.~Li, J.~Thickstun, I.~Gulrajani, P.~S. Liang, and T.~B. Hashimoto,
  ``Diffusion-lm improves controllable text generation,'' \emph{Adv. Neural
  Inf. Process. Syst.}, vol.~35, pp. 4328--4343, 2022.

\bibitem{mittal2021symbolic}
G.~Mittal, J.~Engel, C.~Hawthorne, and I.~Simon, ``Symbolic music generation
  with diffusion models,'' \emph{Proc. Int. Society Music Inf. Retr. Conf.},
  2021.

\bibitem{huang2022prodiff}
R.~Huang, Z.~Zhao, H.~Liu, J.~Liu, C.~Cui, and Y.~Ren, ``Prodiff: {P}rogressive
  fast diffusion model for high-quality text-to-speech,'' in \emph{Proc. ACM
  Int. Conf. Multimedia}, 2022, pp. 2595--2605.

\bibitem{niu2020permutation}
C.~Niu, Y.~Song, J.~Song, S.~Zhao, A.~Grover, and S.~Ermon, ``Permutation
  invariant graph generation via score-based generative modeling,'' in
  \emph{Proc. Int. Conf. Artif. Intell. Stat.}\hskip 1em plus 0.5em minus
  0.4em\relax PMLR, 2020, pp. 4474--4484.

\bibitem{vignac2022digress}
C.~Vignac, I.~Krawczuk, A.~Siraudin, B.~Wang, V.~Cevher, and P.~Frossard,
  ``Di{G}ress: {D}iscrete denoising diffusion for graph generation,'' in
  \emph{Proc. Int. Conf. Learn. Represent.}, 2022.

\bibitem{chen2023efficient}
X.~Chen, J.~He, X.~Han, and L.-P. Liu, ``Efficient and degree-guided graph
  generation via discrete diffusion modeling,'' in \emph{Proc. Int. Conf. Mach.
  Learn.}, 2023, pp. 4585--4610.

\bibitem{peng2023moldiff}
X.~Peng, J.~Guan, Q.~Liu, and J.~Ma, ``Mol{D}iff: {A}ddressing the atom-bond
  inconsistency problem in 3{D} molecule diffusion generation,'' in \emph{Proc.
  Int. Conf. Mach. Learn.}, 2023, pp. 27\,611--27\,629.

\bibitem{ketata2023diffdock}
M.~A. Ketata, C.~Laue, R.~Mammadov, H.~Stark, M.~Wu, G.~Corso, C.~Marquet,
  R.~Barzilay, and T.~S. Jaakkola, ``Diff{D}ock-{PP}: {R}igid protein-protein
  docking with diffusion models,'' in \emph{Proc. Int. Conf. Learn.
  Represent.}, 2023.

\bibitem{huang2022mdm}
L.~Huang, H.~Zhang, T.~Xu, and K.-C. Wong, ``M{DM}: {M}olecular diffusion model
  for 3{D} molecule generation,'' in \emph{Proc. AAAI Conf. Artif. Intell.},
  vol.~37, no.~4, 2023, pp. 5105--5112.

\bibitem{lee2023codi}
C.~Lee, J.~Kim, and N.~Park, ``Co{D}i: {C}o-evolving contrastive diffusion
  models for mixed-type tabular synthesis,'' in \emph{Proc. Int. Conf. Mach.
  Learn.}, 2023, pp. 18\,940--18\,956.

\bibitem{kotelnikov2022tabddpm}
A.~Kotelnikov, D.~Baranchuk, I.~Rubachev, and A.~Babenko, ``Tab{DDPM}:
  {M}odelling tabular data with diffusion models,'' in \emph{Proc. Int. Conf.
  Mach. Learn.}, 2023, pp. 17\,564--17\,579.

\bibitem{neifar2023diffecg}
N.~Neifar, A.~Ben-Hamadou, A.~Mdhaffar, and M.~Jmaiel, ``Diff{ECG}: {A}
  generalized probabilistic diffusion model for {ECG} signals synthesis,''
  \emph{arXiv preprint arXiv:2306.01875}, 2023.

\bibitem{yang2022diffusion}
L.~Yang, Z.~Zhang, Y.~Song, S.~Hong, R.~Xu, Y.~Zhao, Y.~Shao, W.~Zhang, B.~Cui,
  and M.-H. Yang, ``Diffusion models: {A} comprehensive survey of methods and
  applications,'' \emph{ACM Comput. Surv.}, vol.~56, no.~4, pp. 1--39, 2023.

\bibitem{croitoru2023diffusion}
F.-A. Croitoru, V.~Hondru, R.~T. Ionescu, and M.~Shah, ``Diffusion models in
  vision: {A} survey,'' \emph{IEEE Trans. Pattern Anal. Mach. Intell.}, to
  appear, 2023.

\bibitem{reuss2023goal}
M.~Reuss, M.~Li, X.~Jia, and R.~Lioutikov, ``Goal-conditioned imitation
  learning using score-based diffusion policies,'' \emph{Robotics: Science and
  Systems}, 2023.

\bibitem{10108002}
Y.~Li, Y.~Lu, R.~Zhang, B.~Ai, and Z.~Zhong, ``Deep learning for energy
  efficient beamforming in {MU-MISO} networks: {A} {GAT}-based approach,''
  \emph{IEEE Wireless Commun. Lett.}, vol.~12, no.~7, pp. 1264--1268, 2023.

\bibitem{lu2023contrastive}
C.~Lu, H.~Chen, J.~Chen, H.~Su, C.~Li, and J.~Zhu, ``Contrastive energy
  prediction for exact energy-guided diffusion sampling in offline
  reinforcement learning,'' \emph{Proc. Int. Conf. Mach. Learn.}, 2023.

\bibitem{krishnamoorthy2023diffusion}
S.~Krishnamoorthy, S.~M. Mashkaria, and A.~Grover, ``Diffusion models for
  black-box optimization,'' in \emph{Proc. Int. Conf. Mach. Learn.}\hskip 1em
  plus 0.5em minus 0.4em\relax PMLR, 2023, pp. 17\,842--17\,857.

\bibitem{liu2024graph}
Y.~Liu, C.~Du, T.~Pang, C.~Li, W.~Chen, and M.~Lin, ``Graph diffusion policy
  optimization,'' \emph{{\rm{arXiv preprint arXiv:2402.16302}}}, 2024.

\bibitem{du2023age}
H.~Du, D.~Niyato, J.~Kang, Z.~Xiong, P.~Zhang, S.~Cui, X.~Shen, S.~Mao, Z.~Han,
  A.~Jamalipour \emph{et~al.}, ``The age of generative {AI} and {AI}-generated
  everything,'' \emph{arXiv preprint arXiv:2311.00947}, 2023.

\bibitem{du2023ai}
H.~Du, J.~Wang, D.~Niyato, J.~Kang, Z.~Xiong, and D.~I. Kim, ``{AI}-generated
  incentive mechanism and full-duplex semantic communications for information
  sharing,'' \emph{IEEE J. Sel. Areas Commun.}, to appear, 2023.

\bibitem{du2023yolo}
B.~Du, H.~Du, H.~Liu, D.~Niyato, P.~Xin, J.~Yu, M.~Qi, and Y.~Tang,
  ``{YOLO}-based semantic communication with generative {AI}-aided resource
  allocation for digital twins construction,'' \emph{IEEE Internet Things J.},
  to appear, 2023.

\bibitem{du2023user}
H.~Du, R.~Zhang, D.~Niyato, J.~Kang, Z.~Xiong, S.~Cui, X.~Shen, and D.~I. Kim,
  ``User-centric interactive {AI} for distributed diffusion model-based
  {AI}-generated content,'' \emph{arXiv preprint arXiv:2311.11094}, 2023.

\bibitem{cheng2022integrated}
X.~Cheng, D.~Duan, S.~Gao, and L.~Yang, ``Integrated sensing and communications
  ({ISAC}) for vehicular communication networks ({VCN}),'' \emph{IEEE Internet
  Things J.}, vol.~9, no.~23, pp. 23\,441--23\,451, 2022.

\bibitem{wang2023generative}
J.~Wang, H.~Du, D.~Niyato, J.~Kang, S.~Cui, X.~Shen, and P.~Zhang, ``Generative
  {AI} for integrated sensing and communication: Insights from the physical
  layer perspective,'' \emph{arXiv preprint arXiv:2310.01036}, 2023.

\bibitem{yang2022semantic}
W.~Yang, H.~Du, Z.~Q. Liew, W.~Y.~B. Lim, Z.~Xiong, D.~Niyato, X.~Chi, X.~S.
  Shen, and C.~Miao, ``Semantic communications for future internet:
  {F}undamentals, applications, and challenges,'' \emph{IEEE Commun. Surv.
  Tut.}, to appear, 2023.

\bibitem{du2023generativeica}
H.~Du, G.~Liu, D.~Niyato, J.~Zhang, J.~Kang, Z.~Xiong, B.~Ai, and D.~I. Kim,
  ``Generative {AI}-aided joint training-free secure semantic communications
  via multi-modal prompts,'' in \emph{Proc. IEEE Int. Conf. Acoustics Speech
  Signal Proc.}, 2023.

\bibitem{ang2018deployment}
L.-M. Ang, K.~P. Seng, G.~K. Ijemaru, and A.~M. Zungeru, ``Deployment of {IoV}
  for smart cities: {A}pplications, architecture, and challenges,'' \emph{IEEE
  Access}, vol.~7, pp. 6473--6492, 2018.

\bibitem{zhou2023heterogeneous}
H.~Zhou, H.~Zhou, J.~Li, K.~Yang, J.~An, and X.~Shen, ``Heterogeneous
  ultra-dense networks with traffic hotspots: {A} unified handover analysis,''
  \emph{IEEE Internet Things J.}, to appear, 2023.

\bibitem{lin2023unified}
Y.~Lin, Z.~Gao, H.~Du, D.~Niyato, J.~Kang, A.~Jamalipour, and X.~S. Shen, ``A
  unified framework for integrating semantic communication and ai-generated
  content in metaverse,'' \emph{IEEE Netw.}, 2023.

\bibitem{zhou2014chaincluster}
H.~Zhou, B.~Liu, T.~H. Luan, F.~Hou, L.~Gui, Y.~Li, Q.~Yu, and X.~Shen,
  ``Chaincluster: {E}ngineering a cooperative content distribution framework
  for highway vehicular communications,'' \emph{IEEE Trans. Intell. Transp
  Syst.}, vol.~15, no.~6, pp. 2644--2657, Jun. 2014.

\bibitem{9798257}
R.~Zhang, K.~Xiong, X.~Tian, Y.~Lu, P.~Fan, and K.~B. Letaief, ``Inverse
  reinforcement learning meets power allocation in multi-user cellular
  networks,'' in \emph{IEEE INFOCOM 2022 - IEEE Conference on Computer
  Communications Workshops (INFOCOM WKSHPS)}, 2022, pp. 1--2.

\bibitem{du2023spear}
H.~Du, D.~Niyato, J.~Kang, Z.~Xiong, K.-Y. Lam, Y.~Fang, and Y.~Li, ``Spear or
  shield: {L}everaging generative {AI} to tackle security threats of
  intelligent network services,'' \emph{arXiv preprint arXiv:2306.02384}, 2023.

\bibitem{ajay2022conditional}
A.~Ajay, Y.~Du, A.~Gupta, J.~Tenenbaum, T.~Jaakkola, and P.~Agrawal, ``Is
  conditional generative modeling all you need for decision-making?'' in
  \emph{Proc. Int. Conf. Learn. Represent.}, May 2023.

\bibitem{janner2022planning}
M.~Janner, Y.~Du, J.~B. Tenenbaum, and S.~Levine, ``Planning with diffusion for
  flexible behavior synthesis,'' in \emph{Proc. Int. Conf. Mach. Learn.}, Jul.
  2023, pp. 9902--9915.

\bibitem{wang2022diffusion}
Z.~Wang, J.~J. Hunt, and M.~Zhou, ``Diffusion policies as an expressive policy
  class for offline reinforcement learning,'' in \emph{Proc. Int. Conf. Learn.
  Represent.}, May 2023.

\bibitem{chen2022offline}
H.~Chen, C.~Lu, C.~Ying, H.~Su, and J.~Zhu, ``Offline reinforcement learning
  via high-fidelity generative behavior modeling,'' in \emph{Proc. Int. Conf.
  Learn. Represent.}, May 2023.

\bibitem{wang2023diffusion}
H.-C. Wang, S.-F. Chen, and S.-H. Sun, ``Diffusion model-augmented behavioral
  cloning,'' \emph{Proc. Int. Conf. Mach. Learn. Workshop}, 2023.

\bibitem{kazerouni2022diffusion}
A.~Kazerouni, E.~K. Aghdam, M.~Heidari, R.~Azad, M.~Fayyaz, I.~Hacihaliloglu,
  and D.~Merhof, ``Diffusion models for medical image analysis: {A}
  comprehensive survey,'' \emph{arXiv preprint arXiv:2211.07804}, 2022.

\bibitem{zhang2023text}
C.~Zhang, C.~Zhang, M.~Zhang, and I.~S. Kweon, ``Text-to-image diffusion model
  in generative {AI}: {A} survey,'' \emph{arXiv preprint arXiv:2303.07909},
  2023.

\bibitem{ulhaq2022efficient}
A.~Ulhaq, N.~Akhtar, and G.~Pogrebna, ``Efficient diffusion models for vision:
  {A} survey,'' \emph{arXiv preprint arXiv:2210.09292}, 2022.

\bibitem{zou2023diffusion}
H.~Zou, Z.~M. Kim, and D.~Kang, ``Diffusion models in {NLP}: {A} survey,''
  \emph{arXiv preprint arXiv:2305.14671}, 2023.

\bibitem{li2023diffusion}
Y.~Li, K.~Zhou, W.~X. Zhao, and J.-R. Wen, ``Diffusion models for
  non-autoregressive text generation: {A} survey,'' \emph{Int. Joint Conf.
  Artif. Intell.}, 2023.

\bibitem{lin2023diffusion}
L.~Lin, Z.~Li, R.~Li, X.~Li, and J.~Gao, ``Diffusion models for time series
  applications: {A} survey,'' \emph{Frontiers of Information Technology \&
  Electronic Engineering}, pp. 1--23, Dec. 2023.

\bibitem{luo2023comprehensive}
W.~Luo, ``A comprehensive survey on knowledge distillation of diffusion
  models,'' \emph{arXiv preprint arXiv:2304.04262}, 2023.

\bibitem{zhang2023survey}
M.~Zhang, M.~Qamar, T.~Kang, Y.~Jung, C.~Zhang, S.-H. Bae, and C.~Zhang, ``A
  survey on graph diffusion models: {G}enerative {AI} in science for molecule,
  protein and material,'' \emph{arXiv preprint arXiv:2304.01565}, 2023.

\bibitem{zhang2023audio}
C.~Zhang, C.~Zhang, S.~Zheng, M.~Zhang, M.~Qamar, S.-H. Bae, and I.~S. Kweon,
  ``Audio diffusion model for speech synthesis: {A} survey on text to speech
  and speech enhancement in generative {AI},'' \emph{arXiv preprint
  arXiv:2303.13336}, 2023.

\bibitem{guo2023diffusion}
Z.~Guo, J.~Liu, Y.~Wang, M.~Chen, D.~Wang, D.~Xu, and J.~Cheng, ``Diffusion
  models in bioinformatics: {A} new wave of deep learning revolution in
  action,'' \emph{arXiv preprint arXiv:2302.10907}, 2023.

\bibitem{fan2023generative}
W.~Fan, C.~Liu, Y.~Liu, J.~Li, H.~Li, H.~Liu, J.~Tang, and Q.~Li, ``Generative
  diffusion models on graphs: {M}ethods and applications,'' \emph{arXiv
  preprint arXiv:2302.02591}, 2023.

\bibitem{du2023generative}
H.~Du, Z.~Li, D.~Niyato, J.~Kang, Z.~Xiong, H.~Huang, and S.~Mao,
  ``Diffusion-based reinforcement learning for edge-enabled {AI}-generated
  content services,'' \emph{IEEE Trans. Mobile Comput.}, to appear, 2024.

\bibitem{lou2023reflected}
A.~Lou and S.~Ermon, ``Reflected diffusion models,'' \emph{Proc. Int. Conf.
  Mach. Learn.}, 2023.

\bibitem{zhang2023diffcollage}
Q.~Zhang, J.~Song, X.~Huang, Y.~Chen, and M.-Y. Liu, ``Diff{C}ollage:
  {P}arallel generation of large content with diffusion models,'' in
  \emph{Proc. IEEE Conf. Comput. Vis. Pattern Recognit.}, 2023, pp.
  10\,188--10\,198.

\bibitem{ni2023conditional}
H.~Ni, C.~Shi, K.~Li, S.~X. Huang, and M.~R. Min, ``Conditional image-to-video
  generation with latent flow diffusion models,'' in \emph{Proc. IEEE Conf.
  Comput. Vis. Pattern Recognit.}, 2023, pp. 18\,444--18\,455.

\bibitem{enkelmann1988investigations}
W.~Enkelmann, ``Investigations of multigrid algorithms for the estimation of
  optical flow fields in image sequences,'' \emph{Computer Vision, Graphics,
  and Image Processing}, vol.~43, no.~2, pp. 150--177, Feb. 1988.

\bibitem{ho2022video}
J.~Ho, T.~Salimans, A.~Gritsenko, W.~Chan, M.~Norouzi, and D.~J. Fleet, ``Video
  diffusion models,'' \emph{arXiv:2204.03458}, 2022.

\bibitem{yu2022latent}
P.~Yu, S.~Xie, X.~Ma, B.~Jia, B.~Pang, R.~Gao, Y.~Zhu, S.-C. Zhu, and Y.~Wu,
  ``Latent diffusion energy-based model for interpretable text modeling.'' in
  \emph{Proc. Int. Conf. Mach. Learn.}, 2022.

\bibitem{gong2022diffuseq}
S.~Gong, M.~Li, J.~Feng, Z.~Wu, and L.~Kong, ``Diffuseq: {S}equence to sequence
  text generation with diffusion models,'' \emph{Proc. Int. Conf. Learn.
  Represent.}, 2022.

\bibitem{zhang2023diffusum}
H.~Zhang, X.~Liu, and J.~Zhang, ``Diffusum: {G}eneration enhanced extractive
  summarization with diffusion,'' \emph{Assoc. Comput. Linguist.}, 2023.

\bibitem{reid2023diffuser}
M.~Reid, V.~J. Hellendoorn, and G.~Neubig, ``Diffuser: {D}iffusion via
  edit-based reconstruction,'' in \emph{Proc. Int. Conf. Learn. Represent.},
  2023.

\bibitem{ruan2023mm}
L.~Ruan, Y.~Ma, H.~Yang, H.~He, B.~Liu, J.~Fu, N.~J. Yuan, Q.~Jin, and B.~Guo,
  ``Mm-diffusion: {L}earning multi-modal diffusion models for joint audio and
  video generation,'' in \emph{Proc. IEEE Conf. Comput. Vis. Pattern
  Recognit.}, 2023, pp. 10\,219--10\,228.

\bibitem{kongdiffwave}
Z.~Kong, W.~Ping, J.~Huang, K.~Zhao, and B.~Catanzaro, ``Diff{W}ave: {A}
  versatile diffusion model for audio synthesis,'' in \emph{Proc. Int. Conf.
  Learn. Represent.}, 2021.

\bibitem{liu2022diffsinger}
J.~Liu, C.~Li, Y.~Ren, F.~Chen, and Z.~Zhao, ``Diffsinger: {S}inging voice
  synthesis via shallow diffusion mechanism,'' in \emph{Proc. AAAI Conf. Artif.
  Intell.}, vol.~36, no.~10, 2022, pp. 11\,020--11\,028.

\bibitem{rouard2021crash}
S.~Rouard and G.~Hadjeres, ``{CRASH}: {R}aw audio score-based generative
  modeling for controllable high-resolution drum sound synthesis,'' \emph{Proc.
  Int. Society Music Inf. Retr. Conf.}, 2021.

\bibitem{ankile2023denoising}
L.~L. Ankile, A.~Midgley, and S.~Weisshaar, ``Denoising diffusion probabilistic
  models as a defense against adversarial attacks,'' \emph{arXiv preprint
  arXiv:2301.06871}, 2023.

\bibitem{ghalebikesabi2023differentially}
S.~Ghalebikesabi, L.~Berrada, S.~Gowal, I.~Ktena, R.~Stanforth, J.~Hayes,
  S.~De, S.~L. Smith, O.~Wiles, and B.~Balle, ``Differentially private
  diffusion models generate useful synthetic images,'' \emph{arXiv preprint
  arXiv:2302.13861}, 2023.

\bibitem{maungmaung2023generative}
A.~MaungMaung and H.~Kiya, ``Generative model-based attack on learnable image
  encryption for privacy-preserving deep learning,'' \emph{arXiv preprint
  arXiv:2303.05036}, 2023.

\bibitem{blasingame2023diffusion}
Z.~Blasingame and C.~Liu, ``Diffusion models for stronger face morphing
  attacks,'' \emph{arXiv preprint arXiv:2301.04218}, 2023.

\bibitem{nichol2022glide}
A.~Q. Nichol, P.~Dhariwal, A.~Ramesh, P.~Shyam, P.~Mishkin, B.~Mcgrew,
  I.~Sutskever, and M.~Chen, ``{GLIDE}: {T}owards photorealistic image
  generation and editing with text-guided diffusion models,'' in \emph{Int.
  Conf. Mach. Learn.}, 2022, pp. 16\,784--16\,804.

\bibitem{dalle2}
OpenAI, ``Dall·e 2,'' \url{https://openai.com/dall-e-2}.

\bibitem{Imagen}
B.~T. Google~Research, ``Imagen,'' \url{https://imagen.research.google/}.

\bibitem{gui2021review}
J.~Gui, Z.~Sun, Y.~Wen, D.~Tao, and J.~Ye, ``A review on generative adversarial
  networks: {A}lgorithms, theory, and applications,'' \emph{IEEE Trans. Knowl.
  Data Eng.}, vol.~35, no.~4, pp. 3313--3332, Apr. 2021.

\bibitem{10172151}
H.~Du, R.~Zhang, D.~Niyato, J.~Kang, Z.~Xiong, D.~I. Kim, X.~S. Shen, and H.~V.
  Poor, ``Exploring collaborative distributed diffusion-based {AI}-generated
  content ({AIGC}) in wireless networks,'' \emph{IEEE Netw.}, no.~99, pp. 1--8,
  2023.

\bibitem{li2024diffusion}
Z.~Li, H.~Yuan, K.~Huang, C.~Ni, Y.~Ye, M.~Chen, and M.~Wang, ``Diffusion model
  for data-driven black-box optimization,'' \emph{{\rm{arXiv preprint
  arXiv:2403.13219}}}, 2024.

\bibitem{sun2024difusco}
Z.~Sun and Y.~Yang, ``Difusco: Graph-based diffusion solvers for combinatorial
  optimization,'' \emph{Adv. Neural Inf. Process. Syst.}, vol.~36, 2024.

\bibitem{zhang2024enhancing}
B.~Zhang, W.~Luo, and Z.~Zhang, ``Enhancing adversarial robustness via
  score-based optimization,'' \emph{Adv. Neural Inf. Process. Syst.}, vol.~36,
  2024.

\bibitem{giannone2024aligning}
G.~Giannone, A.~Srivastava, O.~Winther, and F.~Ahmed, ``Aligning optimization
  trajectories with diffusion models for constrained design generation,''
  \emph{Adv. Neural Inf. Process. Syst.}, vol.~36, 2024.

\bibitem{luo2022antigen}
S.~Luo, Y.~Su, X.~Peng, S.~Wang, J.~Peng, and J.~Ma, ``Antigen-specific
  antibody design and optimization with diffusion-based generative models for
  protein structures,'' \emph{Adv. Neural Inf. Process. Syst.}, vol.~35, pp.
  9754--9767, 2022.

\bibitem{chen2023score}
H.~Chen, C.~Lu, Z.~Wang, H.~Su, and J.~Zhu, ``Score regularized policy
  optimization through diffusion behavior,'' in \emph{Proc. Int. Conf. Learn.
  Represent.}, 2024.

\bibitem{zhou2024adaptive}
S.~Zhou, Y.~Du, S.~Zhang, M.~Xu, Y.~Shen, W.~Xiao, D.-Y. Yeung, and C.~Gan,
  ``Adaptive online replanning with diffusion models,'' \emph{Adv. Neural Inf.
  Process. Syst.}, vol.~36, 2024.

\bibitem{xu2024stage}
K.~Xu, S.~Lu, B.~Huang, W.~Wu, and Q.~Liu, ``Stage-by-stage wavelet
  optimization refinement diffusion model for sparse-view {CT}
  reconstruction,'' \emph{IEEE Trans. Med. Imaging}, to appear, 2024.

\bibitem{wu2023data}
W.~Wu and Y.~Wang, ``Data-iterative optimization score model for stable
  ultra-sparse-view {CT} reconstruction,'' \emph{{\rm{arXiv preprint
  arXiv:2308.14437}}}, 2023.

\bibitem{jiang2024back}
Z.~Jiang, Z.~Zhou, L.~Li, W.~Chai, C.-Y. Yang, and J.-N. Hwang, ``Back to
  optimization: Diffusion-based zero-shot 3{D} human pose estimation,'' in
  \emph{Proc. IEEE/CVF Winter Conf. Appl. Comput. Vis.}, 2024, pp. 6142--6152.

\bibitem{huang2023diffusion}
S.~Huang, Z.~Wang, P.~Li, B.~Jia, T.~Liu, Y.~Zhu, W.~Liang, and S.-C. Zhu,
  ``Diffusion-based generation, optimization, and planning in 3{D} scenes,'' in
  \emph{Proc. IEEE/CVF Comput. Vis. Pattern Recog.}, 2023, pp.
  16\,750--16\,761.

\bibitem{huang2023dreamtime}
Y.~Huang, J.~Wang, Y.~Shi, X.~Qi, Z.-J. Zha, and L.~Zhang, ``Dreamtime: An
  improved optimization strategy for text-to-3{D} content creation,''
  \emph{{\rm{arXiv preprint arXiv:2306.12422}}}, 2023.

\bibitem{urain2023se}
J.~Urain, N.~Funk, J.~Peters, and G.~Chalvatzaki,
  ``{SE}(3)-{D}iffusion{F}ields: Learning smooth cost functions for joint grasp
  and motion optimization through diffusion,'' in \emph{IEEE Int. Conf. Robot.
  Autom.}, 2023, pp. 5923--5930.

\bibitem{maze2023diffusion}
F.~Maz{\'e} and F.~Ahmed, ``Diffusion models beat gans on topology
  optimization,'' in \emph{Proc. AAAI Conf. Artif. Intell.}, vol.~37, no.~8,
  2023, pp. 9108--9116.

\bibitem{park2021neural}
S.~W. Park, K.~Lee, and J.~Kwon, ``Neural markov controlled {SDE}: Stochastic
  optimization for continuous-time data,'' in \emph{Proc. Int. Conf. Learn.
  Represent.}, 2021.

\bibitem{liu2023dipper}
J.~Liu, M.~Stamatopoulou, and D.~Kanoulas, ``{DiPPeR}: Diffusion-based {2D}
  path planner applied on legged robots,'' \emph{{\rm{arXiv preprint
  arXiv:2310.07842}}}, 2023.

\bibitem{zappone2019model}
A.~Zappone, M.~Di~Renzo, M.~Debbah, T.~T. Lam, and X.~Qian, ``Model-aided
  wireless artificial intelligence: {E}mbedding expert knowledge in deep neural
  networks for wireless system optimization,'' \emph{IEEE Veh. Technol. Mag.},
  vol.~14, no.~3, pp. 60--69, Mar. 2019.

\bibitem{lin2006tutorial}
X.~Lin, N.~B. Shroff, and R.~Srikant, ``A tutorial on cross-layer optimization
  in wireless networks,'' \emph{IEEE J. Sel. Areas Commun.}, vol.~24, no.~8,
  pp. 1452--1463, Aug. 2006.

\bibitem{liu2023deep}
Y.~Liu, H.~Du, D.~Niyato, J.~Kang, Z.~Xiong, D.~I. Kim, and A.~Jamalipour,
  ``Deep generative model and its applications in efficient wireless network
  management: {A} tutorial and case study,'' \emph{IEEE Wireless Commun.}, to
  appear, 2024.

\bibitem{osband2016deep}
I.~Osband, C.~Blundell, A.~Pritzel, and B.~Van~Roy, ``Deep exploration via
  bootstrapped {DQN},'' \emph{Adv. Neural Inf. Process. Syst.}, vol.~29, 2016.

\bibitem{haarnoja2018soft}
T.~Haarnoja, A.~Zhou, P.~Abbeel, and S.~Levine, ``Soft actor-critic:
  {O}ff-policy maximum entropy deep reinforcement learning with a stochastic
  actor,'' in \emph{Proc. Int. Conf. Mach. Learn.}\hskip 1em plus 0.5em minus
  0.4em\relax PMLR, 2018, pp. 1861--1870.

\bibitem{schulman2017proximal}
J.~Schulman, F.~Wolski, P.~Dhariwal, A.~Radford, and O.~Klimov, ``Proximal
  policy optimization algorithms,'' \emph{arXiv preprint arXiv:1707.06347},
  2017.

\bibitem{goldsmith2005wireless}
A.~Goldsmith, \emph{Wireless communications}.\hskip 1em plus 0.5em minus
  0.4em\relax Cambridge university press, 2005.

\bibitem{zheng2019intelligent}
B.~Zheng and R.~Zhang, ``Intelligent reflecting surface-enhanced {OFDM}:
  {C}hannel estimation and reflection optimization,'' \emph{IEEE Wireless
  Commun. Lett.}, vol.~9, no.~4, pp. 518--522, Apr. 2019.

\bibitem{desale2015heuristic}
S.~Desale, A.~Rasool, S.~Andhale, and P.~Rane, ``Heuristic and meta-heuristic
  algorithms and their relevance to the real world: {A} survey,'' \emph{Int. J.
  Comput. Eng. Res. Trends}, vol. 351, no.~5, pp. 2349--7084, May 2015.

\bibitem{yu2004iterative}
W.~Yu, W.~Rhee, S.~Boyd, and J.~M. Cioffi, ``Iterative water-filling for
  gaussian vector multiple-access channels,'' \emph{IEEE Trans. Inf. Theory},
  vol.~50, no.~1, pp. 145--152, Jan. 2004.

\bibitem{feriani2021single}
A.~Feriani and E.~Hossain, ``Single and multi-agent deep reinforcement learning
  for {AI}-enabled wireless networks: {A} tutorial,'' \emph{IEEE Commun. Surv.
  Tut.}, vol.~23, no.~2, pp. 1226--1252, Feb. 2021.

\bibitem{yu2019deep}
Y.~Yu, T.~Wang, and S.~C. Liew, ``Deep-reinforcement learning multiple access
  for heterogeneous wireless networks,'' \emph{IEEE J. Sel. Areas Commun.},
  vol.~37, no.~6, pp. 1277--1290, Jun. 2019.

\bibitem{du2023enabling}
H.~Du, Z.~Li, D.~Niyato, J.~Kang, Z.~Xiong, D.~I. Kim \emph{et~al.}, ``Enabling
  {AI}-generated content ({AIGC}) services in wireless edge networks,''
  \emph{arXiv preprint arXiv:2301.03220}, 2023.

\bibitem{zhou2010novel}
H.~Zhou, Y.~Wu, Y.~Hu, and G.~Xie, ``A novel stable selection and reliable
  transmission protocol for clustered heterogeneous wireless sensor networks,''
  \emph{Comput. Commun.}, vol.~33, no.~15, pp. 1843--1849, 2010.

\bibitem{ching2006markov}
W.-K. Ching and M.~K. Ng, ``Markov chains,'' \emph{Models, algorithms and
  applications}, 2006.

\bibitem{xu2020service}
X.~Xu, B.~Shen, S.~Ding, G.~Srivastava, M.~Bilal, M.~R. Khosravi, V.~G. Menon,
  M.~A. Jan, and M.~Wang, ``Service offloading with deep {Q}-network for
  digital twinning-empowered internet of vehicles in edge computing,''
  \emph{iEEE Trans. Industr. Inform.}, vol.~18, no.~2, pp. 1414--1423, Feb.
  2020.

\bibitem{ohira2021novel}
M.~Ohira, K.~Takano, and Z.~Ma, ``A novel deep-{Q}-network-based fine-tuning
  approach for planar bandpass filter design,'' \emph{IEEE Microw. Wireless
  Compon. Lett.}, vol.~31, no.~6, pp. 638--641, Jun. 2021.

\bibitem{iqbal2021double}
A.~Iqbal, M.-L. Tham, and Y.~C. Chang, ``Double deep {Q}-network-based
  energy-efficient resource allocation in cloud radio access network,''
  \emph{IEEE Access}, vol.~9, pp. 20\,440--20\,449, 2021.

\bibitem{hasselt2010double}
H.~Hasselt, ``Double {Q}-learning,'' \emph{Adv. Neural Inf. Process. Syst.},
  vol.~23, 2010.

\bibitem{vimal2020energy}
S.~Vimal, M.~Khari, R.~G. Crespo, L.~Kalaivani, N.~Dey, and M.~Kaliappan,
  ``Energy enhancement using multiobjective ant colony optimization with double
  {Q} learning algorithm for {IoT} based cognitive radio networks,''
  \emph{Comput. Commun.}, vol. 154, pp. 481--490, 2020.

\bibitem{luong2019applications}
N.~C. Luong, D.~T. Hoang, S.~Gong, D.~Niyato, P.~Wang, Y.-C. Liang, and D.~I.
  Kim, ``Applications of deep reinforcement learning in communications and
  networking: {A} survey,'' \emph{IEEE Commun. Surv. Tut.}, vol.~21, no.~4, pp.
  3133--3174, Apr. 2019.

\bibitem{xu2019load}
Y.~Xu, W.~Xu, Z.~Wang, J.~Lin, and S.~Cui, ``Load balancing for ultradense
  networks: {A} deep reinforcement learning-based approach,'' \emph{IEEE
  Internet Things J.}, vol.~6, no.~6, pp. 9399--9412, Jun. 2019.

\bibitem{zhang2022lad}
E.~Zhang, Y.~Lu, W.~Wang, and A.~Zhang, ``{LAD}: {L}anguage augmented diffusion
  for reinforcement learning,'' \emph{Adv. Neural Inf. Process. Syst.
  Workshop}, 2022.

\bibitem{wang2023pdpp}
H.~Wang, Y.~Wu, S.~Guo, and L.~Wang, ``{PDPP}: {P}rojected diffusion for
  procedure planning in instructional videos,'' in \emph{Proc. Comput. Vis.
  Pattern Recog.}, 2023, pp. 14\,836--14\,845.

\bibitem{brehmer2023edgi}
J.~Brehmer, J.~Bose, P.~De~Haan, and T.~Cohen, ``{EDGI}: {E}quivariant
  diffusion for planning with embodied agents,'' \emph{Proc. Adv. Neural Inf.
  Process. Syst.}, vol.~36, 2024.

\bibitem{cao2023multi}
Y.~Cao, E.~Rizk, S.~Vlaski, and A.~H. Sayed, ``Multi-agent adversarial training
  using diffusion learning,'' in \emph{ICASSP 2023-2023 IEEE International
  Conference on Acoustics, Speech and Signal Processing (ICASSP)}.\hskip 1em
  plus 0.5em minus 0.4em\relax IEEE, 2023, pp. 1--5.

\bibitem{liang2023adaptdiffuser}
Z.~Liang, Y.~Mu, M.~Ding, F.~Ni, M.~Tomizuka, and P.~Luo, ``Adapt{D}iffuser:
  {D}iffusion models as adaptive self-evolving planners,'' in \emph{Proc. Int.
  Conf. Mach. Learn.}, 2023, pp. 20\,725--20\,745.

\bibitem{pearce2023imitating}
T.~Pearce, T.~Rashid, A.~Kanervisto, D.~Bignell, M.~Sun, R.~Georgescu, S.~V.
  Macua, S.~Z. Tan, I.~Momennejad, K.~Hofmann \emph{et~al.}, ``Imitating human
  behaviour with diffusion models,'' in \emph{Proc. Int. Conf. Learn.
  Represent.}, 2023.

\bibitem{9195488}
R.~Zhang, K.~Xiong, W.~Guo, X.~Yang, P.~Fan, and K.~B. Letaief,
  ``Q-learning-based adaptive power control in wireless {RF} energy harvesting
  heterogeneous networks,'' \emph{IEEE Syst. J.}, vol.~15, no.~2, pp.
  1861--1872, 2021.

\bibitem{10172220}
X.~Tian, K.~Xiong, R.~Zhang, P.~Fan, D.~Niyato, and K.~B. Letaief, ``Sum rate
  maximization in muti-cell muti-user networks: An inverse reinforcement
  learning-based approach,'' \emph{IEEE Wireless Commun. Lett.}, pp. 1--1,
  2023.

\bibitem{9575181}
R.~Zhang, K.~Xiong, Y.~Lu, B.~Gao, P.~Fan, and K.~B. Letaief, ``Joint
  coordinated beamforming and power splitting ratio optimization in {MU-MISO}
  {SWIPT}-enabled hetnets: {A} multi-agent {DDQN}-based approach,'' \emph{IEEE
  J. Sel. Areas Commun.}, vol.~40, no.~2, pp. 677--693, 2022.

\bibitem{mnih2015human}
V.~Mnih, K.~Kavukcuoglu, D.~Silver, A.~A. Rusu, J.~Veness, M.~G. Bellemare,
  A.~Graves, M.~Riedmiller, A.~K. Fidjeland, G.~Ostrovski \emph{et~al.},
  ``Human-level control through deep reinforcement learning,'' \emph{Nature},
  vol. 518, no. 7540, pp. 529--533, 2015.

\bibitem{schaul2015prioritized}
T.~Schaul, J.~Quan, I.~Antonoglou, and D.~Silver, ``Prioritized experience
  replay,'' \emph{arXiv preprint arXiv:1511.05952}, 2015.

\bibitem{hausknecht2015deep}
M.~Hausknecht and P.~Stone, ``Deep recurrent {Q}-learning for partially
  observable {MDPs},'' in \emph{AAAI Fall Symposium Series}, 2015.

\bibitem{10032267}
R.~Zhang, K.~Xiong, Y.~Lu, P.~Fan, D.~W.~K. Ng, and K.~B. Letaief, ``Energy
  efficiency maximization in {RIS}-assisted {SWIPT} networks with rsma: {A}
  {PPO}-based approach,'' \emph{IEEE J. Sel. Areas Commun.}, vol.~41, no.~5,
  pp. 1413--1430, 2023.

\bibitem{williams1992simple}
R.~J. Williams, ``Simple statistical gradient-following algorithms for
  connectionist reinforcement learning,'' \emph{Reinforcement Learning}, pp.
  5--32, 1992.

\bibitem{hessel2018rainbow}
M.~Hessel, J.~Modayil, H.~Van~Hasselt, T.~Schaul, G.~Ostrovski, W.~Dabney,
  D.~Horgan, B.~Piot, M.~Azar, and D.~Silver, ``Rainbow: {C}ombining
  improvements in deep reinforcement learning,'' in \emph{Proc. AAAI Conf.
  Artif. Intell.}, vol.~32, no.~1, 2018.

\bibitem{wang2021incorporating}
C.~Wang, L.~Liu, C.~Jiang, S.~Wang, P.~Zhang, and S.~Shen, ``Incorporating
  distributed {DRL} into storage resource optimization of space-air-ground
  integrated wireless communication network,'' \emph{IEEE J. Sel. Top. Signal
  Process.}, vol.~16, no.~3, pp. 434--446, Mar. 2021.

\bibitem{li2019deep}
Y.~Li, X.~Hu, Y.~Zhuang, Z.~Gao, P.~Zhang, and N.~El-Sheimy, ``Deep
  reinforcement learning ({DRL}): {A}nother perspective for unsupervised
  wireless localization,'' \emph{IEEE Internet Things J.}, vol.~7, no.~7, pp.
  6279--6287, Jul. 2019.

\bibitem{tang2022constructing}
J.~Tang, A.~Mihailovic, and H.~Aghvami, ``Constructing a {DRL} decision making
  scheme for multi-path routing in all-{IP} access network,'' in \emph{Proc.
  IEEE Global Commun. Conf.}\hskip 1em plus 0.5em minus 0.4em\relax IEEE, 2022,
  pp. 3623--3628.

\bibitem{10107766}
Y.~Zhang, Y.~Lu, R.~Zhang, B.~Ai, and D.~Niyato, ``Deep reinforcement learning
  for secrecy energy efficiency maximization in ris-assisted networks,''
  \emph{IEEE Trans. Veh. Technol.}, pp. 1--6, 2023.

\bibitem{ashraf2021optimizing}
N.~M. Ashraf, R.~R. Mostafa, R.~H. Sakr, and M.~Rashad, ``Optimizing
  hyperparameters of deep reinforcement learning for autonomous driving based
  on whale optimization algorithm,'' \emph{Plos one}, vol.~16, no.~6, p.
  e0252754, 2021.

\bibitem{fu2020d4rl}
J.~Fu, A.~Kumar, O.~Nachum, G.~Tucker, and S.~Levine, ``D4{RL}: {D}atasets for
  deep data-driven reinforcement learning,'' \emph{arXiv preprint
  arXiv:2004.07219}, 2020.

\bibitem{vargas2023expressiveness}
F.~Vargas, T.~Reu, and A.~Kerekes, ``Expressiveness remarks for denoising
  diffusion models and samplers,'' \emph{arXiv preprint arXiv:2305.09605},
  2023.

\bibitem{liu2016balanced}
R.~Liu, H.~Liu, D.~Kwak, Y.~Xiang, C.~Borcea, B.~Nath, and L.~Iftode,
  ``Balanced traffic routing: {D}esign, implementation, and evaluation,''
  \emph{Ad Hoc Networks}, vol.~37, pp. 14--28, 2016.

\bibitem{watson2021learning}
D.~Watson, W.~Chan, J.~Ho, and M.~Norouzi, ``Learning fast samplers for
  diffusion models by differentiating through sample quality,'' in \emph{Int.
  Conf. Learn. Represent.}, 2021.

\bibitem{hong2022improving}
S.~Hong, G.~Lee, W.~Jang, and S.~Kim, ``Improving sample quality of diffusion
  models using self-attention guidance,'' in \emph{Proc. IEEE Conf. Comput.
  Vis. Pattern Recognit.}, June 2023, pp. 7462--7471.

\bibitem{lyu2022accelerating}
Z.~Lyu, X.~Xu, C.~Yang, D.~Lin, and B.~Dai, ``Accelerating diffusion models via
  early stop of the diffusion process,'' \emph{arXiv preprint
  arXiv:2205.12524}, 2022.

\bibitem{dai2022psaccf}
M.~Dai, L.~Luo, J.~Ren, H.~Yu, and G.~Sun, ``{PSACCF}: {P}rioritized online
  slice admission control considering fairness in {5G/B5G} networks,''
  \emph{IEEE Trans. Netw. Sci. Eng.}, vol.~9, no.~6, pp. 4101--4114, Jun. 2022.

\bibitem{gu2017deep}
S.~Gu, E.~Holly, T.~Lillicrap, and S.~Levine, ``Deep reinforcement learning for
  robotic manipulation with asynchronous off-policy updates,'' in \emph{Proc.
  IEEE Int. Conf. Robot. Autom.}, 2017, pp. 3389--3396.

\bibitem{khaneja2005optimal}
N.~Khaneja, T.~Reiss, C.~Kehlet, T.~Schulte-Herbr{\"u}ggen, and S.~J. Glaser,
  ``Optimal control of coupled spin dynamics: design of {NMR} pulse sequences
  by gradient ascent algorithms,'' \emph{J. Magn. Reson.}, vol. 172, no.~2, pp.
  296--305, Feb. 2005.

\bibitem{zhao2020domain}
S.~Zhao, M.~Gong, T.~Liu, H.~Fu, and D.~Tao, ``Domain generalization via
  entropy regularization,'' \emph{Adv. Neural Inf. Process. Syst.}, vol.~33,
  pp. 16\,096--16\,107, 2020.

\bibitem{tesauro1995temporal}
G.~Tesauro \emph{et~al.}, ``Temporal difference learning and {TD}-{G}ammon,''
  \emph{Commun. ACM}, vol.~38, no.~3, pp. 58--68, 1995.

\bibitem{pmlr-v119-cobbe20a}
K.~Cobbe, C.~Hesse, J.~Hilton, and J.~Schulman, ``Leveraging procedural
  generation to benchmark reinforcement learning,'' in \emph{Proc. Int. Conf.
  Mach. Learn.}, vol. 119.\hskip 1em plus 0.5em minus 0.4em\relax PMLR, 13--18
  Jul 2020, pp. 2048--2056.

\bibitem{rl-zoo}
A.~Raffin, ``Rl baselines zoo,''
  \url{https://github.com/araffin/rl-baselines-zoo}, 2018.

\bibitem{liu2023blockchainempowered}
Y.~Liu, H.~Du, D.~Niyato, J.~Kang, Z.~Xiong, C.~Miao, Xuemin, Shen, and
  A.~Jamalipour, ``Blockchain-empowered lifecycle management for
  {AI}-{G}enerated {C}ontent ({AIGC}) products in edge networks,'' \emph{IEEE
  Wireless Commun.}, to appear, 2023.

\bibitem{crowdout}
E.~Aubry, T.~Silverston, A.~Lahmadi, and O.~Festor, ``Crowdout: A mobile
  crowdsourcing service for road safety in digital cities,'' in \emph{2014 IEEE
  International Conference on Pervasive Computing and Communication Workshops
  (PERCOM WORKSHOPS)}, 2014, pp. 86--91.

\bibitem{8664132}
Y.~Liu, K.~Wang, Y.~Lin, and W.~Xu, ``$\mathsf{LightChain}$: {A} lightweight
  blockchain system for industrial {I}nternet of {T}hings,'' \emph{IEEE Trans.
  Indust. Inform.}, vol.~15, no.~6, pp. 3571--3581, June. 2019.

\bibitem{zeng2021comprehensive}
R.~Zeng, C.~Zeng, X.~Wang, B.~Li, and X.~Chu, ``A comprehensive survey of
  incentive mechanism for federated learning,'' \emph{arXiv preprint
  arXiv:2106.15406}, 2021.

\bibitem{yang2013coping}
D.~Yang, G.~Xue, J.~Zhang, A.~Richa, and X.~Fang, ``Coping with a smart jammer
  in wireless networks: {A} stackelberg game approach,'' \emph{IEEE Trans.
  Wireless Commun.}, vol.~12, no.~8, pp. 4038--4047, Aug. 2013.

\bibitem{9773059}
X.~Chen, Y.~Deng, G.~Zhu, D.~Wang, and Y.~Fang, ``From resource auction to
  service auction: {A}n auction paradigm shift in wireless networks,''
  \emph{IEEE Wirel. Commun.}, vol.~29, no.~2, pp. 185--191, Apr. 2022.

\bibitem{contractkang}
J.~Kang, Z.~Xiong, D.~Niyato, D.~Ye, D.~I. Kim, and J.~Zhao, ``Toward secure
  blockchain-enabled internet of vehicles: Optimizing consensus management
  using reputation and contract theory,'' \emph{IEEE Trans. Veh. Technol.},
  vol.~68, no.~3, pp. 2906--2920, 2019.

\bibitem{complexity}
A.~Kumar, ``Model complexity,''
  \url{https://vitalflux.com/model-complexity-overfitting-in-machine-learning/}.

\bibitem{performance}
Microsoft, ``The relationshio between model size and performance,''
  \url{https://learn.microsoft.com/en-us/semantic-kernel/prompt-engineering/llm-models}.

\bibitem{lin2023semantic}
Y.~Lin, Z.~Gao, H.~Du, D.~Niyato, J.~Kang, R.~Deng, and X.~S. Shen, ``{A
  unified blockchain-semantic framework for wireless edge intelligence enabled
  web 3.0},'' \emph{IEEE Wirel Commun}, 2023.

\bibitem{du2023semantic}
H.~Du, J.~Wang, D.~Niyato, J.~Kang, Z.~Xiong, J.~Zhang, and X.~Shen, ``Semantic
  communications for wireless sensing: {RIS}-aided encoding and self-supervised
  decoding,'' \emph{IEEE J. Sel. Areas Commun.}, to appear, 2023.

\bibitem{liang2023generative}
C.~Liang, H.~Du, Y.~Sun, D.~Niyato, J.~Kang, D.~Zhao, and M.~A. Imran,
  ``Generative {AI}-driven semantic communication networks: Architecture,
  technologies and applications,'' \emph{arXiv preprint arXiv:2401.00124},
  2023.

\bibitem{du2022rethinking}
H.~Du, J.~Wang, D.~Niyato, J.~Kang, Z.~Xiong, M.~Guizani, and D.~I. Kim,
  ``Rethinking wireless communication security in semantic internet of
  things,'' \emph{IEEE Wireless Commun. Mag.}, to appear, 2023.

\bibitem{kang2022personalized}
J.~Kang, H.~Du, Z.~Li, Z.~Xiong, S.~Ma, D.~Niyato, and Y.~Li, ``Personalized
  saliency in task-oriented semantic communications: {I}mage transmission and
  performance analysis,'' \emph{IEEE J. Sel. Areas Commun.}, vol.~41, no.~1,
  pp. 186--201, Jan. 2022.

\bibitem{van2023generative}
N.~Van~Huynh, J.~Wang, H.~Du, D.~T. Hoang, D.~Niyato, D.~N. Nguyen, D.~I. Kim,
  and K.~B. Letaief, ``Generative {AI} for physical layer communications: {A}
  survey,'' \emph{IEEE Trans. on Cogn. Commun. Netw.}, to appear, 2024.

\bibitem{lin2023commag}
Y.~Lin, Z.~Gao, Y.~Tu, H.~Du, D.~Niyato, J.~Kang, and H.~Yang, ``{A
  Blockchain-based Semantic Exchange Framework for Web 3.0 toward Participatory
  Economy},'' \emph{IEEE Commun Mag}, 2023.

\bibitem{lin2023blockchain}
Y.~Lin, H.~Du, D.~Niyato, J.~Nie, J.~Zhang, Y.~Cheng, and Z.~Yang,
  ``{Blockchain-aided secure semantic communication for AI-generated content in
  metaverse},'' \emph{IEEE Open J. Comput. Soc.}, vol.~4, pp. 72--83, 2023.

\bibitem{zhang2022deep}
H.~Zhang, S.~Shao, M.~Tao, X.~Bi, and K.~B. Letaief, ``Deep learning-enabled
  semantic communication systems with task-unaware transmitter and dynamic
  data,'' \emph{IEEE J. Sel. Areas Commun.}, vol.~41, no.~1, pp. 170--185,
  2022.

\bibitem{alemi2016deep}
A.~A. Alemi, I.~Fischer, J.~V. Dillon, and K.~Murphy, ``Deep variational
  information bottleneck,'' \emph{Proc. Int. Conf. Learn. Represent.}, 2016.

\bibitem{wang2023unified}
J.~Wang, H.~Du, D.~Niyato, J.~Kang, Z.~Xiong, D.~Rajan, S.~Mao \emph{et~al.},
  ``A unified framework for guiding generative {AI} with wireless perception in
  resource constrained mobile edge networks,'' \emph{IEEE Trans. Mobile
  Comput.}, to appear, 2024.

\bibitem{du2023attention}
H.~Du, J.~Liu, D.~Niyato, J.~Kang, Z.~Xiong, J.~Zhang, and D.~I. Kim,
  ``Attention-aware resource allocation and {QoE} analysis for metaverse
  {xURLLC} services,'' \emph{IEEE J. Sel. Areas Commun.}, to appear, 2023.

\bibitem{liu2019deep}
Y.~Liu, H.~Yu, S.~Xie, and Y.~Zhang, ``Deep reinforcement learning for
  offloading and resource allocation in vehicle edge computing and networks,''
  \emph{IEEE Trans. Veh. Technol.}, vol.~68, no.~11, pp. 11\,158--11\,168,
  2019.

\bibitem{yan2022qoe}
L.~Yan, Z.~Qin, R.~Zhang, Y.~Li, and G.~Y. Li, ``Qoe-aware resource allocation
  for semantic communication networks,'' in \emph{Proc. IEEE Global Commun.
  Conf.}\hskip 1em plus 0.5em minus 0.4em\relax IEEE, 2022, pp. 3272--3277.

\bibitem{zhou2020evolutionary}
H.~Zhou, W.~Xu, J.~Chen, and W.~Wang, ``Evolutionary {V2X} technologies toward
  the internet of vehicles: {C}hallenges and opportunities,'' \emph{Proc.
  IEEE}, vol. 108, no.~2, pp. 308--323, Feb. 2020.

\bibitem{zhang2023generative}
R.~Zhang, K.~Xiong, H.~Du, D.~Niyato, J.~Kang, X.~Shen, and H.~V. Poor,
  ``Generative {AI}-enabled vehicular networks: {F}undamentals, framework, and
  case study,'' \emph{IEEE Netw.}, to appear, 2024.

\bibitem{9136587}
W.~Duan, J.~Gu, M.~Wen, G.~Zhang, Y.~Ji, and S.~Mumtaz, ``Emerging technologies
  for 5{G}-{IoV} networks: {A}pplications, trends and opportunities,''
  \emph{IEEE Net.}, vol.~34, no.~5, pp. 283--289, 2020.

\bibitem{8967260}
H.~Zhou, W.~Xu, J.~Chen, and W.~Wang, ``Evolutionary v2x technologies toward
  the {I}nternet of {V}ehicles: {C}hallenges and opportunities,'' \emph{Proc.
  IEEE}, vol. 108, no.~2, pp. 308--323, 2020.

\bibitem{7913583}
L.~Liang, G.~Y. Li, and W.~Xu, ``Resource allocation for {D}2{D}-enabled
  vehicular communications,'' \emph{IEEE Trans. Commun.}, vol.~65, no.~7, pp.
  3186--3197, 2017.

\bibitem{Huang_2023_CVPR}
S.~Huang, Z.~Wang, P.~Li, B.~Jia, T.~Liu, Y.~Zhu, W.~Liang, and S.-C. Zhu,
  ``Diffusion-based generation, optimization, and planning in 3{D} scenes,'' in
  \emph{Proc. IEEE Conf. Comput. Vis. Pattern Recognit.}, June 2023, pp.
  16\,750--16\,761.

\bibitem{8792382}
L.~Liang, H.~Ye, and G.~Y. Li, ``Spectrum sharing in vehicular networks based
  on multi-agent reinforcement learning,'' \emph{IEEE J. Sel. Areas Commun.},
  vol.~37, no.~10, pp. 2282--2292, 2019.

\bibitem{dovelos2021channel}
K.~Dovelos, M.~Matthaiou, H.~Q. Ngo, and B.~Bellalta, ``Channel estimation and
  hybrid combining for wideband terahertz massive mimo systems,'' \emph{IEEE J.
  Sel. Areas Commun.}, vol.~39, no.~6, pp. 1604--1620, Jun. 2021.

\bibitem{liu2014channel}
Y.~Liu, Z.~Tan, H.~Hu, L.~J. Cimini, and G.~Y. Li, ``Channel estimation for
  {OFDM},'' \emph{IEEE Commun. Surv. Tutor.}, vol.~16, no.~4, pp. 1891--1908,
  Apr. 2014.

\bibitem{nawaz2012superimposed}
S.~J. Nawaz, K.~I. Ahmed, M.~N. Patwary, and N.~M. Khan, ``Superimposed
  training-based compressed sensing of sparse multipath channels,'' \emph{IET
  Commun.}, vol.~6, no.~18, pp. 3150--3156, 2012.

\bibitem{liao2019chanestnet}
Y.~Liao, Y.~Hua, X.~Dai, H.~Yao, and X.~Yang, ``Chanestnet: A deep learning
  based channel estimation for high-speed scenarios,'' in \emph{ICC 2019-2019
  IEEE international conference on communications (ICC)}.\hskip 1em plus 0.5em
  minus 0.4em\relax IEEE, 2019, pp. 1--6.

\bibitem{arvinte2022mimo}
M.~Arvinte and J.~I. Tamir, ``{MIMO} channel estimation using score-based
  generative models,'' \emph{IEEE Trans. Wireless Commun.}, 2022.

\bibitem{song2020improved}
Y.~Song and S.~Ermon, ``Improved techniques for training score-based generative
  models,'' \emph{Adv. Neural Inf. Process. Syst.}, vol.~33, pp.
  12\,438--12\,448, 2020.

\bibitem{lin2017refinenet}
G.~Lin, A.~Milan, C.~Shen, and I.~Reid, ``Refinenet: Multi-path refinement
  networks for high-resolution semantic segmentation,'' in \emph{Proceedings of
  the IEEE conference on computer vision and pattern recognition}, 2017, pp.
  1925--1934.

\bibitem{balevi2020high}
E.~Balevi, A.~Doshi, A.~Jalal, A.~Dimakis, and J.~G. Andrews, ``High
  dimensional channel estimation using deep generative networks,'' \emph{IEEE
  J. Sel. Areas Commun.}, vol.~39, no.~1, pp. 18--30, 2020.

\bibitem{schniter2014channel}
P.~Schniter and A.~Sayeed, ``Channel estimation and precoder design for
  millimeter-wave communications: The sparse way,'' in \emph{2014 48th Asilomar
  conference on signals, systems and computers}.\hskip 1em plus 0.5em minus
  0.4em\relax IEEE, 2014, pp. 273--277.

\bibitem{jalal2021robust}
A.~Jalal, M.~Arvinte, G.~Daras, E.~Price, A.~G. Dimakis, and J.~Tamir, ``Robust
  compressed sensing mri with deep generative priors,'' \emph{Adv. Neural Inf.
  Process. Syst.}, vol.~34, pp. 14\,938--14\,954, 2021.

\bibitem{biglieri2005coding}
E.~Biglieri, \emph{Coding for wireless channels}.\hskip 1em plus 0.5em minus
  0.4em\relax Springer Science \& Business Media, 2005.

\bibitem{choukroun2022error}
Y.~Choukroun and L.~Wolf, ``Error correction code transformer,'' \emph{Adv.
  Neural Inf. Process. Syst.}, vol.~35, pp. 38\,695--38\,705, 2022.

\bibitem{choukroun2022denoising}
------, ``Denoising diffusion error correction codes,'' in \emph{Proc. Int.
  Conf. Mach. Learn.}, Jul. 2023.

\bibitem{bao2024improving}
Q.~Bao, Z.~Hui, R.~Zhu, P.~Ren, X.~Xie, and W.~Yang, ``Improving
  diffusion-based image restoration with error contraction and error
  correction,'' in \emph{Proc. AAAI Conf. Artif. Intell.}, vol.~38, no.~2,
  2024, pp. 756--764.

\bibitem{wu2024cddm}
T.~Wu, Z.~Chen, D.~He, L.~Qian, Y.~Xu, M.~Tao, and W.~Zhang, ``{CDDM}: Channel
  denoising diffusion models for wireless semantic communications,'' \emph{IEEE
  Trans. Wireless Commun.}, to appear, 2024.

\bibitem{kim2023learning}
M.~Kim, R.~Fritschek, and R.~F. Schaefer, ``Learning end-to-end channel coding
  with diffusion models,'' in \emph{International ITG Workshop on Smart
  Antennas and 13th Conference on Systems, Communications, and Coding}, 2023,
  pp. 1--6.

\bibitem{cheng2018air}
N.~Cheng, W.~Xu, W.~Shi, Y.~Zhou, N.~Lu, H.~Zhou, and X.~Shen, ``Air-ground
  integrated mobile edge networks: {A}rchitecture, challenges, and
  opportunities,'' \emph{IEEE Commun. Mag.}, vol.~56, no.~8, pp. 26--32, Aug.
  2018.

\bibitem{cao2021hap}
X.~Cao, B.~Yang, C.~Yuen, and Z.~Han, ``Hap-reserved communications in
  space-air-ground integrated networks,'' \emph{IEEE Trans. Veh. Tech.},
  vol.~70, no.~8, pp. 8286--8291, Aug. 2021.

\bibitem{du2022performance}
H.~Du, D.~Niyato, Y.-A. Xie, Y.~Cheng, J.~Kang, and D.~I. Kim, ``Performance
  analysis and optimization for jammer-aided multiantenna {UAV} covert
  communication,'' \emph{IEEE J. Sel. Areas Commun.}, vol.~40, no.~10, pp.
  2962--2979, Oct. 2022.

\bibitem{li2022age}
D.~Li, S.~Wu, J.~Jiao, N.~Zhang, and Q.~Zhang, ``Age-oriented transmission
  protocol design in space-air-ground integrated networks,'' \emph{IEEE Trans.
  Wireless Commun.}, vol.~21, no.~7, pp. 5573--5585, Jul. 2022.

\bibitem{jia2021toward}
Z.~Jia, M.~Sheng, J.~Li, and Z.~Han, ``Toward data collection and transmission
  in {6G} space-air-ground integrated networks: {C}ooperative {HAP} and {LEO}
  satellite schemes,'' \emph{IEEE Internet Things J.}, vol.~9, no.~13, pp.
  10\,516--10\,528, Sept. 2021.

\bibitem{zhang2023generativeSAGIN}
R.~Zhang, H.~Du, D.~Niyato, J.~Kang, Z.~Xiong, A.~Jamalipour, P.~Zhang, and
  D.~I. Kim, ``Generative {AI} for space-air-ground integrated networks
  ({SAGIN}),'' \emph{arXiv preprint arXiv:2311.06523}, 2023.

\bibitem{cui2022space}
H.~Cui, J.~Zhang, Y.~Geng, Z.~Xiao, T.~Sun, N.~Zhang, J.~Liu, Q.~Wu, and
  X.~Cao, ``Space-air-ground integrated network ({SAGIN}) for {6G}:
  {R}equirements, architecture and challenges,'' \emph{China Commun.}, vol.~19,
  no.~2, pp. 90--108, Feb. 2022.

\bibitem{cheng2020comprehensive}
N.~Cheng, W.~Quan, W.~Shi, H.~Wu, Q.~Ye, H.~Zhou, W.~Zhuang, X.~Shen, and
  B.~Bai, ``A comprehensive simulation platform for space-air-ground integrated
  network,'' \emph{IEEE Wireless Commun.}, vol.~27, no.~1, pp. 178--185, Jan.
  2020.

\bibitem{liu2018space}
J.~Liu, Y.~Shi, Z.~M. Fadlullah, and N.~Kato, ``Space-air-ground integrated
  network: {A} survey,'' \emph{IEEE Commun. Surveys Tuts.}, vol.~20, no.~4, pp.
  2714--2741, Apr. 2018.

\bibitem{ye2020space}
J.~Ye, S.~Dang, B.~Shihada, and M.-S. Alouini, ``Space-air-ground integrated
  networks: {O}utage performance analysis,'' \emph{IEEE Trans. Wireless
  Commun.}, vol.~19, no.~12, pp. 7897--7912, Dec. 2020.

\bibitem{10058144}
W.~Mao, K.~Xiong, Y.~Lu, P.~Fan, and Z.~Ding, ``Energy consumption minimization
  in secure multi-antenna {UAV}-assisted {MEC} networks with channel
  uncertainty,'' \emph{IEEE Trans. Wireless Commun.}, pp. 1--1, 2023.

\bibitem{wang2024tutorial}
Z.~Wang, J.~Zhang, H.~Du, D.~Niyato, S.~Cui, B.~Ai, M.~Debbah, K.~B. Letaief,
  and H.~V. Poor, ``A tutorial on extremely large-scale {MIMO} for {6G}:
  Fundamentals, signal processing, and applications,'' \emph{IEEE Commun. Surv.
  Tutor.}, 2024.

\bibitem{wang2023extremely}
Z.~Wang, J.~Zhang, H.~Du, E.~Wei, B.~Ai, D.~Niyato, and M.~Debbah, ``Extremely
  large-scale {MIMO}: {F}undamentals, challenges, solutions, and future
  directions,'' \emph{IEEE Wireless Commun.}, 2023.

\bibitem{du2021millimeter}
H.~Du, J.~Zhang, J.~Cheng, and B.~Ai, ``Millimeter wave communications with
  reconfigurable intelligent surfaces: {P}erformance analysis and
  optimization,'' \emph{IEEE Trans. Commun.}, vol.~69, no.~4, pp. 2752--2768,
  Apr. 2021.

\bibitem{wang2022uplink}
Z.~Wang, J.~Zhang, B.~Ai, C.~Yuen, and M.~Debbah, ``Uplink performance of
  cell-free massive {MIMO} with multi-antenna users over jointly-correlated
  {R}ayleigh fading channels,'' \emph{IEEE Trans. Wireless Commun.}, vol.~21,
  no.~9, pp. 7391--7406, Sept. 2022.

\bibitem{du2022performancethz}
H.~Du, J.~Zhang, K.~Guan, D.~Niyato, H.~Jiao, Z.~Wang, and T.~K{\"u}rner,
  ``Performance and optimization of reconfigurable intelligent surface aided
  {THz} communications,'' \emph{IEEE Trans. Commun.}, vol.~70, no.~5, pp.
  3575--3593, May 2022.

\bibitem{wang2023uplink}
Z.~Wang, J.~Zhang, H.~Q. Ngo, B.~Ai, and M.~Debbah, ``Uplink precoding design
  for cell-free massive {MIMO} with iteratively weighted {MMSE},'' \emph{IEEE
  Trans. Commun.}, vol.~71, no.~3, pp. 1646--1664, Mar. 2023.

\bibitem{cui2021integrating}
Y.~Cui, F.~Liu, X.~Jing, and J.~Mu, ``Integrating sensing and communications
  for ubiquitous {IoT}: {A}pplications, trends, and challenges,'' \emph{IEEE
  Netw.}, vol.~35, no.~5, pp. 158--167, May. 2021.

\bibitem{wang2020twpalo}
J.~Wang, Z.~Tian, X.~Yang, and M.~Zhou, ``{TWP}alo: {T}hrough-the-wall passive
  localization of moving human with {Wi-Fi},'' \emph{Computer Commun.}, vol.
  157, pp. 284--297, 2020.

\bibitem{njima2021indoor}
W.~Njima, M.~Chafii, A.~Chorti, R.~M. Shubair, and H.~V. Poor, ``Indoor
  localization using data augmentation via selective generative adversarial
  networks,'' \emph{IEEE Access}, vol.~9, pp. 98\,337--98\,347, 2021.

\bibitem{chen2020fido}
X.~Chen, H.~Li, C.~Zhou, X.~Liu, D.~Wu, and G.~Dudek, ``Fido: Ubiquitous
  fine-grained wifi-based localization for unlabelled users via domain
  adaptation,'' in \emph{Proc. Web Conf.}, 2020, pp. 23--33.

\bibitem{xiao2019csigan}
C.~Xiao, D.~Han, Y.~Ma, and Z.~Qin, ``{CsiGAN}: {R}obust channel state
  information-based activity recognition with gans,'' \emph{IEEE Internet
  Things J.}, vol.~6, no.~6, pp. 10\,191--10\,204, 2019.

\bibitem{new2023fluid}
W.~K. New, K.-K. Wong, H.~Xu, K.-F. Tong, and C.-B. Chae, ``Fluid antenna
  system: {N}ew insights on outage probability and diversity gain,'' \emph{IEEE
  Trans. Wireless Commun.}, to appear, 2023.

\bibitem{khammassi2023new}
M.~Khammassi, A.~Kammoun, and M.-S. Alouini, ``A new analytical approximation
  of the fluid antenna system channel,'' \emph{IEEE Trans. Wireless Commun.},
  to appear, 2023.

\bibitem{shojaeifard2022mimo}
A.~Shojaeifard, K.-K. Wong, K.-F. Tong, Z.~Chu, A.~Mourad, A.~Haghighat,
  I.~Hemadeh, N.~T. Nguyen, V.~Tapio, and M.~Juntti, ``{MIMO} evolution beyond
  5g through reconfigurable intelligent surfaces and fluid antenna systems,''
  \emph{Proc. IEEE}, vol. 110, no.~9, pp. 1244--1265, Sept. 2022.

\bibitem{tlebaldiyeva2022enhancing}
L.~Tlebaldiyeva, G.~Nauryzbayev, S.~Arzykulov, A.~Eltawil, and T.~Tsiftsis,
  ``Enhancing {QoS} through fluid antenna systems over correlated nakagami-m
  fading channels,'' in \emph{Proc. IEEE Wireless Commun. Netw. Conf.}\hskip
  1em plus 0.5em minus 0.4em\relax IEEE, 2022, pp. 78--83.

\bibitem{10038824}
Y.~Zhao, F.~Zhou, L.~Feng, W.~Li, and P.~Yu, ``Madrl-based 3d deployment and
  user association of cooperative mmwave aerial base stations for capacity
  enhancement,'' \emph{Chinese J. Electron.}, vol.~32, no.~2, pp. 283--294,
  2023.

\bibitem{lin2023drl}
Y.~Lin, Z.~Gao, H.~Du, J.~Kang, D.~Niyato, Q.~Wang, J.~Ruan, and S.~Wan,
  ``{DRL}-based adaptive sharding for blockchain-based federated learning,''
  \emph{IEEE Trans. Commun.}, 2023.

\end{thebibliography}
\end{document}